\documentclass[%
 aip,
 amsmath,amssymb,
preprint,%
]{revtex4-1}
\usepackage{graphicx}
\usepackage[export]{adjustbox}
\usepackage{caption}

\usepackage{dcolumn}
\usepackage{bm}
\usepackage[version=4]{mhchem}
\usepackage{color}
\usepackage[utf8]{inputenc}
\usepackage[decimalsymbol=comma,expproduct=\cdot]{siunitx}
\usepackage[utf8]{inputenc}
\usepackage{amsmath}
\usepackage{amssymb}
\usepackage{xr-hyper}
\usepackage{hyperref}
\usepackage[retainorgcmds]{IEEEtrantools}
\usepackage[mathscr]{euscript}
\usepackage{mathtools}
\usepackage{mhchem}
\usepackage{siunitx}
\usepackage{enumitem}
\usepackage{array,booktabs}
\newcolumntype{C}{>{$\displaystyle}c<{$}} 
\usepackage{subfigure}

\newcommand{\eqnref}[1]{(\ref{#1})}

\begin{document}
\preprint{}
\title[Jo \textit{et al.}]{Rovibrational Internal Energy Transfer and Dissociation of High-Temperature Oxygen Mixture}
\author{Sung Min Jo}
\affiliation{Center for Hypersonics and Entry Systems Studies (CHESS), \\ University of Illinois at Urbana-Champaign, Urbana, IL 61801, USA}
\author{Simone Venturi}
\affiliation{Center for Hypersonics and Entry Systems Studies (CHESS), \\ University of Illinois at Urbana-Champaign, Urbana, IL 61801, USA}
\author{Jae Gang Kim}
\affiliation{Department of Aerospace System Engineering, Sejong University, Seoul 05006, Republic of Korea}
\author{Marco Panesi}
\email{mpanesi@illinois.edu}
\affiliation{Center for Hypersonics and Entry Systems Studies (CHESS), \\ University of Illinois at Urbana-Champaign, Urbana, IL 61801, USA}
\date{\today}
\begin{abstract}
This work constructs a rovibrational state-to-state model for the $\text{O}_2$+$\text{O}_2$ system leveraging high-fidelity potential energy surfaces and quasi-classical trajectory calculations. The model is used to investigate internal energy transfer and non-equilibrium reactive processes in dissociating environment using a master equation approach, whereby the kinetics of each internal rovibrational state is explicitly computed. To cope with the exponentially large number of elementary processes that characterize reactive bimolecular collisions, the internal states of the collision partner are assumed to follow a Boltzmann distribution at a prescribed internal temperature. This procedure makes the problem tractable, reducing the computational cost to a comparable scale with the $\text{O}_2$+O system. The constructed rovibrational-specific kinetic database covers the temperature range of 7500-\SI{20000}{\kelvin}. The reaction rate coefficients included in the database are parameterized in the function of kinetic and internal temperatures. The analysis of the energy transfer and dissociation process in isochoric and isothermal conditions reveals that significant departures from the equilibrium Boltzmann distribution occur during the energy transfer and dissociation phase. Comparing the population distribution of the $\text{O}_2$ molecules against the $\text{O}_2$+O case demonstrates a more significant extent of non-equilibrium characterized by a more diffuse distribution whereby the vibrational strands are more clearly identifiable. This is partly due to a less efficient mixing of the rovibrational states, which results in more diffuse rovibrational distributions in the quasi-steady-state distribution of $\text{O}_2$+$\text{O}_2$.
The master equation analysis for the combined $\text{O}_3$+$\text{O}_4$ system reveals that the $\text{O}_2$+$\text{O}_2$ governs the early stage of energy transfer, while the $\text{O}_2$+O takes control of the dissociation dynamics. The findings achieved by the present work will provide strong physical foundations exploitable to construct an improved reduced-order model for oxygen chemistry.
\end{abstract}
\maketitle

\section{Introduction}\label{sec:intro}
Behind strong shock waves, diatom-diatom collisions are essential as they initiate the dissociation process and determine the chemical production rates of atomic species. Once a significant number of atoms are created, atom-diatom processes govern the thermochemical non-equilibrium state of the gas \cite{PANESI_2013_BOXRVC,Venturi2020,Jo_N2O,Sharma_HCN}. In hypersonic flows within Earth's atmosphere, $\text{O}_2$+$\text{O}_2$ \cite{Jo_2022_Scitech,GROVER_O3O4,Streicher_2021}, $\text{O}_2$+$\text{N}_2$ \cite{Torres_2022_Scitech,Andrienko_O2N2}, and $\text{N}_2$+$\text{N}_2$ \cite{MACDONALD_MEQCT} are the relevant diatom-diatom systems. Among them, $\text{O}_2$+$\text{O}_2$ interaction provides the focus of the present work, owing to its dominant contribution to the early stage of thermochemical relaxation, justified by the weak bond energy of molecular oxygen.

The $\text{O}_2$+$\text{O}_2$ system has been investigated extensively, both experimentally\cite{Ibraguimova_2013,Streicher_2021} and numerically\cite{LINODASILVA201228,ANDRIENKO201774,Varandas_O4_Chapter,GROVER_O3O4,Paukku_O4_1,Paukku_O4_2,Jo_2022_Scitech}. Recently, shock tube experiments were carried out by Ibraguimova \emph{et al.} \cite{Ibraguimova_2013} and Streicher \emph{et al.} \cite{Streicher_2021} to measure vibrational relaxation times and non-equilibrium dissociation rate coefficients. Those measurements provided are mainly used as a reference for validating models. Theoretical calculations \cite{LINODASILVA201228,ANDRIENKO201774,Varandas_O4_Chapter,GROVER_O3O4,Paukku_O4_1,Paukku_O4_2,Jo_2022_Scitech} of the $\text{O}_2$+$\text{O}_2$ system include approximate and high-fidelity estimates. Lino da Silva \emph{et al.} \cite{LINODASILVA201228} proposed a vibrational-specific (VS) state-to-state (StS) model constructed using the Forced Harmonic Oscillator (FHO) theory and one-dimensional Morse inter-molecular potential. Andrienko and Boyd \cite{ANDRIENKO201774} devised another VS model using the quasi-classical trajectory (QCT) method. In their work \cite{ANDRIENKO201774}, a double many-body expansion (DMBE) potential energy surface (PES) developed by Varandas and Pais \cite{Varandas_O4_Chapter} was used to describe molecular interactions. The major shortcoming of these studies \cite{LINODASILVA201228,ANDRIENKO201774} is the crude treatment of rotational degrees of freedom, assumed to be in equilibrium at the translational energy. The previous works \cite{Jo_2021,PANESI_2013_BOXRVC,Venturi2020} demonstrated that the rotational non-equilibrium is essential to model hypersonic shock layers accurately. Grover \emph{et al.} \cite{GROVER_O3O4} performed the direct molecular simulation (DMS) using the \emph{ab-initio} PESs proposed by Paukku \emph{et al.} \cite{Paukku_O4_1,Paukku_O4_2}. The results of Grover \emph{et al.} accounted for the rotational non-equilibrium and showed good agreement with the available experimental data. Although the DMS simulation captured general aspects of non-equilibrium kinetics in the $\text{O}_2$+$\text{O}_2$ system, the method cannot capture the details of the rovibrational-specific population dynamics, which is still unknown to the author's best knowledge. Understanding the rovibrational distributions and their dynamics in time is crucial to developing an advanced reduced-order model for non-equilibrium thermochemistry \cite{SAHAI_ADAPTIVE,Venturi2020}. 

The numerical simulation of the rovibrational-specific molecular interactions in the $\text{O}_2$+$\text{O}_2$ system is computationally prohibitive. Several efforts have been made to model the rovibrational energy transfer and non-equilibrium dissociation in 4-atom systems. Kim and Boyd \cite{Kim_POF_2012} employed an ordinary Kriging model to estimate fully rovibrational-specific StS rate coefficients in the $\text{H}_2$+$\text{H}_2$. In another paper, Kim \emph{et al.} \cite{Kim_JTHT_2010} proposed an approximated treatment of the internal states, assumed to be populated according to a Boltzmann distribution of one of the molecules in the $\text{H}_2$+$\text{H}_2$ system to reduce the computational complexity. Macdonald \emph{et al.} \cite{MACDONALD_MEQCT} developed a coarse-grain QCT (CG-QCT) method to investigate non-equilibrium rovibrational energy transfer and dissociation in $\text{N}_2$+$\text{N}_2$ by relying on a binning strategy among internal states. Following Kim \emph{et al.} \cite{Kim_JTHT_2010}, the present work proposes to model rovibrational energy transfer and dissociation in the $\text{O}_2$+$\text{O}_2$ system, assuming a thermal distribution for a collision partner.

Toward this end, the present study proposes a hybrid strategy to model the rovibrational-specific interactions in the $\text{O}_2$+$\text{O}_2$ system by extending the modeling concepts by Kim \emph{et al.} \cite{Kim_JTHT_2010}. This procedure imposes a single energy bin on the colliding molecule that is Boltzmann distributed at a prescribed internal temperature. At the same time, the target $\text{O}_2$ contains complete information on the rovibrational states. The present model is built on the framework consisting of the QCT method and master equation analyses. The QCT simulations employ the \emph{ab-initio} PESs by Paukku \emph{et al.} \cite{Paukku_O4_1,Paukku_O4_2} to construct a rovibrational-specific StS kinetic database. Then the master equation analysis was performed in the kinetic temperature range of 7500-\SI{20000}{\kelvin}. The rovibrational-specific population dynamics, energy transfer, and dissociation are thoroughly investigated to understand the fundamental physics behind the $\text{O}_2$+$\text{O}_2$ collisional system. A direct comparison with a VS model, which this work has also constructed, is made. In addition, the underlying physics in the rovibrational distributions of the $\text{O}_2$+$\text{O}_2$ system is directly compared with that of $\text{O}_2$+O to make clear differences between them. Finally, the master equation study is extended to the combined $\text{O}_3$ and $\text{O}_4$ systems (\emph{i.e.}, $\text{O}_2$+O and $\text{O}_2$+$\text{O}_2$) to identify their relative contributions.

This paper is organized as follows: The cost-effective modeling strategy to evaluate rovibrational distributions in the $\text{O}_2$+$\text{O}_2$ system is proposed in Sec. \ref{sec:phys}. In Sec. \ref{sec:kinetics}, the constructed StS kinetic database is presented. Then the rovibrational energy transfer and non-equilibrium dissociation are discussed in Sec. \ref{sec:energy} and Sec. \ref{sec:all}, respectively. Section \ref{sec:O3O4} provides the master equation results for the combined $\text{O}_3$ and $\text{O}_4$ systems. In Sec. \ref{sec:Comp_Literature}, the present macroscopic results are compared with existing data from the literature. Finally, Sec. \ref{sec:conc} provides the conclusions from the present study.

\section{Physical modeling}\label{sec:phys}
\subsection{Modeling of rovibrational energy transfer and dissociation of $\text{O}_2\left(\text{X}^3\Sigma_g^-\right)$+$\text{O}_2\left(\text{X}^3\Sigma_g^-\right)$ system}\label{sec:formulation}
This work assumes that all species relevant to the $\text{O}_2$+$\text{O}_2$ system remain in their electronic ground state during the collisional interaction. Then the rovibrational energy transfer and dissociation processes of the $\text{O}_2$+$\text{O}_2$ system can be expressed as
\begin{IEEEeqnarray}{rCl}
	\text{O}_2(i)+\text{O}_2(m) &\leftrightarrow& \text{O}_2(j)+\text{O}_2(l), \label{eq:O4_EnergyTransfer_StS} \\
	\text{O}_2(i)+\text{O}_2(m) &\leftrightarrow& \text{O}+\text{O}+\text{O}_2(l), \label{eq:O4_TargetDiss_StS} \\
	\text{O}_2(i)+\text{O}_2(m) &\leftrightarrow& \text{O}_2(j)+\text{O}+\text{O}, \label{eq:O4_ColliderDiss_StS} \\
	\text{O}_2(i)+\text{O}_2(m) &\leftrightarrow& \text{O}+\text{O}+\text{O}+\text{O}, \label{eq:O4_DoubleDiss_StS}
\end{IEEEeqnarray}
\noindent
where the indices $(i,\, m,\, j,\, l)$ denote the rovibrational states of $\text{O}_2\left(\text{X}^3\Sigma_g^-\right)$ that are stored by increasing energy in the set $\mathcal{I}_{\mathrm{O}_2}$. Each rovibrational level corresponds to a unique vibrational ($v$) and rotational ($J$) quantum number (\emph{e.g.}, $i{=}i(v, \, J)$). The reaction in Eq. \eqnref{eq:O4_EnergyTransfer_StS} includes both the inelastic and exchange kinetics. The processes in Eqs. \eqnref{eq:O4_TargetDiss_StS}-\eqnref{eq:O4_DoubleDiss_StS} contain dissociation from the direct and exchanged pairs. In the present master equation study, the double-dissociation reaction in Eq. \eqnref{eq:O4_DoubleDiss_StS} is neglected since it was found to have little influence for the conditions of present interest \cite{Andrienko_2021}. Then the system of master equations describing the evolution of the rovibrational states of $\text{O}_2$ and the concentration of O atom reads:
\begin{align}
	\frac{dn_i}{dt} &= \sum_j\sum_l\sum_m\left(-k_{im \rightarrow jl}n_in_m+k_{jl \rightarrow im}n_jn_l\right)\nonumber\\
	&+ \sum_l\sum_m\left(-k_{im \rightarrow cl}n_in_m+k_{cl \rightarrow im}n_{\text{O}}^2n_l\right)\nonumber\\
	&+ \sum_j\sum_m\left(-k_{im \rightarrow jc}n_in_m+k_{jc \rightarrow im}n_jn_{\text{O}}^2\right),
	\label{eq:ME-FullStS_O2}
\end{align}
\begin{align}
	\frac{dn_{\text{O}}}{dt} &= 2\sum_i\sum_l\sum_m\left(k_{im \rightarrow cl}n_in_m-k_{cl \rightarrow im}n_{\text{O}}^2n_l\right)\nonumber\\
	&+ 2\sum_i\sum_j\sum_m\left(k_{im \rightarrow jc}n_in_m-k_{jc \rightarrow im}n_jn_{\text{O}}^2\right).
	\label{eq:ME-FullStS_O}
\end{align}
\noindent
Here $n_X$ identifies the number density of $X$ species/level, and $t$ denotes time. The rate coefficient $k$ describes the corresponding bound-bound (\emph{e.g.}, $k_{im \rightarrow jl}$) and bound-free (\emph{e.g.}, $k_{im \rightarrow cl}$) transitions where the subscript $c$ denotes the dissociated state. It is important to note that the rate coefficient $k$ is a dependent variable of the kinetic temperature $T$, although its dependency is omitted for brevity.

The description of the rovibrational energy transfer and dissociation of $\text{O}_2$ using Eqs. \eqnref{eq:ME-FullStS_O2} and \eqnref{eq:ME-FullStS_O} requires a prohibitive amount of computational cost since $\text{O}_2\left(\text{X}^3\Sigma_g^-\right)$ has 6115 different rovibrational levels (see Sec. \ref{sec:qct}) resulting in about 6115${}^4$ collisional processes to be considered. This constraint in the computational cost motivated the development of coarse-graining strategies \cite{Munafo_EPJD_2012,SAHAI_ADAPTIVE,MACDONALD_MEQCT,Sharma_2020,Venturi2020}. However, those order reduction strategies provide an approximated description of the StS population dynamics, rather than the complete rovibrational-specific distributions which are crucial ingredients in developing advanced reduced order models\cite{SAHAI_ADAPTIVE,Sahai_2019_PRF,Venturi2020}. This work thus presents a novel and simple reduced order model, which allows investigating the rovibrational distribution in the $\text{O}_2$+$\text{O}_2$ system by combining the concept of coarse-grain QCT method \cite{MACDONALD_MEQCT} with the model proposed by Kim \emph{et al.} \cite{Kim_JTHT_2010} assumes a Boltzmann distribution among the collision partner's internal states. 

Derivation of the proposed model can start from Eqs. \eqnref{eq:O4_EnergyTransfer_StS}-\eqnref{eq:O4_DoubleDiss_StS}. We enforced the rovibrational states $m$ of the collision partner being populated as a Boltzmann distribution at a prescribed internal temperature $T_{\text{int}}$ by introducing a state-specific distribution function $\mathcal{F}_m$:
\begin{equation}
	\label{eq:DistFunction_TC}
	\mathcal{F}_m\left(T_{\text{int}}\right)=\frac{n_m}{n_{\text{O}_2}}=\frac{Q_m\left(T_{\text{int}}\right)}{Q_{\text{O}_2}\left(T_{\text{int}}\right)}.
\end{equation}
\noindent
The partition functions $Q_m$ and $Q_{\text{O}_2}$ are defined as
\begin{equation}
	\label{eq:PartitionFunction}
	Q_m\left(T_{\text{int}}\right)=g_m\exp\left(-\frac{e_m}{k_BT_{\text{int}}}\right), \quad Q_{\text{O}_2}\left(T_{\text{int}}\right)=\sum_m^{\text{O}_2}Q_m\left(T_{\text{int}}\right),
\end{equation}
\noindent
where $e_m$ and $k_B$ denote the $m$-th state energy and the Boltzmann constant, respectively. The rovibrational degeneracy $g_m{=}g_{nuc}\left(2J+1\right)$ includes the nuclear spin contribution as $g_{nuc}{=}0.5$ for all the levels based on a semiclassical approximation \cite{Venturi2020}. Then Eqs. \eqnref{eq:O4_EnergyTransfer_StS}-\eqnref{eq:O4_DoubleDiss_StS} can be recast as:
\begin{IEEEeqnarray}{rCl}
	\text{O}_2(i)+\text{O}_2(T_{\text{int}}) &\leftrightarrow& \text{O}_2(j)+\sum_l^{\text{O}_2}\text{O}_2(l), \label{eq:O4_EnergyTransfer_TC} \\
	\text{O}_2(i)+\text{O}_2(T_{\text{int}}) &\leftrightarrow& \text{O}+\text{O}+\sum_l^{\text{O}_2}\text{O}_2(l), \label{eq:O4_TargetDiss_TC} \\
	\text{O}_2(i)+\text{O}_2(T_{\text{int}}) &\leftrightarrow& \text{O}_2(j)+\text{O}+\text{O}, \label{eq:O4_ColliderDiss_TC} \\
	\text{O}_2(i)+\text{O}_2(T_{\text{int}}) &\leftrightarrow& \text{O}+\text{O}+\text{O}+\text{O}, \label{eq:O4_DoubleDiss_TC}
\end{IEEEeqnarray}
\noindent
where $\text{O}_2(T_{\text{int}})$ denotes the collision partner Boltzmann-populated at $T_{\text{int}}$. It is important to note that Figure \ref{Suppfig:O4_Thermal_DissRate_Decomp} in the Supplementary Material verifies the insignificant impact of the double-dissociation in Eq. \eqnref{eq:O4_DoubleDiss_TC} by comparing the thermal dissociation rate coefficients from the mechanisms in Eqs. \eqnref{eq:O4_TargetDiss_TC}-\eqnref{eq:O4_DoubleDiss_TC}. By neglecting the double-dissociation, the set of master equations corresponding to Eqs. \eqnref{eq:O4_EnergyTransfer_TC}-\eqnref{eq:O4_ColliderDiss_TC} leads to:
\begin{align}
	\frac{dn_i}{dt} &= \sum_j\left(-k_{i \rightarrow j}\left(T_{\text{int}}\right)n_in_{\text{O}_2}+k_{j \rightarrow i}\left(T_{\text{int}}\right)n_jn_{\text{O}_2}\right)\nonumber\\
	&- k_{i \rightarrow c}\left(T_{\text{int}}\right)n_in_{\text{O}_2}+k_{c \rightarrow i}\left(T_{\text{int}}\right)n_{\text{O}}^2n_{\text{O}_2}\nonumber\\
	&+ \sum_j\left(-k_{i \rightarrow j}^{\star}\left(T_{\text{int}}\right)n_in_{\text{O}_2}+k_{j \rightarrow i}^{\star}\left(T_{\text{int}}\right)n_jn_{\text{O}}^2\right),
	\label{eq:ME-TC_O2}
\end{align}
\begin{align}
	\frac{dn_{\text{O}}}{dt} &= 2\sum_i\left(k_{i \rightarrow c}\left(T_{\text{int}}\right)n_in_{\text{O}_2}-k_{c \rightarrow i}\left(T_{\text{int}}\right)n_{\text{O}}^2n_{\text{O}_2}\right)\nonumber\\
	&+2\sum_i\sum_j\left(k_{i \rightarrow j}^{\star}\left(T_{\text{int}}\right)n_in_{\text{O}_2}-k_{j \rightarrow i}^{\star}\left(T_{\text{int}}\right)n_jn_{\text{O}}^2\right).
	\label{eq:ME-TC_O}
\end{align}
\noindent
The form of rate coefficients $k_{i \rightarrow j}$, $k_{i \rightarrow c}$, and $k_{i \rightarrow j}^{\star}$ can be defined as
\begin{IEEEeqnarray}{rCl}
	k_{i \rightarrow j}\left(T_{\text{int}}\right)&=&\sum_l\sum_mk_{im \rightarrow jl}\mathcal{F}_m\left(T_{\text{int}}\right), \label{eq:Bound-bound_Rate_TC} \\
	k_{i \rightarrow c}\left(T_{\text{int}}\right)&=&\sum_l\sum_mk_{im \rightarrow cl}\mathcal{F}_m\left(T_{\text{int}}\right), \label{eq:Bound-free_Rate_TC} \\
	k_{i \rightarrow j}^{\star}\left(T_{\text{int}}\right)&=&\sum_mk_{im \rightarrow jc}\mathcal{F}_m\left(T_{\text{int}}\right), \label{eq:Bound-free_Col_Rate_TC}
\end{IEEEeqnarray}
\noindent
where $k_{i \rightarrow j}$, $k_{i \rightarrow c}$, and $k_{i \rightarrow j}^{\star}$ correspond to the reactions in Eqs. \eqnref{eq:O4_EnergyTransfer_TC}, \eqnref{eq:O4_TargetDiss_TC}, and \eqnref{eq:O4_ColliderDiss_TC}, respectively.
It is important to note that dependency of the rate coefficients to collision partner's initial and final states (\emph{i.e.}, $m$ and $l$) is eliminated by prescribing their internal distributions. This implies that the computational cost for $\text{O}_2\left(\text{X}^3\Sigma_g^-\right)$+$\text{O}_2\left(\text{X}^3\Sigma_g^-\right)$ system for the given $T_{\text{int}}$ and $T$ can be reduced to the same order of the one for $\text{O}_2\left(\text{X}^3\Sigma_g^-\right)$+$\text{O}\left({}^3\text{P}_2\right)$. Consequently, solutions from Eqs. \eqnref{eq:ME-TC_O2} and \eqnref{eq:ME-TC_O} describe the rovibrational energy transfer and dissociation in $\text{O}_2\left(\text{X}^3\Sigma_g^-\right)$+$\text{O}_2\left(\text{X}^3\Sigma_g^-\right)$ system with much less computational expense compared with the exact formulation in Eqs. \eqnref{eq:ME-FullStS_O2} and \eqnref{eq:ME-FullStS_O}. The aforementioned order reduction strategy can be interpreted as a coarse-grain QCT model: The target species is fully-resolved to its rovibrational states, whereas the collision partner is grouped as a single bin in which all the levels are lumped into.

The rovibrational-specific state-to-state rate coefficients in Eqs. \eqnref{eq:Bound-bound_Rate_TC}-\eqnref{eq:Bound-free_Col_Rate_TC} are computed by means of the QCT method discussed in Sec. \ref{sec:qct}. At a given $T$, the QCT calculations are carried out for several $T_{\text{int}}$ that range from \SI{300}{\kelvin} to $T$ (See Table \ref{table:QCT_T_Tint_range} for more details). To integrate the set of master equations in Eqs. \eqnref{eq:ME-TC_O2} and \eqnref{eq:ME-TC_O}, the rate coefficients $k$ and $k^{\star}$ depending on the $T_{\text{int}}$ are updated at each time step by interpolating them in $T_{\text{int}}$-space. This strategy is adopted to perform the master equation study at $T$=\SI{10000}{\kelvin}, whereas for other $T$ the rate coefficients are averaged over $T_{\text{int}}$-space and kept as constant values for further reduction of computational costs. It is important to note that the latter averaging approach does not affect to the key findings of the present work. The aforementioned reduced-order modeling strategy is denoted as thermal collider (TC) approach in the following sections.

The integration of the master equations is carried out using \textsc{plato} (PLAsmas in Thermodynamic
nOn-equilibrium) \cite{MUNAFO_HEGEL}, a library for non-equilibrium plasmas developed within The Center for Hypersonics and Entry System Studies (CHESS) at University of Illinois at Urbana-Champaign. In the master equation analysis to be presented, the chemical composition is initialized to have 100\% of $\text{O}_2$. The initial gas pressure and internal temperature are set to 1000 Pa and \SI{300}{\kelvin}, respectively. Those conditions are applied for all simulations unless otherwise stated. As per the heat-bath temperatures, the values of \SI{7500}{\kelvin}, \SI{10000}{\kelvin}, \SI{15000}{\kelvin}, and \SI{20000}{\kelvin} are considered.

\subsection{Quasi-classical trajectory calculations}\label{sec:qct}
To compute the rovibrational-specific StS rate coefficients for Eqs. \eqnref{eq:O4_EnergyTransfer_TC}-\eqnref{eq:O4_DoubleDiss_TC}, QCT calculations are performed using \textsc{CoarseAIR} \cite{Venturi2020_ML,Venturi2020}. It is an in-house QCT code that is a modernized version of the original \textsc{VVTC} code developed at NASA Ames Research Center by Schwenke \cite{SCHWENKE_VVTC_1988}. The three different \emph{ab-initio} PESs proposed by Paukku \emph{et al.} \cite{Paukku_O4_1,Paukku_O4_2} are employed to consider the adiabatic interactions occurring in the electronic ground state of $\text{O}_2$+$\text{O}_2$ system through the singlet, triplet, and quintet spin couplings. The statistical weights for each spin interaction are approximated as 1/9, 3/9, and 5/9, respectively, based on the high-temperature limit. The 6115 rovibrational energy levels are obtained by solving Schr$\ddot{\text{o}}$dinger's equation based on the WKB semi-classical approximation \cite{Truhlar1979,SCHWENKE_VVTC_1988}. Resultant dissociation energy of the rovibrational ground state is found as 5.113 eV. The initial state of thermal collision partner $\text{O}_2\left(T_{\text{int}}\right)$ is sampled from the prescribed distribution function $\mathcal{F}_m\left(T_{\text{int}}\right)$. Consequently, the rate coefficients in Eqs. \eqnref{eq:Bound-bound_Rate_TC}-\eqnref{eq:Bound-free_Col_Rate_TC} are computed by summing up the statistics of individual trajectories with respect to the final state of collision partner.

For a given bound-bound transition, both endothermic and exothermic trajectories can exist as the result of QCT calculations. In this case, one of them must be selected as a reference direction and the other one needs to be reconstructed based on the micro-reversibility to guarantee appropriate equilibration of the initial heat-bath simulations. In the present QCT calculations, the sampled trajectories are investigated to decide which direction (\emph{i.e.}, endothermic or exothermic) has better statistics as presented in Sec. \ref{sec:Sampled_Trajectories} where detailed discussion is provided. As a result, we utilize the endothermic sampled trajectories as the reference forward direction with higher priority than the exothermic ones to construct the StS kinetic database described in Sec. \ref{sec:kinetics}. Exceptionally, the exothermic trajectory is used in a case that the given reaction channel does not have the endothermic one. The reverse rate coefficients are then reconstructed based on the micro-reversibility.

\section{Results}\label{sec:results}
\subsection{State-to-state kinetic database}\label{sec:kinetics}
The constructed StS kinetic database in this work covers the kinetic temperature range 7500-\SI{20000}{\kelvin}. The combination of $T$ and $T_{\text{int}}$ for the QCT calculations are summarized in Table \ref{table:QCT_T_Tint_range}. The present analysis at $T$=\SI{10000}{\kelvin} ensures the highest fidelity in terms of accuracy since it covers the largest number of QCT databases along the $T_{\text{int}}$-space. The results in this work at other $T$, such as \SI{7500}{\kelvin}, \SI{15000}{\kelvin}, and \SI{20000}{\kelvin} demonstrate the extension of the present modeling strategy to various kinetic temperature range. This section investigates influences of the sampling temperature $T_{\text{int}}$ on the constructed StS kinetic database, especially for the inelastic and target particle's dissociation of Eqs. \eqnref{eq:Bound-bound_Rate_TC} and \eqnref{eq:Bound-free_Rate_TC}, respectively.
\begin{table}[h]
	\begin{center}
		\caption{Covered kinetic and internal temperatures in the present StS kinetic database.}
		\label{table:QCT_T_Tint_range}
		\begin{tabular}{c|c} \hline
			$T$ & $T_{\text{int}}$\\ \hline
			\SI{7500}{\kelvin} & \SI{300}{\kelvin}, \SI{3500}{\kelvin}, \SI{7500}{\kelvin}\\ 
			\SI{10000}{\kelvin} & \SI{300}{\kelvin}, \SI{2500}{\kelvin}, \SI{5000}{\kelvin}, \SI{7500}{\kelvin}, \SI{10000}{\kelvin}\\ 
			\SI{15000}{\kelvin} & \SI{300}{\kelvin}, \SI{15000}{\kelvin}\\ 
			\SI{20000}{\kelvin} & \SI{300}{\kelvin}, \SI{20000}{\kelvin}\\ \hline
		\end{tabular}
	\end{center}
\end{table}

\begin{figure}[h!]
	\centering
	\includegraphics[width=0.75\textwidth]{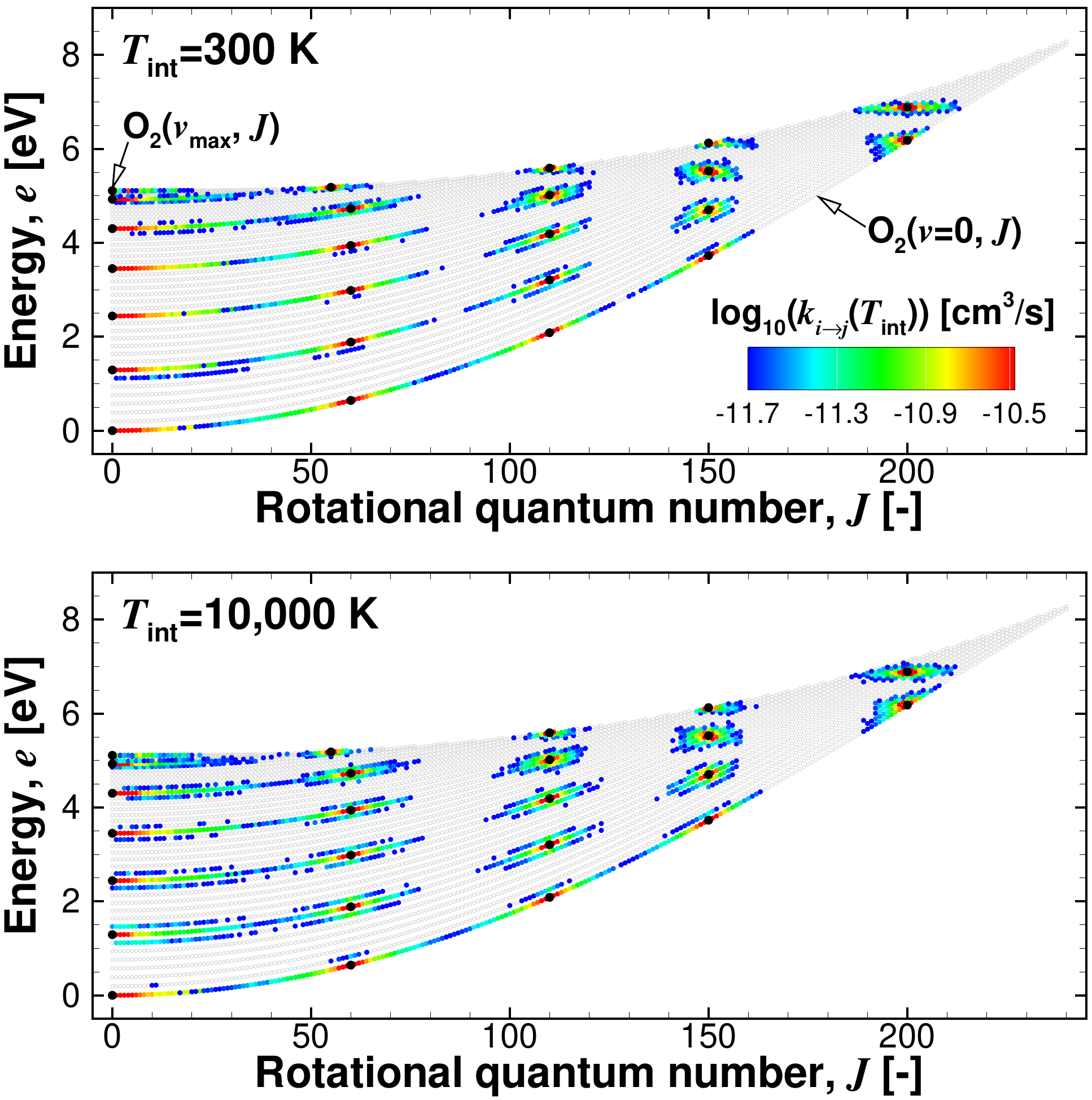}
	\caption{Distributions of the inelastic rate coefficients in Eq. \eqnref{eq:O4_EnergyTransfer_TC} at $T_{\text{int}}$=\SI{300}{\kelvin} (top) and $T_{\text{int}}$=\SI{10000}{\kelvin} (bottom) at $T$=\SI{10000}{\kelvin}, overlayed on the $\text{O}_2$ diatomic potential. The gray dots indicate that the internal states having the rate coefficients lower than a cutoff value of $k_{i{\rightarrow}j}\left(T_{\text{int}}\right)$=1.58{$\times$}10$^{-12}$ cm${}^3$/s. The black dots represent the initial target states.}
	\label{fig:Inel_dist}
\end{figure}

\begin{figure}[h]
	\centering
	\subfigure[]
	{
		\includegraphics[width=0.48\textwidth]{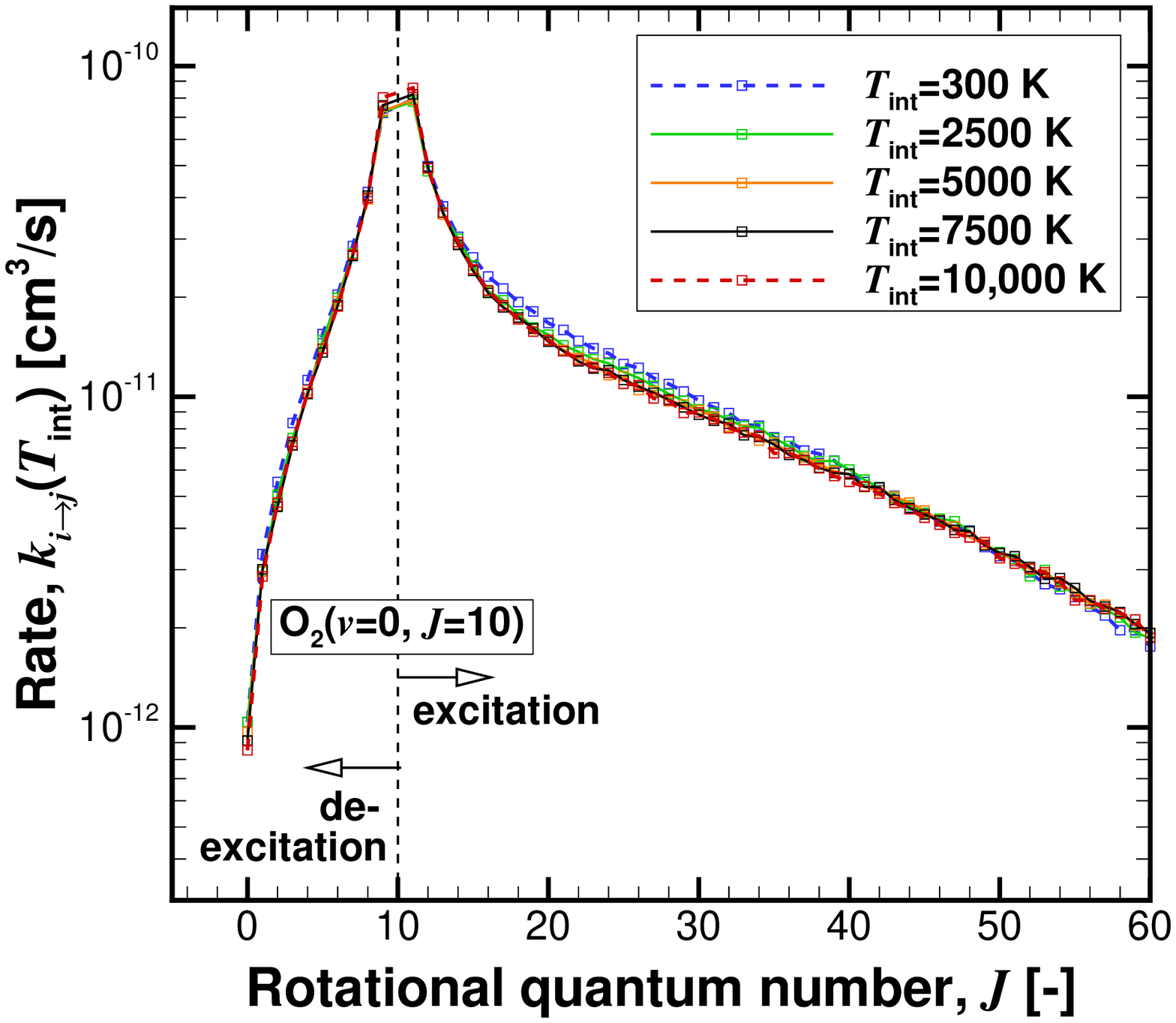}
		\label{fig:Inel_along-J}
	}
	\subfigure[]
	{
		\includegraphics[width=0.48\textwidth]{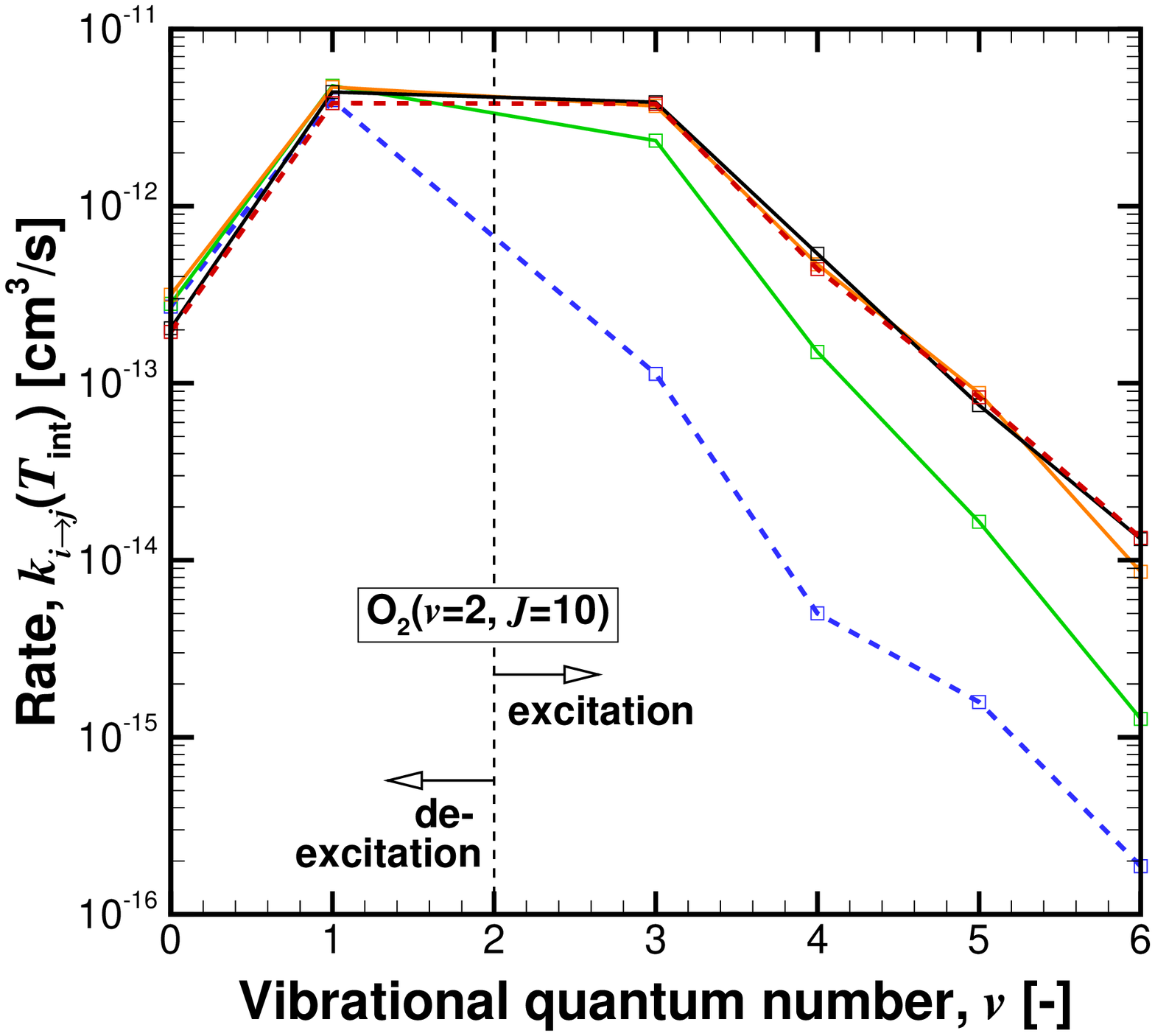}
		\label{fig:Inel_along-v}
	}
	\caption{Distributions of the inelastic rate coefficients in Eq. \eqnref{eq:O4_EnergyTransfer_TC} at $T$=\SI{10000}{\kelvin}. The initial target states are: (a) $\text{O}_2(v{=}0, J{=}10)$ and (b) $\text{O}_2(v{=}2, J{=}10)$.}
	\label{fig:Inel_line}
\end{figure}

Figure \ref{fig:Inel_dist} shows the inelastic rate coefficient distributions at $T$=\SI{10000}{\kelvin}. The magnitude of rates is overlayed on the diatomic potential of $\text{O}_2\left(\text{X}^3\Sigma_g^-\right)$. Overall trends observed from both $T_{\text{int}}$ are qualitatively similar to each other: In the low-lying states, transition pathways are mostly aligned to the same vibrational strands, whereas they are spread across different vibrational levels in the high-lying energy regime. This may imply that the assumption for separation between rotational and vibrational states are less feasible in the high-energy area. A distinct difference according to the variation of $T_{\text{int}}$ is the rate coefficient for vibrational ladders. At the higher $T_{\text{int}}$, the vibrational transition rates are larger particularly in the low-$v$ and low-$J$ region that governs majority of the rovibrational energy transfer. This difference is quantitatively highlighted in Fig. \ref{fig:Inel_line}. The rotational transition shown in Fig. \ref{fig:Inel_along-J} is not sensitive to the variation of $T_{\text{int}}$. Magnitudes of the rotational transition rates are almost identical to each other. On the other hand, the vibrational excitation rates change more than an order of magnitude for the variation of $T_{\text{int}}$ as observed in Fig. \ref{fig:Inel_along-v}. As $T_{\text{int}}$ increases, the vibrational energy transfer is significantly raised. This increase is started to saturate at $T_{\text{int}}$=\SI{5000}{\kelvin}. From the investigations in Figs. \ref{fig:Inel_dist} and \ref{fig:Inel_line}, it is found that the sampling temperature $T_{\text{int}}$ has significant influence on the vibrational transition rates, whereas the impact to the rotational transitions is insignificant.

\begin{figure}[h]
	\centering
	\includegraphics[width=0.75\textwidth]{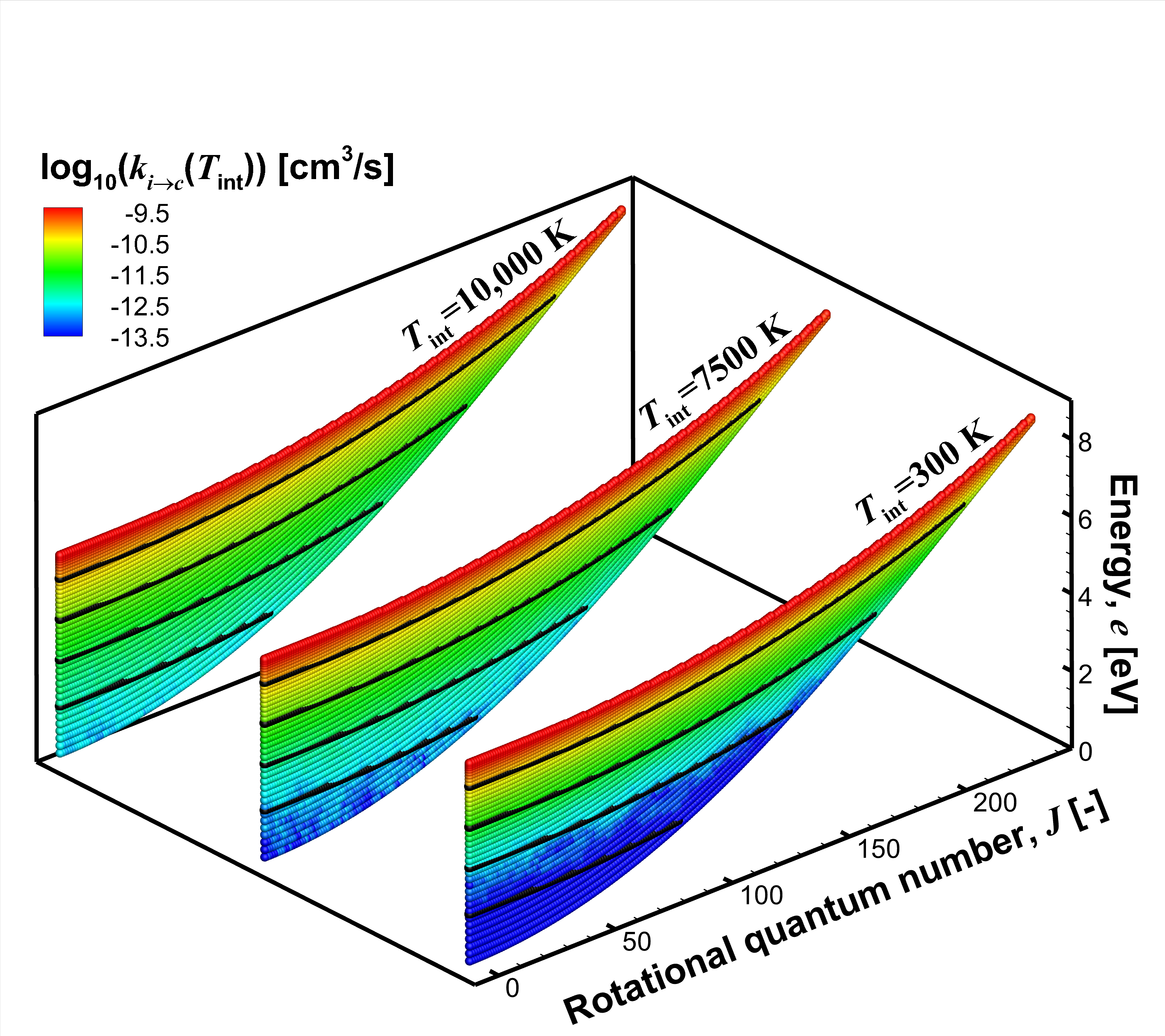}
	\caption{Distributions of the dissociation rate coefficients in Eq. \eqnref{eq:Bound-free_Rate_TC} at $T$=\SI{10000}{\kelvin} along the $T_{\text{int}}$ range, overlayed on the $\text{O}_2$ diatomic potential. The black lines indicate the isolines from the centrifugal barrier (\emph{i.e.}, isolines of energy deficit).}
	\label{fig:O4_DissTar_T10000K_3D}
\end{figure}

The target particle's dissociation rate coefficient distributions are presented in Fig. \ref{fig:O4_DissTar_T10000K_3D}. The figure compares the rates calculated at the three-different $T_{\text{int}}$. The dependence of the rate coefficient is a function of the distance from the centrifugal barrier. This behavior was observed by Venturi \emph{et al.} \cite{Venturi2020}. In the case of $T_{\text{int}}$ around \SI{10000}{\kelvin}, the rate coefficients for $\text{O}_2$+$\text{O}_2$ are indistinguishable from the ones computed for the $\text{O}_2$+O system (See Fig. \ref{Suppfig:DissRate_O3_O4_T10000K})

Influence of the $T_{\text{int}}$ variation is confined to the low-lying energy levels which do not have a significant impact on the dissociation kinetics. By relating the rates with the concept of energy deficit \cite{Venturi2020}, defined as a distance from the centrifugal barrier in energy space, it is clear that the $T_{\text{int}}$ of colliding molecule works as a \emph{stretching} factor for the magnitude of rates, especially for the states far away from the centrifugal barrier. This is because as collision partner has higher internal energy (\emph{i.e.}, higher $T_{\text{int}}$), the low-lying states of target $\text{O}_2$ can more easily gain the required barrier energy to dissociate during collisions.

It is important to note that in the high-lying energy states the rates are very close to each other and consequently a difference of the thermal rate coefficients between $T_{\text{int}}$=\SI{300}{\kelvin} and \SI{10000}{\kelvin} is less than a factor of 1.5 at $T$=\SI{10000}{\kelvin}. Figure \ref{Suppfig:DissTar_line} supports this aspect by showing more quantitative distributions of the rates for some selected quantum pairs. 
%

\subsection{Internal energy transfer processes}\label{sec:energy}
In this section, internal energy transfer by the inelastic and exchange kinetics is studied in detail by neglecting the dissociation mechanisms while solving the set of master equations. The influence of sampling temperature $T_{\text{int}}$ on the rovibrational population dynamics by the inelastic process is firstly investigated at $T$=\SI{10000}{\kelvin}. Then the master equation analysis is extended to other kinetic temperatures along with examination of the exchange process.

\begin{figure}[h]
	\centering
	\includegraphics[width=0.75\textwidth]{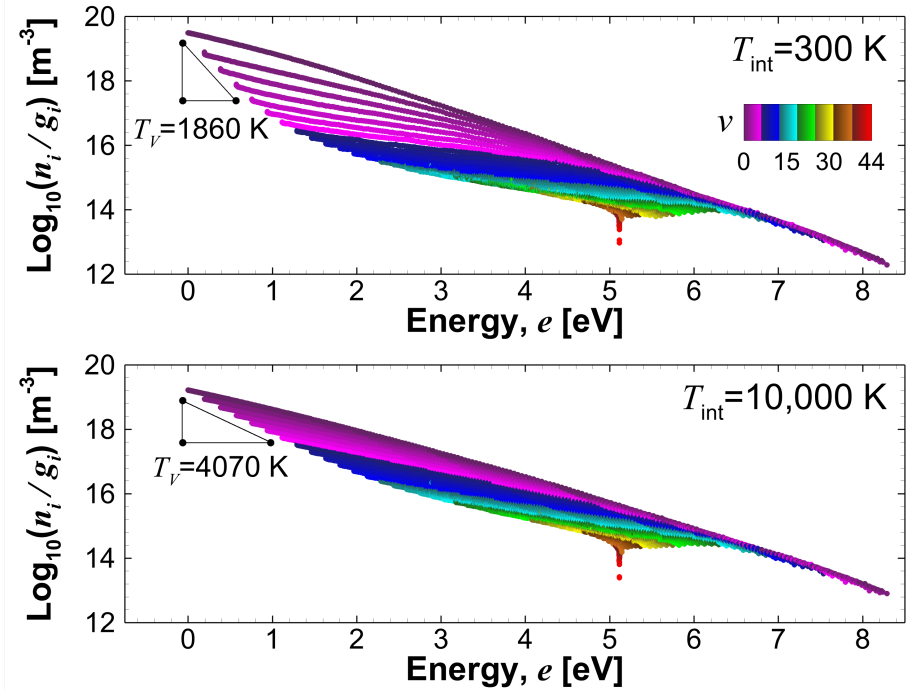}
	\caption{Rovibrational distributions at $T_{\text{int}}$=\SI{300}{\kelvin} (top) and $T_{\text{int}}$=\SI{10000}{\kelvin} (bottom) at $T$=\SI{10000}{\kelvin} by the inelastic kinetics only. The distributions are extracted from the identical time instant $t$=10$^{-7}$ s and colored by the vibrational quantum number.}
	\label{fig:CompPop_Inel_Dist_T10000K}
\end{figure}

Figure \ref{fig:CompPop_Inel_Dist_T10000K} presents snapshots of the rovibrational distributions by the inelastic process at a time instant $t$=10$^{-7}$ s and $T$=\SI{10000}{\kelvin}. The time instant 10$^{-7}$ s is selected to compare the distributions in the early stage of energy transfer. Overall structures at the different $T_{\text{int}}$ values are similar to each other. At the higher $T_{\text{int}}$, the excited vibrational levels are more populated that can be verified from the slope among the heads of vibrational strands which implies the vibrational temperature $T_V$. At the same time instant, the rate of average vibrational energy transfer at $T_{\text{int}}$=\SI{10000}{\kelvin} occurs about 2 times faster compared to that in $T_{\text{int}}$=\SI{300}{\kelvin}. 

%

\begin{figure}[h]
	\centering
	\subfigure[]
	{
		\includegraphics[width=0.48\textwidth]{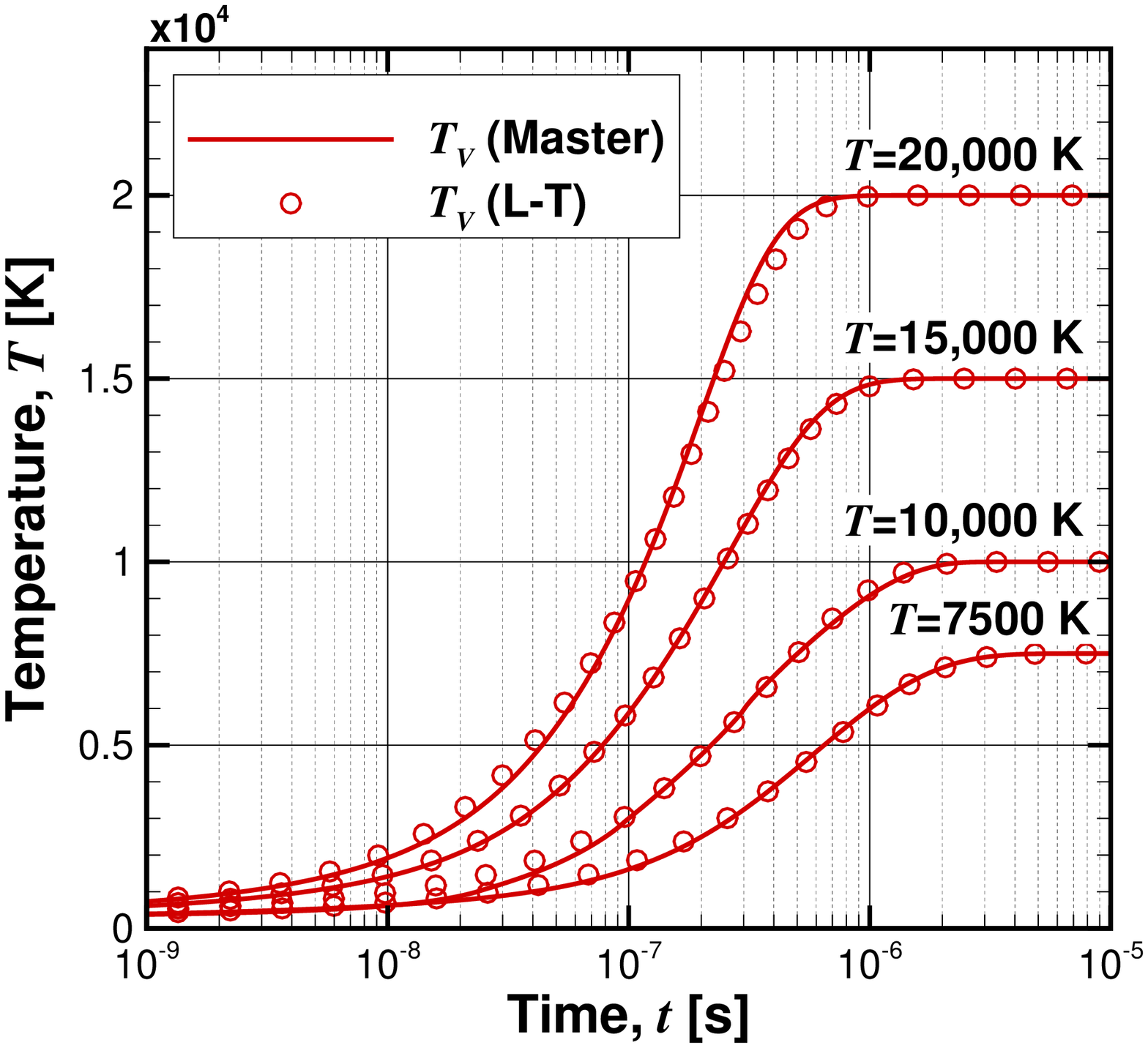}
		\label{fig:Temp_Master_LT_TV}
	}
	\subfigure[]
	{
		\includegraphics[width=0.48\textwidth]{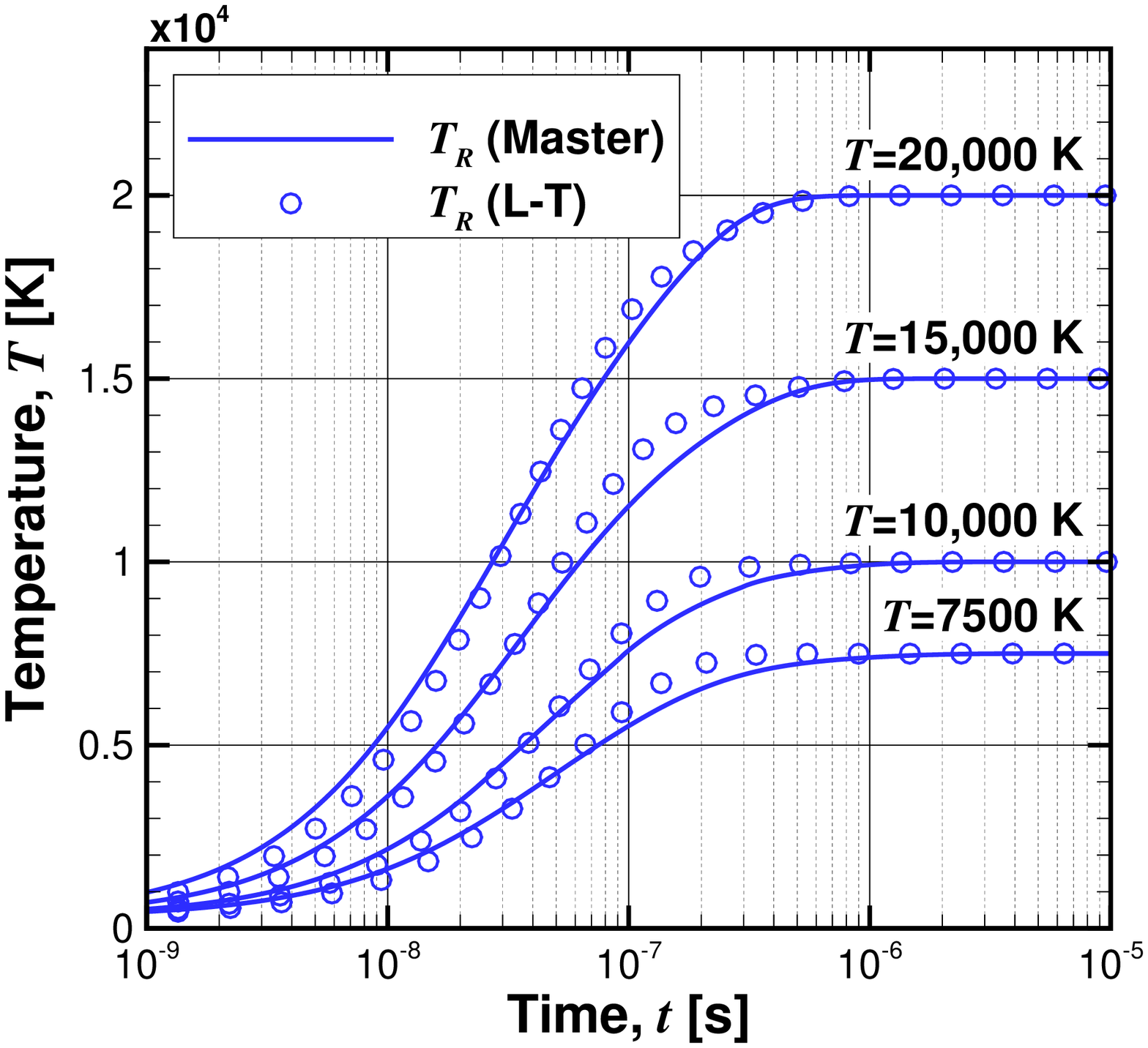}
		\label{fig:Temp_Master_LT_TR}
	}
	\caption{Temporal evolution of (a) vibrational and (b) rotational temperatures of $\text{O}_2$ at the different heat-bath temperatures (dissociation kinetics excluded). The solid lines are obtained from the master equation analysis while the symbols are computed based on the Landau-Teller relaxation model.}
	\label{fig:Temp_Master_LT}
\end{figure}

The master equation study is extended to other kinetic temperatures to investigate global trend of rovibrational energy transfers in the $\text{O}_2$+$\text{O}_2$ system. Figure \ref{fig:Temp_Master_LT} shows the temporal evolution of vibrational ($T_V$) and rotational ($T_R$) temperatures along the variation of $T$. From the master equation results, the characteristic rotational-translational ($\tau_{RT}$) and vibrational-translational ($\tau_{VT}$) relaxation times are determined using the \emph{e-folding} method \cite{Park_book}. Then they are fed into the linear Landau-Teller (L-T) relaxation equation \cite{LT_1936} to compare with the internal energy transfers from the master equation study. The L-T relaxation equation can be expressed as 
\begin{equation}
	\frac{dE_V}{dt}=\frac{E_{V,eq}-E_V}{\tau_{VT}}, \label{eq:L-T_Vibration}
\end{equation}
\begin{equation}
	\frac{dE_R}{dt}=\frac{E_{R,eq}-E_R}{\tau_{RT}}, \label{eq:L-T_Rotation}
\end{equation}
\noindent
where $E_V$ and $E_R$ are the average energy of vibration and rotation, respectively. The subscript $eq$ indicates the quantity evaluated at the equilibrium temperature $T$. It is important to note that the L-T model reasonably reproduces the master equation results in the considered $T$ range as shown in the figure. Although the difference between the vibrational and rotational characteristic relaxation times is reduced as $T$ increases (See Fig. \ref{fig:Tau}), the rotational energy transfer is distinctly faster than the vibration in the considered temperature range. 

\begin{figure}[h]
	\centering
	\includegraphics[width=0.75\textwidth]{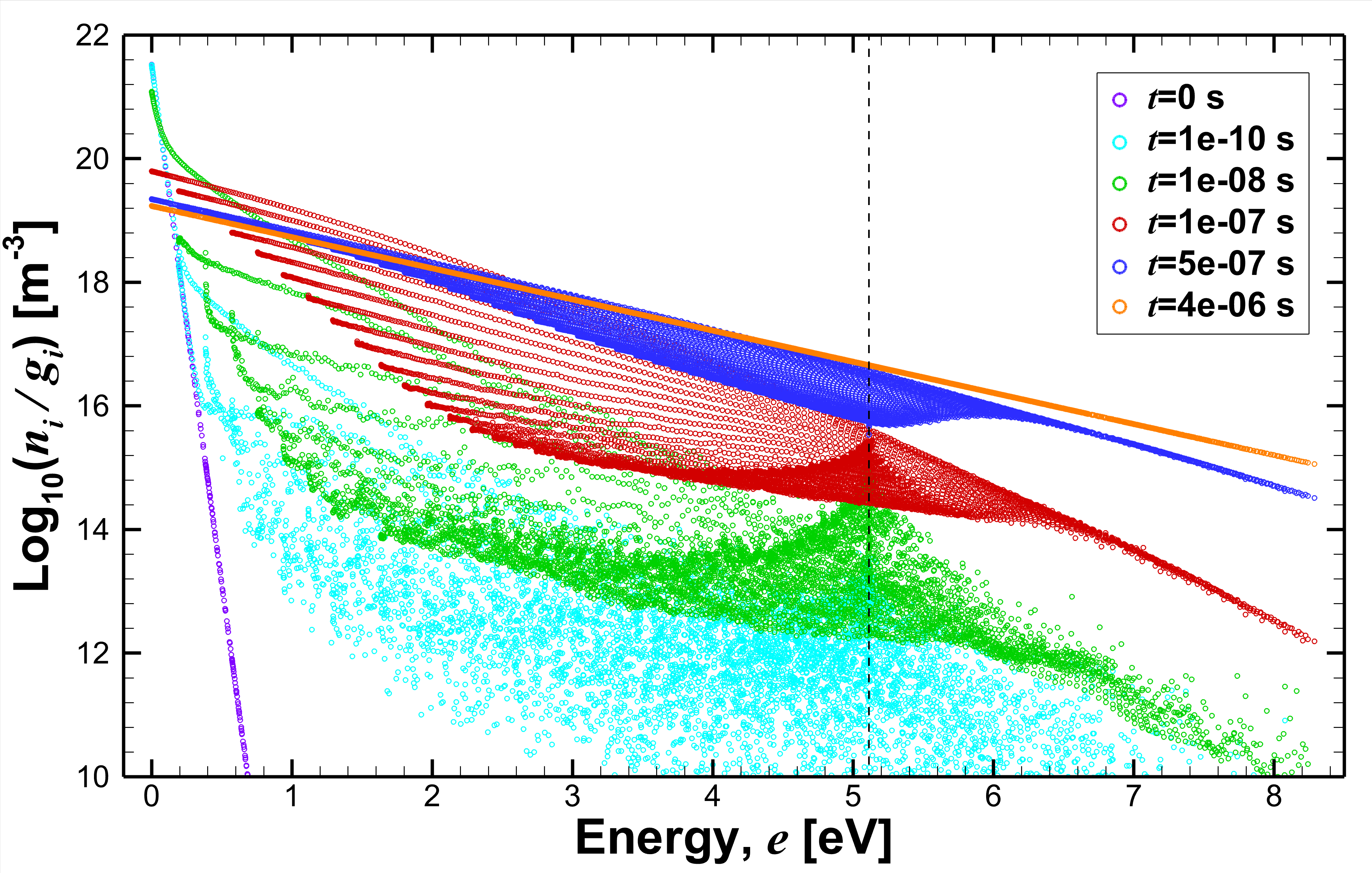}
	\caption{Rovibrational distributions along $t$ at $T$=\SI{10000}{\kelvin} (dissociation kinetics excluded). The dashed vertical line indicates the average dissociation energy of $\text{O}_2$, 5.113 eV.}
	\label{fig:CompPop_Evolution_T10000K}
\end{figure}

\begin{figure}[h]
	\centering
	\subfigure[]
	{
		\includegraphics[width=0.48\textwidth]{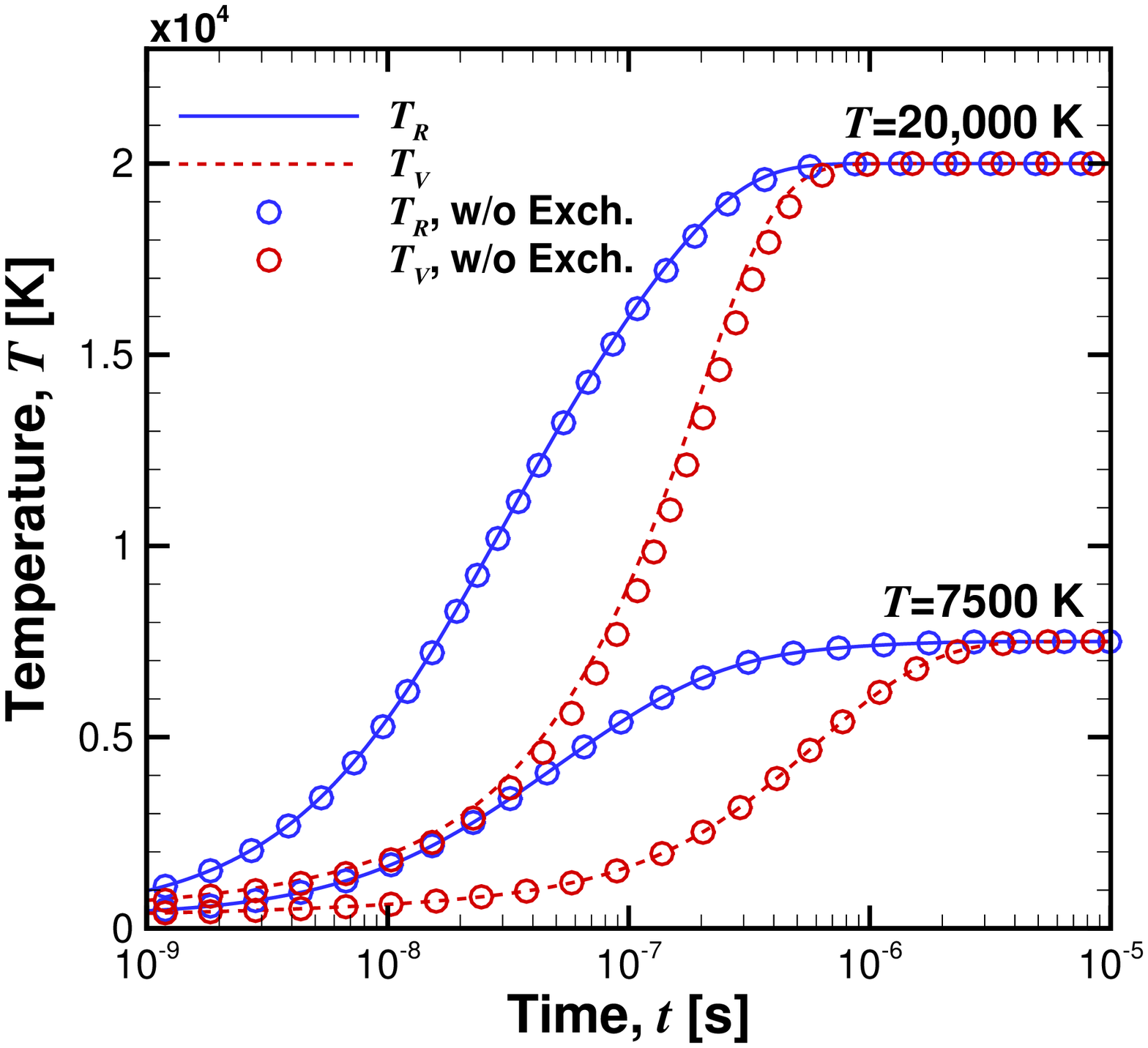}
		\label{fig:Effect_of_Exch_O4_TC}
	}
	\subfigure[]
	{
		\includegraphics[width=0.48\textwidth]{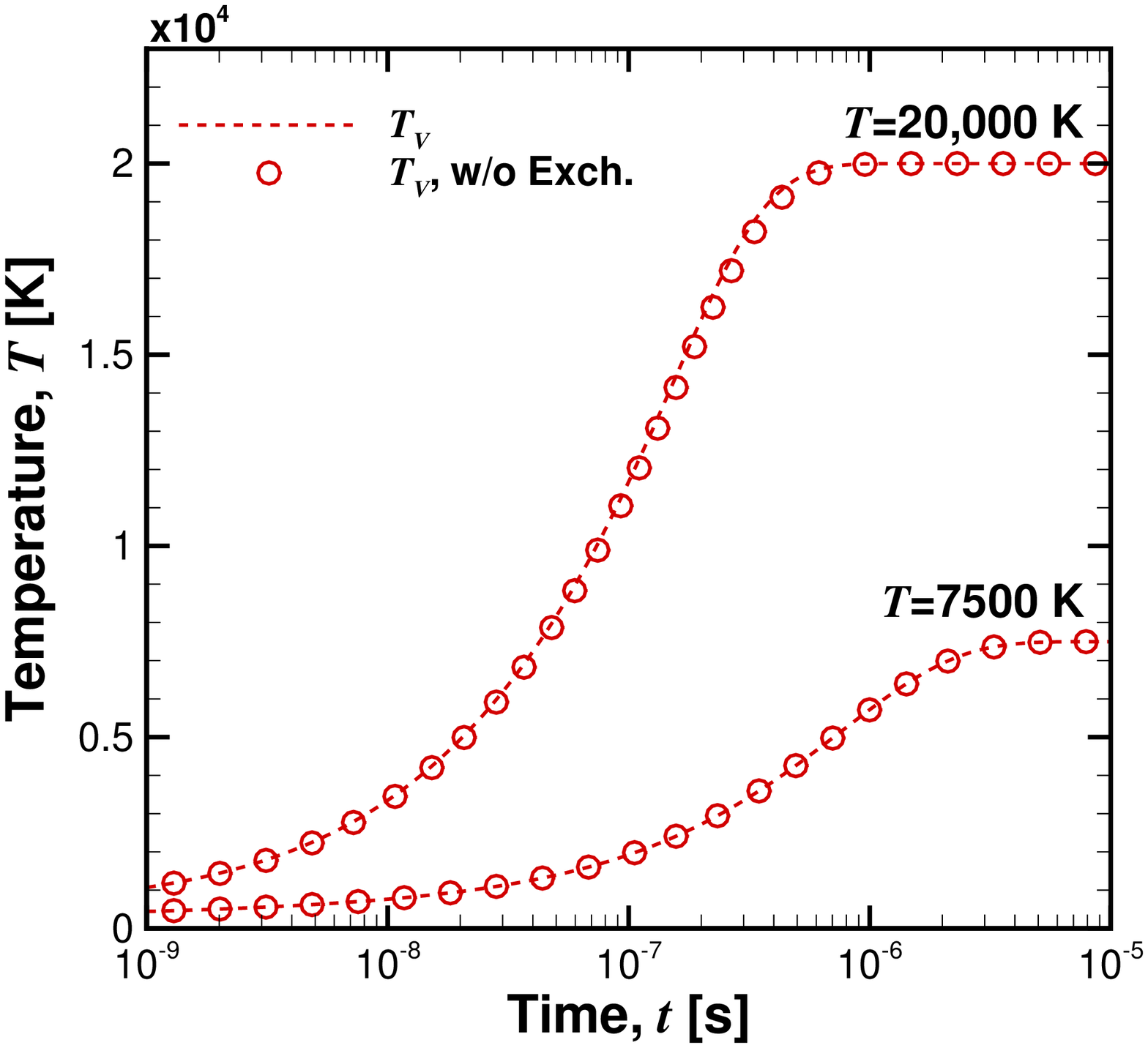}
		\label{fig:Effect_of_Exch_O4_VSM}
	}
	\subfigure[]
	{
		\includegraphics[width=0.48\textwidth]{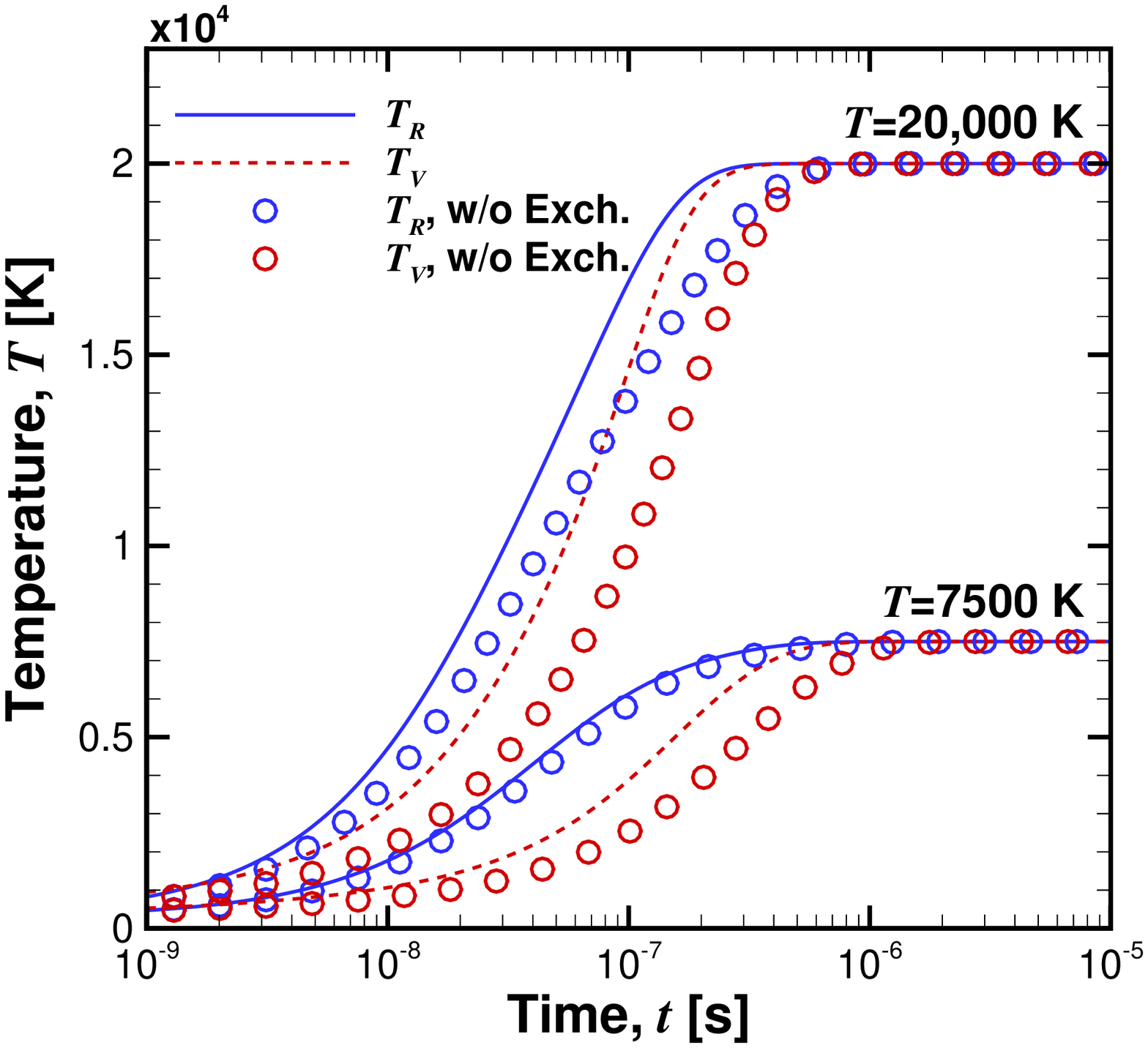}
		\label{fig:Effect_of_Exch_O3_StS}
	}
	\caption{Influence of exchange kinetics on the internal energy transfer. (a) $\text{O}_2$+$\text{O}_2$ with the TC model, (b) $\text{O}_2$+$\text{O}_2$ with the VS model, and (c) $\text{O}_2$+O with the fully-StS approach. The dissociation kinetics is excluded.}
	\label{fig:Temp_EnergyTransfer}
\end{figure}

\begin{figure}[h]
	\centering
	\includegraphics[width=0.75\textwidth]{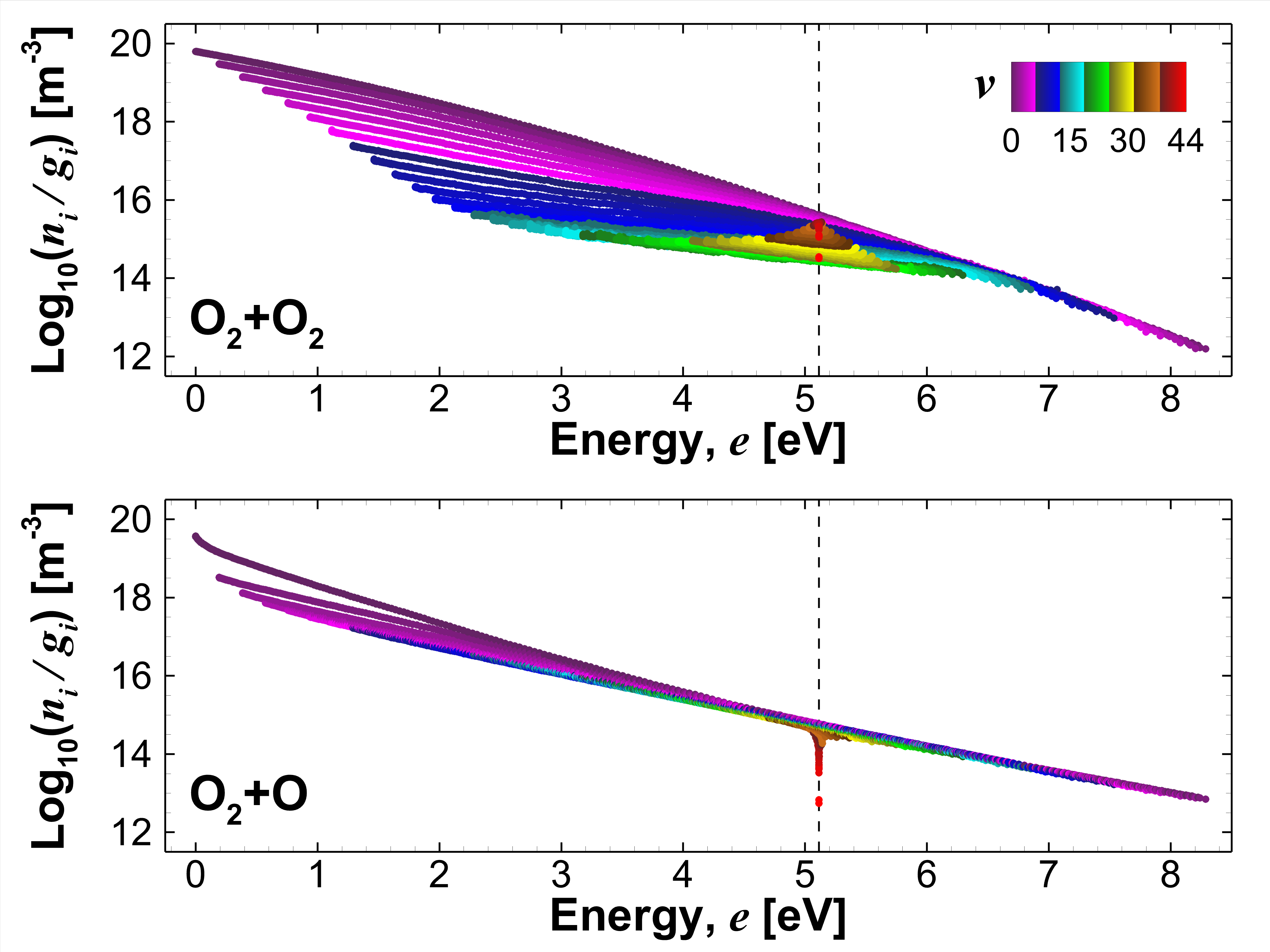}
	\caption{Rovibrational distributions in $\text{O}_2$+$\text{O}_2$ (top) and $\text{O}_2$+O (bottom) systems at 20\% of vibrational relaxation at $T$=\SI{10000}{\kelvin} (dissociation kinetics excluded). The distributions are colored by the vibrational quantum number.}
	\label{fig:CompPop_O3O4_20PercVib_T10000K}
\end{figure}

\begin{figure}[h]
	\centering
	\subfigure[]
	{
		\includegraphics[width=0.48\textwidth]{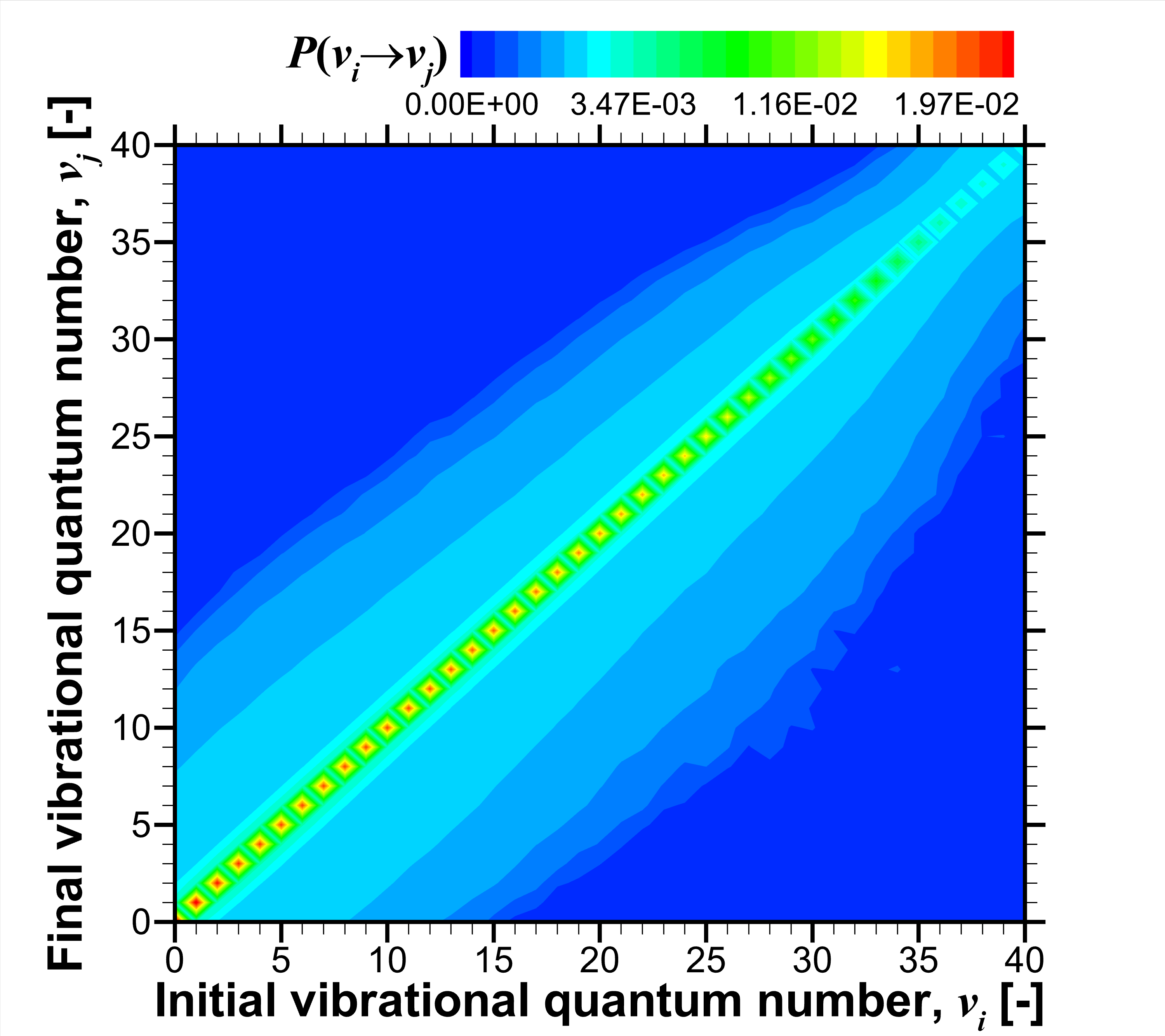}
		\label{fig:Vibrational_Transition_Probability_O4}
	}
	\subfigure[]
	{
		\includegraphics[width=0.48\textwidth]{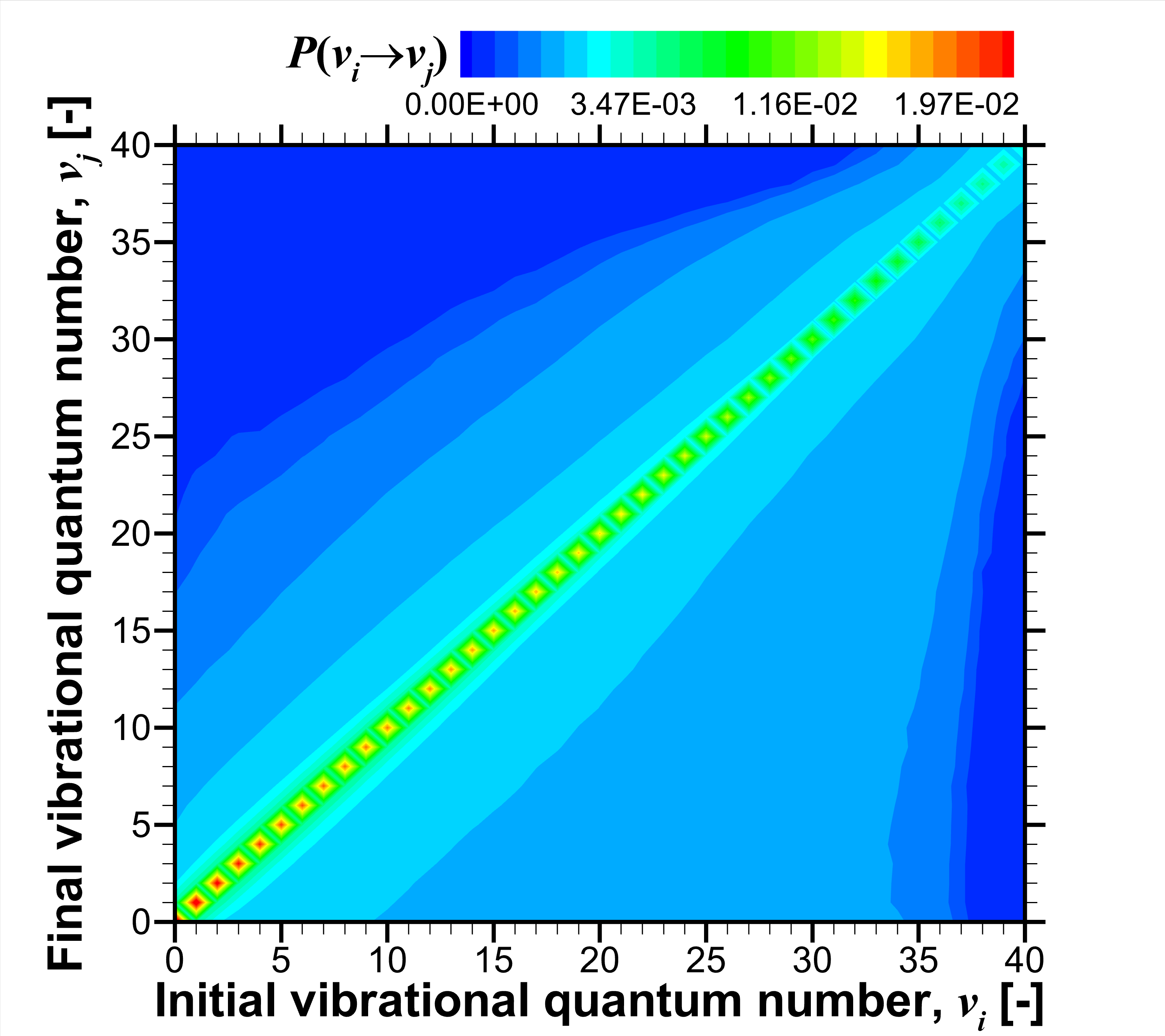}
		\label{fig:Vibrational_Transition_Probability_O3}
	}
	\caption{Vibrational transition probability distributions due to the inelastic process for (a) $\text{O}_2$+$\text{O}_2$ at $T_{\text{int}}$=\SI{300}{\kelvin} and (b) $\text{O}_2$+O at $T$=\SI{10000}{\kelvin}.}
	\label{fig:Vibrational_Transition_Probability_10000K}
\end{figure}

The internal energy transfer processes in $\text{O}_2$+$\text{O}_2$ system are analyzed in detail by investigating time evolution of rovibrational distributions as shown in Fig. \ref{fig:CompPop_Evolution_T10000K}. The figure presents the distributions at $T$=\SI{10000}{\kelvin}. In the early times, the low-lying levels form several distinct vibrational strands whereas the high-lying states near the dissociation limit are preferentially excited than the others around (See $t$=1e-08 s and $t$=1e-07 s) due to the exchange kinetics. The quasi-bound levels having higher energy than 6.5 eV are quickly equilibrated to each other that forms a Boltzmann distribution in the tail region. From $t$=1e-08 s to $t$=1e-07 s, the excitation of high-lying states is saturated and thermalization among the low-lying levels is slowly followed. One of important findings of the present study is that in $\text{O}_2$+$\text{O}_2$ system the thermalization among different vibrational levels is relatively slower than that of $\text{O}_2$+O. This finding is attributed to the two facts: In $\text{O}_2$+$\text{O}_2$, (1) the inelastic process occurs in more \emph{rotation-favorable} ways rather than the vibration, and (2) the exchange kinetics is much less effective in comparison with those of $\text{O}_2$+O. The aforementioned points are discussed in detail through Figs. \ref{fig:Temp_EnergyTransfer}-\ref{fig:Vibrational_Transition_Probability_10000K}.

Figure \ref{fig:Temp_EnergyTransfer} highlights the influence of exchange kinetics on the internal energy transfer of the $\text{O}_2$+$\text{O}_2$ and $\text{O}_2$+O systems. The lower and upper bounds of $T$ are considered. Results obtained from a present VS model for the $\text{O}_2$+$\text{O}_2$ system are also presented as comparison group. The rovibrational-specific kinetic database for the $\text{O}_2$+O system used in the present study is taken from the work by Venturi \emph{et al.} \cite{Venturi2020}. Obviously, the exchange kinetics has little impact on the internal energy transfer of the $\text{O}_2$+$\text{O}_2$ system, whereas it significantly hastens the relaxation of $\text{O}_2$+O. This fact results in the different rate of thermalization among the internal energy states. In $\text{O}_2$+$\text{O}_2$, the equilibration among different vibrational strands slowly occurs as observed from Fig. \ref{fig:CompPop_Evolution_T10000K}. On the contrary, the exchange process advances the \emph{mixing} of the internal levels in $\text{O}_2$+O system. Figure \ref{Suppfig:O3_Effect_of_Exch_Pop} presents more details of it in terms of the rovibrational distributions of $\text{O}_2$+O. The exchange reaction more favorably occurs in $\text{O}_2$+O compared to the $\text{O}_2$+$\text{O}_2$, since the 3-atom system has lower effective barrier for bond breaking required for the particle exchange.

Figure \ref{fig:CompPop_O3O4_20PercVib_T10000K} provides an alternative interpretation regarding the results shown in Fig. \ref{fig:Temp_EnergyTransfer} on a more microscopic scale. The figure compares the rovibrational distributions in the $\text{O}_2$+$\text{O}_2$ and $\text{O}_2$+O systems at $T$=\SI{10000}{\kelvin}. Both distributions are extracted from the 20\% of vibrational relaxation to make the direct comparison. The two systems commonly have the strand-like structures in the low-lying energy regime. In the $\text{O}_2$+$\text{O}_2$, impact of the exchange process is only confined to levels near the dissociation limit that can be inferred by comparing Figs. \ref{fig:CompPop_Inel_Dist_T10000K} and \ref{fig:CompPop_O3O4_20PercVib_T10000K}(top). On the other hand, in $\text{O}_2$+O most of the vibrational strands are already thermalized, except for some low-lying levels having the state energy less than 1.5 eV. The result in Fig. \ref{fig:CompPop_O3O4_20PercVib_T10000K} implies that fundamental structure of the rovibrational distributions between the two systems is different to each other. It is important to note that the difference observed from Fig. \ref{fig:CompPop_O3O4_20PercVib_T10000K} is not only from the exchange kinetics but also attributed to the inelastic process. Even though with the inelastic kinetics only, the rovibrational distributions of $\text{O}_2$+$\text{O}_2$ are more broadly spread than those of $\text{O}_2$+O, especially in the low-lying energy region in which the energy transfer is dominated by the inelastic kinetics. 

The different natures in the inelastic kinetics can be more effectively distinguished by investigating the vibrational transition probability $P\left(v_i{\rightarrow}v_j\right)$ due to the inelastic where $v_i$ and $v_j$ are the initial and final vibrational quantum numbers of the target $\text{O}_2$. Figure \ref{fig:Vibrational_Transition_Probability_10000K} shows the distributions of $P\left(v_i{\rightarrow}v_j\right)$ in $\text{O}_2$+$\text{O}_2$ and $\text{O}_2$+O systems at $T$=\SI{10000}{\kelvin}. The probability is computed by summing up the individual probabilities, which are obtained from the QCT calculations, along with the initial and final rotational states. In $\text{O}_2$+$\text{O}_2$, the $P\left(v_i{\rightarrow}v_j\right)$ distribution is more clustered around the diagonal line (\emph{i.e.}, $v_i$=$v_j$) compared to $\text{O}_2$+O, implying that the vibrational transition is less favorable. In the same vein, distributions of the rotational transition probability $P\left(J_i{\rightarrow}J_j\right)$ are also provided to Fig. \ref{Suppfig:Rotational_Transition_Probability_10000K}, confirming that the $\text{O}_2$+O system has more diagonally-clustered probability, resulting in the faster rate of vibrational transitions through the inelastic process.

\subsection{Dissociation/recombination processes}\label{sec:all}
\begin{figure}[h]
	\centering
	\subfigure[]
	{
		\includegraphics[width=0.48\textwidth]{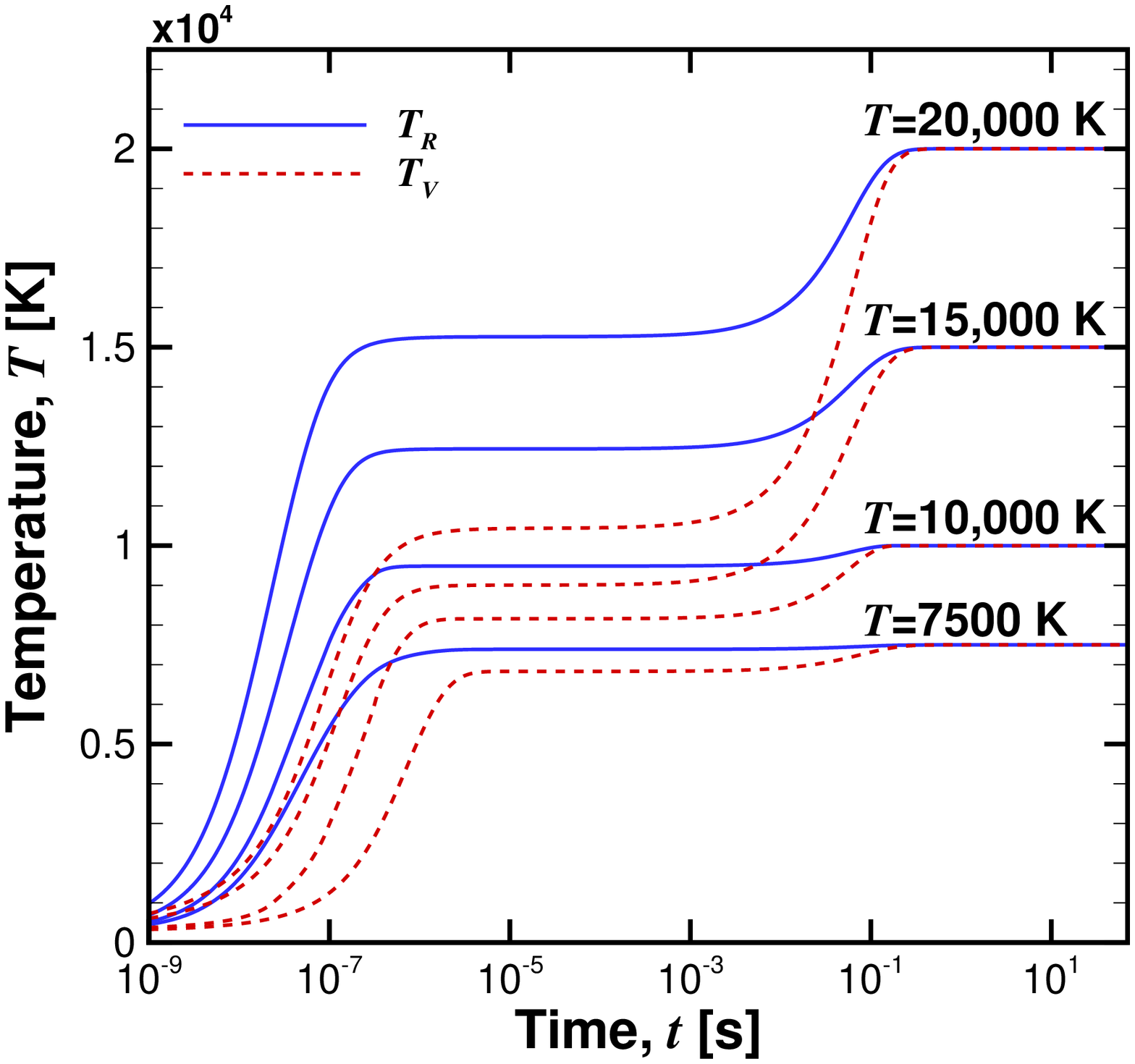}
		\label{fig:Temp_All_TC}
	}
	\subfigure[]
	{
		\includegraphics[width=0.48\textwidth]{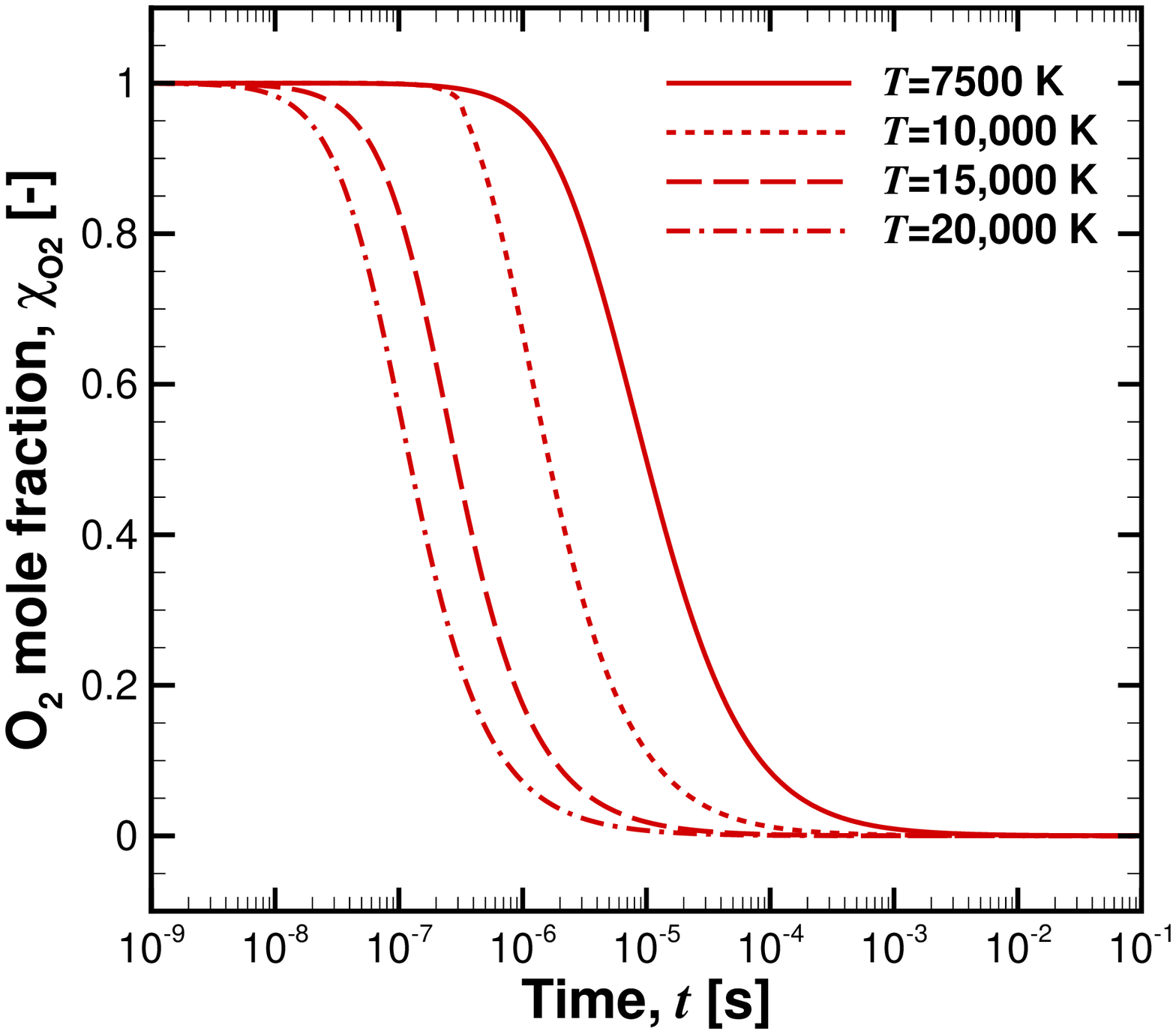}
		\label{fig:O2MoleFrac_TC}
	}
	\caption{Temporal evolution of (a) rotational and vibrational temperature and (b) mole fraction of $\text{O}_2$ at the different heat-bath temperatures.}
	\label{fig:All_Temp_MoleFrac}
\end{figure}

In this section, dissociation and recombination processes in $\text{O}_2$+$\text{O}_2$ system are investigated. Figure \ref{fig:All_Temp_MoleFrac} shows temporal evolution of the internal temperatures and $\text{O}_2$ mole fraction along the variation of $T$. In the considered $T$ range, the plateau region of the temperatures that implies the quasi-steady-state (QSS) period is clearly observed as shown in Fig. \ref{fig:Temp_All_TC}. In the QSS region, the rotational and vibrational temperatures are distinctly separated from each other, especially at $T$=\SI{15000}{\kelvin} and \SI{20000}{\kelvin}. The slow thermalization among the vibrational strands discussed in Fig. \ref{fig:CompPop_O3O4_20PercVib_T10000K} contributes to that separation by allowing that the two energy modes have certainly different gradients of population distributions (See Fig. \ref{Suppfig:O4_QSS_Pop_T20000K} for more details). 
As shown in Fig. \ref{fig:O2MoleFrac_TC}, it is confirmed that $\text{O}_2$ is completely dissociated in the given $T$ range. It is important to note that in Eq. \eqnref{eq:O4_ColliderDiss_TC} influence of combined inelastic and dissociation process is negligible, whereas combined exchange and dissociation kinetics has substantial contributions. This is because when exchange occurs in $\text{O}_2$+$\text{O}_2$ collision bond breaking happens that ends up the higher probability of the dissociation. In Fig. \ref{Suppfig:Temp_All_TC_VS}, the $\text{O}_2$ composition profiles by the TC model are compared with the VS approach. This demonstrates that the two methods represent reasonable agreement along the considered $T$ range. %

Accuracy of the present modeling strategy for $\text{O}_2$+$\text{O}_2$ dissociation can be verified by comparing QSS vibrational populations with the existing result obtained from the DMS by Grover \emph{et al.} \cite{GROVER_O3O4}. The normalized cumulative vibrational population $f_v$ is defined as
\begin{equation}
	\label{eq:NormCumPop}
	f_v=\frac{\sum_{J=0}^{J_{\text{max}}(v)}n_i\left(v,J\right)}{n_{\text{O}_2}},
\end{equation}
\noindent
where $J_{\text{max}}$ implies the maximum rotational quantum number at the given $v$. It is important to note that comparison of $f_v$ can function as one of necessary conditions, but not as a sufficient one \cite{Venturi2020,MACDONALD_MEQCT_Comp}.

\begin{figure}[h]
	\centering
	\includegraphics[width=0.55\textwidth]{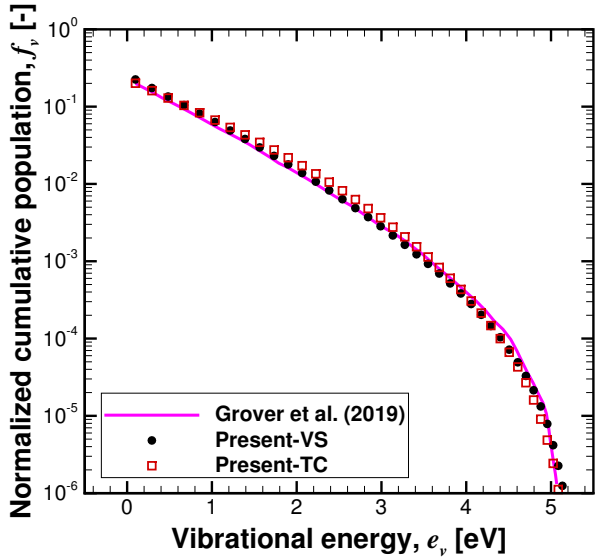}
	\caption{Comparison of normalized vibrational distributions at QSS of $T$=\SI{10000}{\kelvin}.}
	\label{fig:CompNormCumVib_T10000K}
\end{figure}

Figure \ref{fig:CompNormCumVib_T10000K} shows the comparison of $f_v$ at QSS of $T$=\SI{10000}{\kelvin}. The distributions are computed from the three different models. In the QSS region, the low-lying vibrational levels form a Boltzmann distribution whereas the high-lying ones are in strong non-Boltzmann state. The dissociation process is mostly governed by the high energy levels that confirms the $\text{O}_2$+$\text{O}_2$ dissociation occurs in a non-equilibrium manner during the QSS period. As shown in the figure, the results from the different models are in reasonable agreement that demonstrate the capability of the present TC approach to predict the QSS vibrational distribution of the molecule.

\begin{figure}[h]
	\centering
	\subfigure[]
	{
		\includegraphics[width=0.48\textwidth]{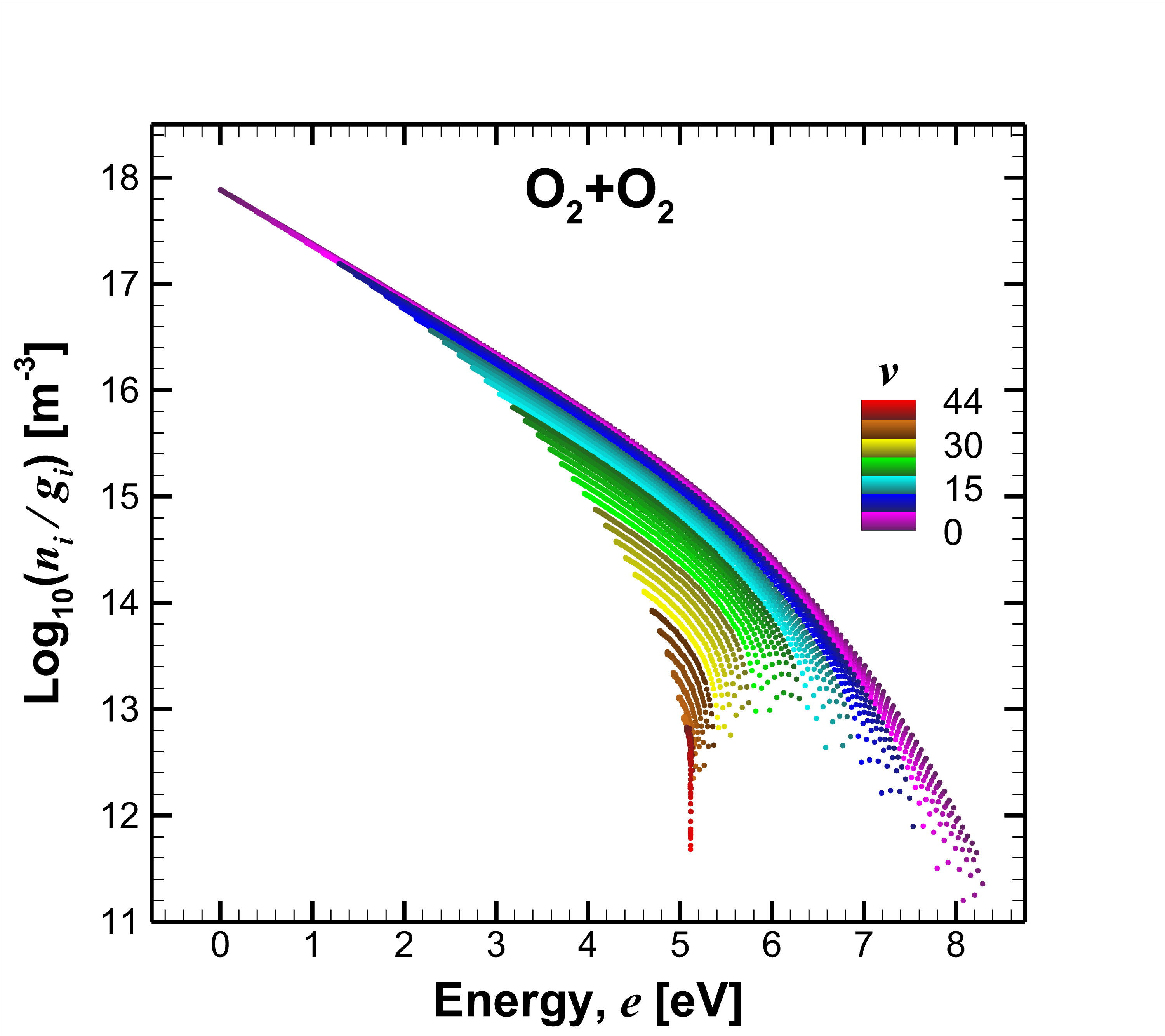}
		\label{fig:QSS_Pop_O4_TC_T10000K}
	}
	\subfigure[]
	{
		\includegraphics[width=0.48\textwidth]{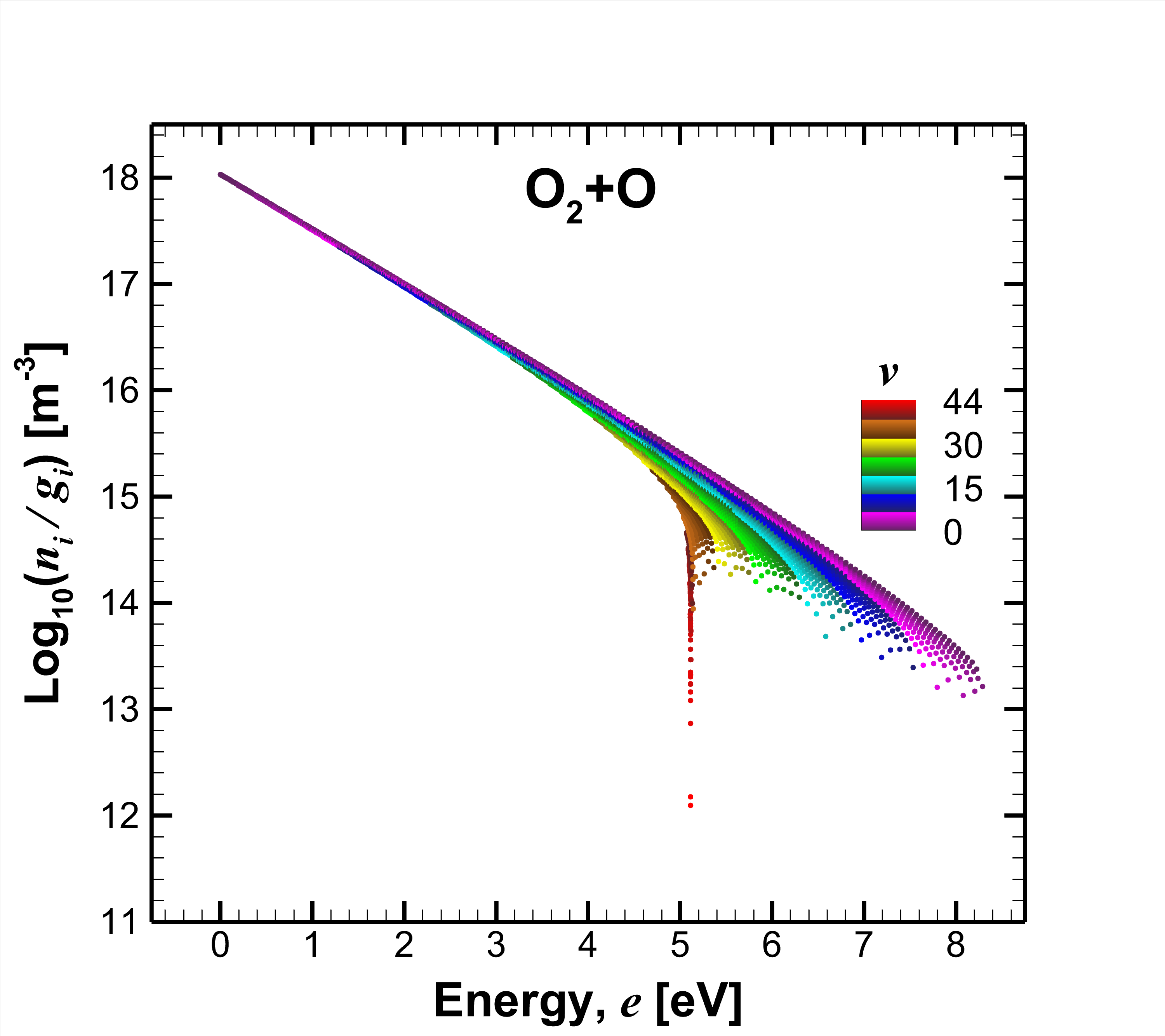}
		\label{fig:QSS_Pop_O3_FullyStS_T10000K}
	}
	\subfigure[]
	{
		\includegraphics[width=0.48\textwidth]{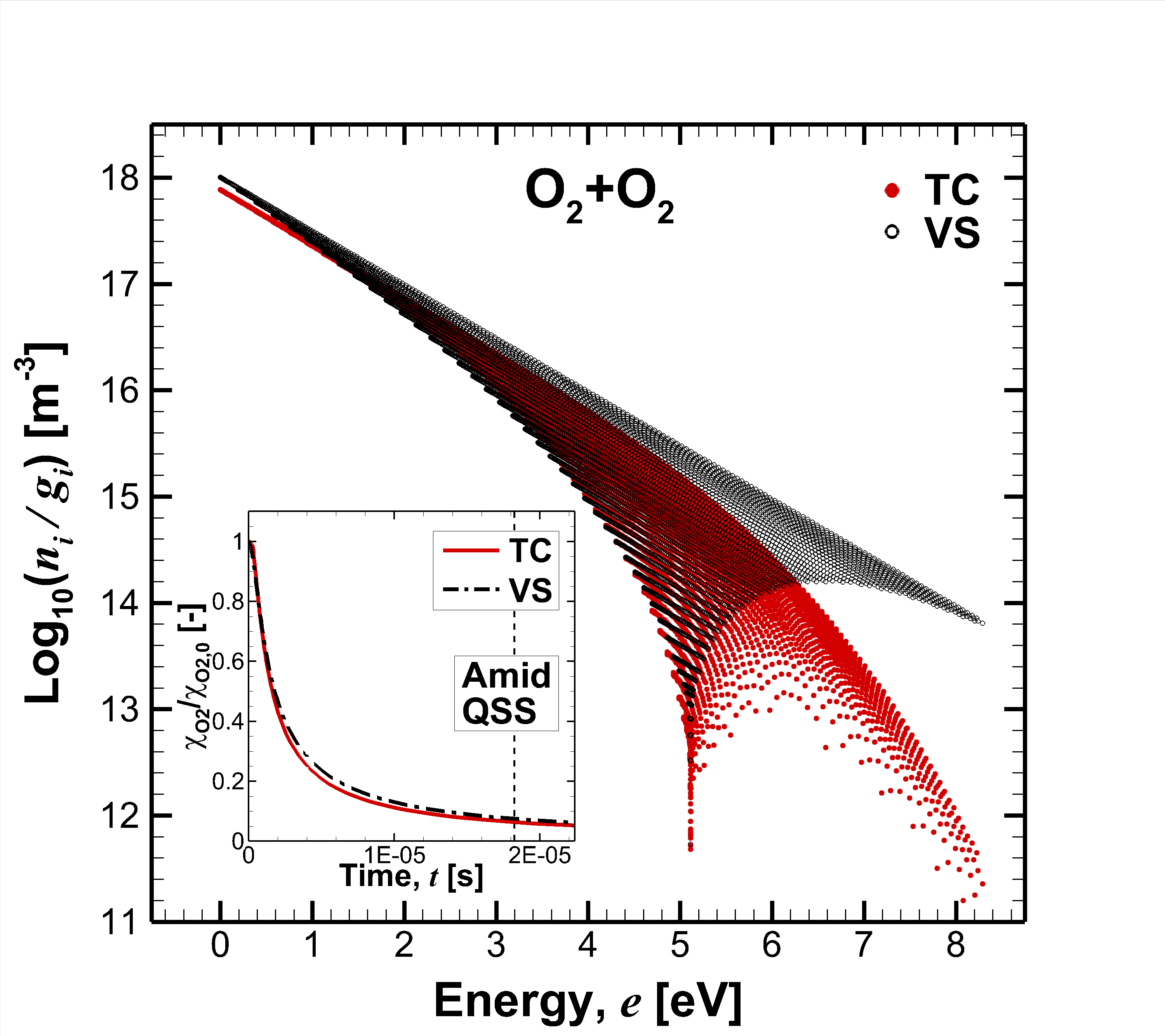}
		\label{fig:QSS_Pop_O4_TC_T10000K_CompVS}
	}
	\subfigure[]
	{
		\includegraphics[width=0.48\textwidth]{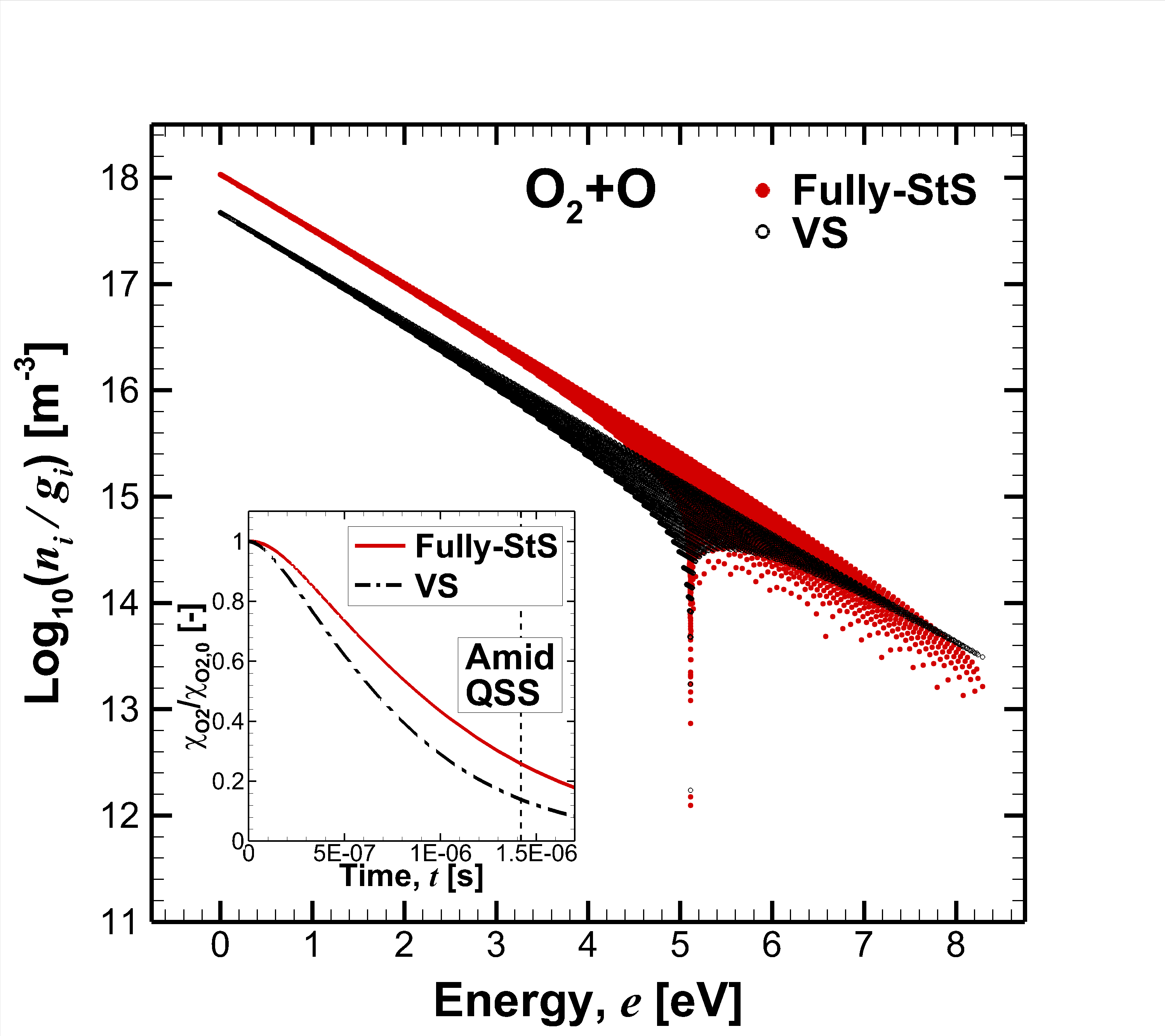}
		\label{fig:QSS_Pop_O3_FullyStS_T10000K_CompVS}
	}
	\caption{QSS rovibrational distributions of $\text{O}_2$+$\text{O}_2$ and $\text{O}_2$+O systems at $T$=\SI{10000}{\kelvin}. (a) and (b): The distributions are colored by the vibrational quantum number. (c) and (d): Comparison with the VS model. The inset figures indicate normalized $\text{O}_2$ mole fraction $\chi_{\text{O}_2}$/$\chi_{\text{O}_2,0}$.}
	\label{fig:QSS_Pop_T10000K}
\end{figure}

The rovibrational distribution at the molecular QSS period is one of the most important ingredients in studying physics behind dissociation kinetics. In this work, the QSS rovibrational distribution of $\text{O}_2$+$\text{O}_2$ is compared with that of $\text{O}_2$+O as shown in Fig. \ref{fig:QSS_Pop_T10000K} for $T$=\SI{10000}{\kelvin}. By comparing Figs. \ref{fig:QSS_Pop_O4_TC_T10000K} and \ref{fig:QSS_Pop_O3_FullyStS_T10000K}, it is found that structures of the distributions are different to each other; overall, $\text{O}_2$+$\text{O}_2$ system has a more scattered rovibrational population. This difference is mainly attributed to the dissimilarity in the behavior of energy transfers (\emph{i.e.}, inelastic and exchange), rather than the dissociation dynamics. As discussed in Sec. \ref{sec:energy}, $\text{O}_2$+$\text{O}_2$ system represents the significantly slower thermalization rate among the internal states, especially for the vibrational strands. This ends up the more scattered population distribution at the time the system enters the QSS period in which majority of $\text{O}_2$ molecule dissociates.

In Figs. \ref{fig:QSS_Pop_O4_TC_T10000K_CompVS} and \ref{fig:QSS_Pop_O3_FullyStS_T10000K_CompVS}, the QSS distributions are compared with those from the VS model. The rotational populations by the VS model are reconstructed by assuming the Boltzmann distribution at $T$. The comparisons are made in the middle of QSS regions, determined by the TC model in $\text{O}_2$+$\text{O}_2$ and by the Fully-StS approach in $\text{O}_2$+O, respectively. The inset figures present normalized $\text{O}_2$ mole fractions where $\chi_{\text{O}_2,0}$ implies the initial value of $\chi_{\text{O}_2}$. In $\text{O}_2$+$\text{O}_2$, the QSS distributions are well matched for the bound levels, whereas substantial discrepancy exists for the quasi-bound states. The VS model overestimates the quasi-bound populations. Albeit that difference, both models predict almost identical mole fraction profiles. In $\text{O}_2$+O, the mole fraction evolution calculated by the VS model is distinctly different from the Fully-StS approach: The VS model overestimates the amount of dissociation. In the previous study by Venturi \emph{et al.} \cite{Venturi2020}, the similar issue was addressed and an improved coarse-grain model for the $\text{O}_2$+O dissociation kinetics was designed by leveraging the physics behind the diatomic potential and the centrifugal barrier. 
Since the faster thermalization quickly squeezes the rovibrational distribution, the high-lying levels around the dissociation limit have relatively larger contributions to the amount of dissociation in $\text{O}_2$+O, compared with $\text{O}_2$+$\text{O}_2$ in which the equilibration among the vibrational states is much slower in nature. 
%

\begin{figure}[h]
	\centering
	\subfigure[]
	{
		\includegraphics[width=0.48\textwidth]{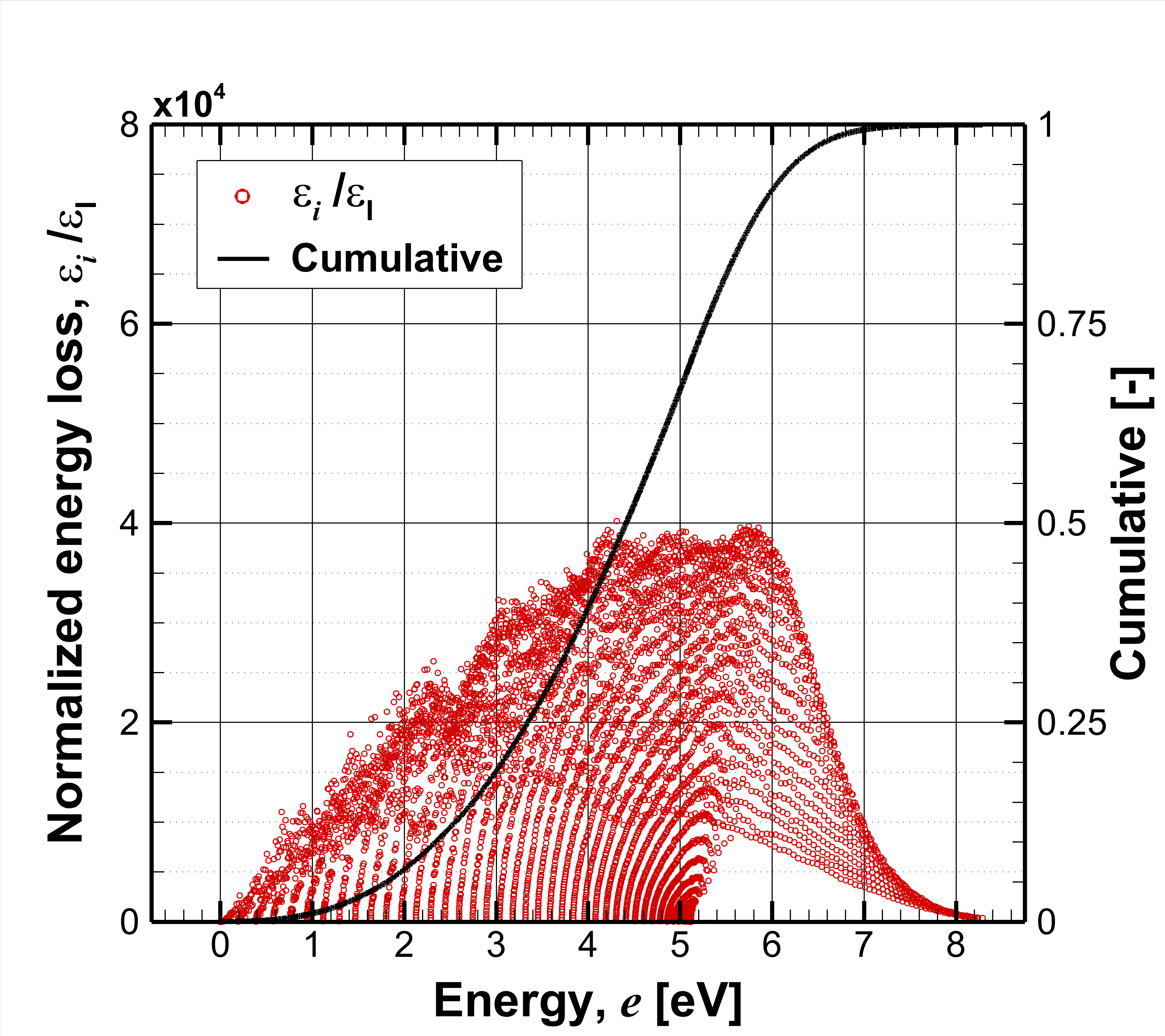}
		\label{fig:NormEintLoss_O4DissTar_T10000K}
	}
	\subfigure[]
	{
		\includegraphics[width=0.48\textwidth]{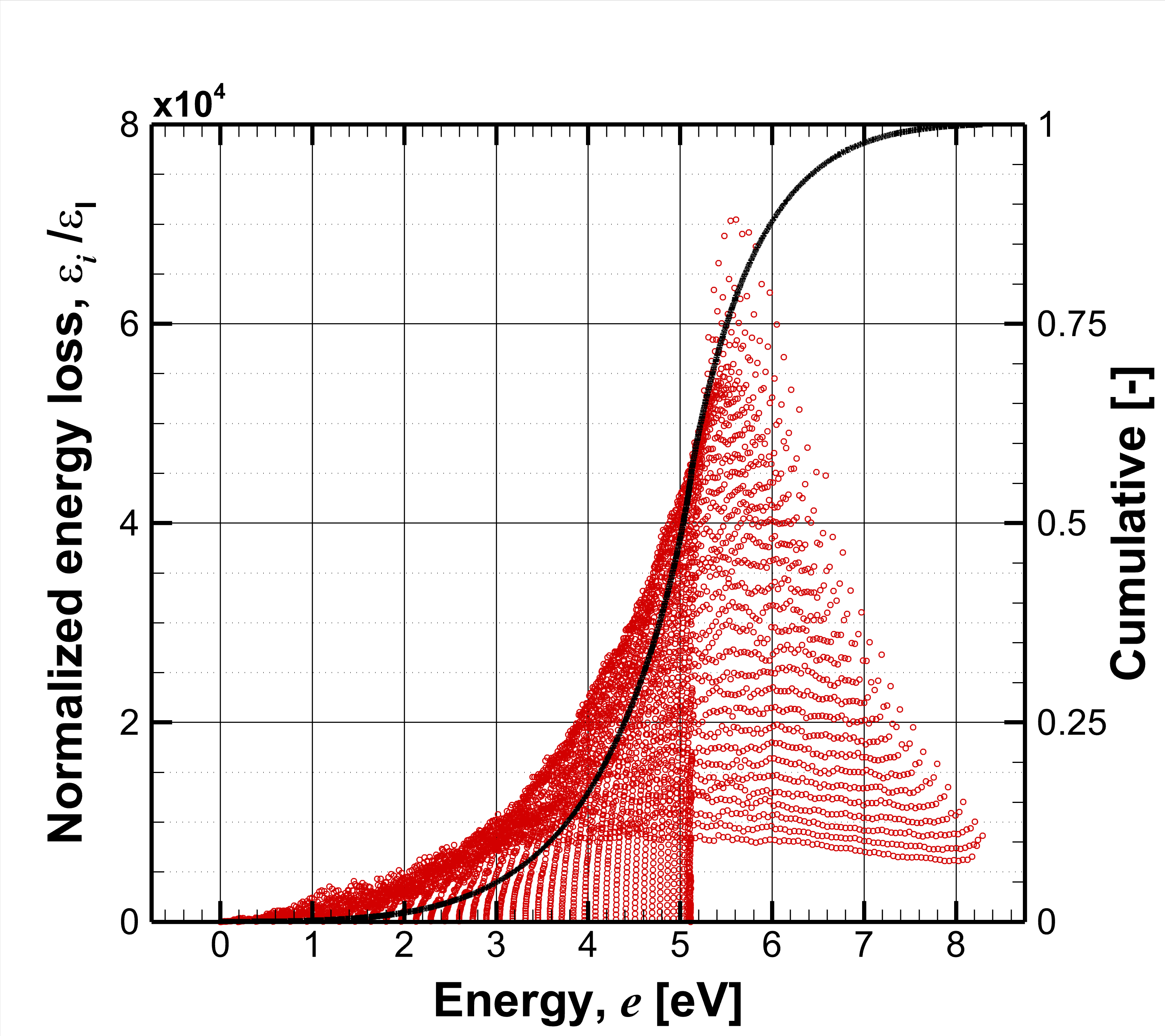}
		\label{fig:NormEintLoss_O3_T10000K}
	}
	\caption{Comparison of normalized rovibrational energy loss ratio $\epsilon_i$/$\epsilon_{\text{I}}$ at the QSS of $T$=\SI{10000}{\kelvin}. (a) $\text{O}_2$+$\text{O}_2$ dissociation through Eq. \eqnref{eq:O4_TargetDiss_TC}. (b) $\text{O}_2$+O.}
	\label{fig:NormEintLoss_T10000K}
\end{figure}

In Fig. \ref{fig:NormEintLoss_T10000K}, normalized rovibrational energy loss ratio $\epsilon_i$/$\epsilon_{\text{I}}$ is compared between $\text{O}_2$+$\text{O}_2$ and $\text{O}_2$+O systems. The state-specific internal energy loss ratio $\epsilon_i$ is calculated using the definition proposed by Panesi \emph{et al.} \cite{PANESI_2013_BOXRVC} where $\epsilon_{\text{I}}$ is accumulated quantity of $\epsilon_i$ over the entire internal states. By using the present TC model for the $\text{O}_2$+$\text{O}_2$ system, the $\epsilon_i$ can only be defined for the dissociation process in Eq. \eqnref{eq:O4_TargetDiss_TC}. This fact is one of the drawbacks of the present TC approach. In $\text{O}_2$+$\text{O}_2$, the distribution has more gradual shape, whereas it shows a \emph{spike}-like structure in $\text{O}_2$+O centered around the dissociation limit. From the comparison, it is found that in $\text{O}_2$+$\text{O}_2$ the quasi-bound states have much less contribution to the dissociation-energy coupling than that of $\text{O}_2$+O. This is attributed to the lower amount of rovibrational populations of the quasi-bound levels in $\text{O}_2$+$\text{O}_2$ as discussed in Fig. \ref{fig:QSS_Pop_T10000K}.

\subsection{Application to combined $\text{O}_3$+$\text{O}_4$ system}\label{sec:O3O4}
In this section, the master equation analysis is extended to complete oxygen mixture that includes both $\text{O}_2$+$\text{O}_2$ and $\text{O}_2$+O interactions. Purpose of this extension is to investigate relative contributions from those two collisional systems to the internal energy transfer and dissociation of $\text{O}_2$ in the full oxygen mixture. The set of master equations corresponding to the combined $\text{O}_3$+$\text{O}_4$ system is presented in App. \ref{sec:master_O3O4}. The present results are compared with those of existing study by Grover \emph{et al.} \cite{GROVER_O3O4} obtained from the DMS method. For this comparison, the chemical components are initialized to have 100\% of $\text{O}_2$ and initial number density of 2.414$\times$10$^{19}$ cm$^{-3}$. The initial rotational and vibrational temperatures are set to 1394.3 K and 985.1 K, respectively. Then the kinetic temperature $T$ is suddenly raised to \SI{10000}{\kelvin} and kept constant to study the non-equilibrium energy transfer and dissociation processes.

\begin{figure}[h]
	\centering
	\subfigure[]
	{
		\includegraphics[width=0.48\textwidth]{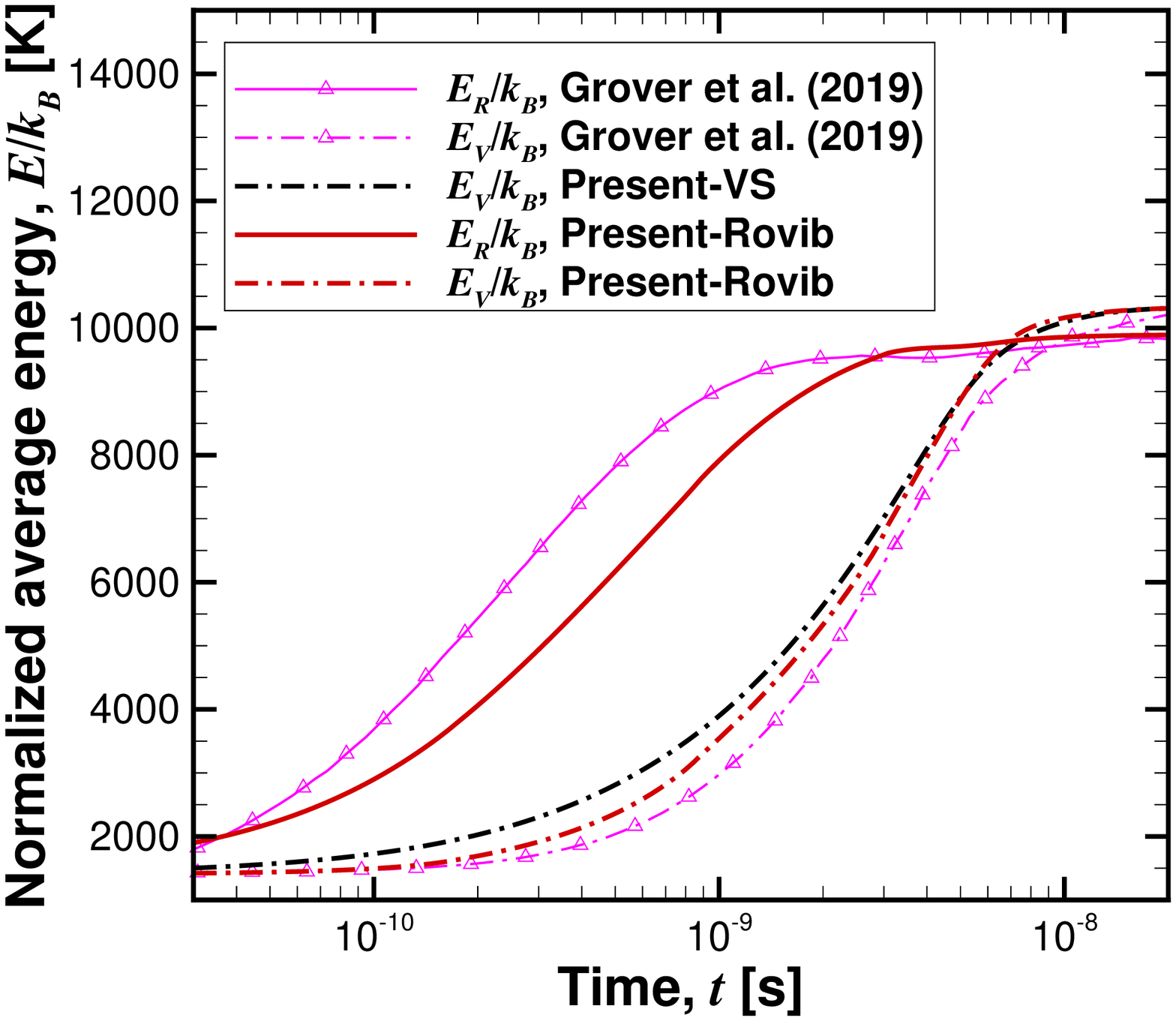}
		\label{fig:DMS_Case_Energy}
	}
	\subfigure[]
	{
		\includegraphics[width=0.48\textwidth]{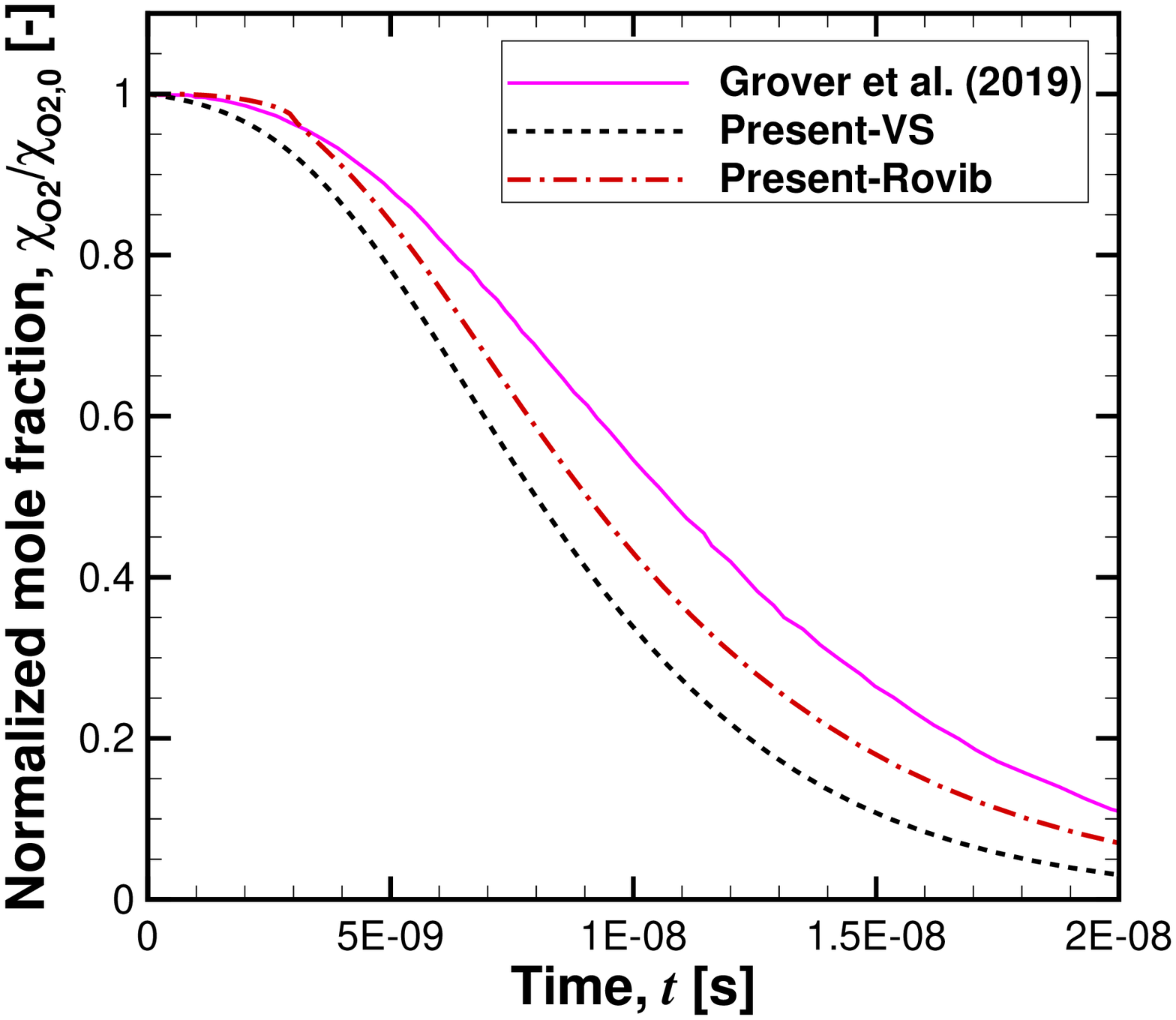}
		\label{fig:DMS_Case_MoleFrac}
	}
	\caption{Temporal evolution of (a) normalized average energy and (b) $\text{O}_2$ mole fraction. Comparison with the previous result by Grover \emph{et al.} \cite{GROVER_O3O4}.}
	\label{fig:DMS_Case}
\end{figure}

Figure \ref{fig:DMS_Case} presents normalized average internal energy and $\text{O}_2$ mole fraction profiles. It is important to note that for the present results shown in the figure the recombination process is turned off for consistency with the DMS result by Grover \emph{et al.} \cite{GROVER_O3O4}. Among the results, \emph{Present-Rovib} implies a rovibrational-specific approach that is a hybrid of the TC model for $\text{O}_2$+$\text{O}_2$ and the Fully-StS method for $\text{O}_2$+O. For both average vibrational energy and $\text{O}_2$ mole fraction profiles, the present result by the rovibrational-specific approach represents better agreement than the VS model compared to the reference DMS data \cite{GROVER_O3O4}. This improvement on the accuracy is mostly attributed to the rovibrational-specific treatment of the $\text{O}_2$+O rather than $\text{O}_2$+$\text{O}_2$, since we observed that the VS and the TC approaches provide very similar macroscopic mole fraction profiles as shown in Fig. \ref{fig:QSS_Pop_O4_TC_T10000K_CompVS}. The present average rotational energy profile shows a larger discrepancy from the DMS result than the vibration during the non-equilibrium energy transfer. According to the previous studies by Venturi \emph{et al.} \cite{Venturi2020} and Macdonald \emph{et al.} \cite{Macdonald_2020}, it was demonstrated that the StS master equation approach yields very similar results with those from the DMS provided that the identical PES is adopted. Therefore, the discrepancy of the average rotational energy profile in Fig. \ref{fig:DMS_Case_Energy} requires further investigation.

\begin{figure}[h!]
	\centering
	\subfigure[]
	{
		\includegraphics[width=0.7\textwidth]{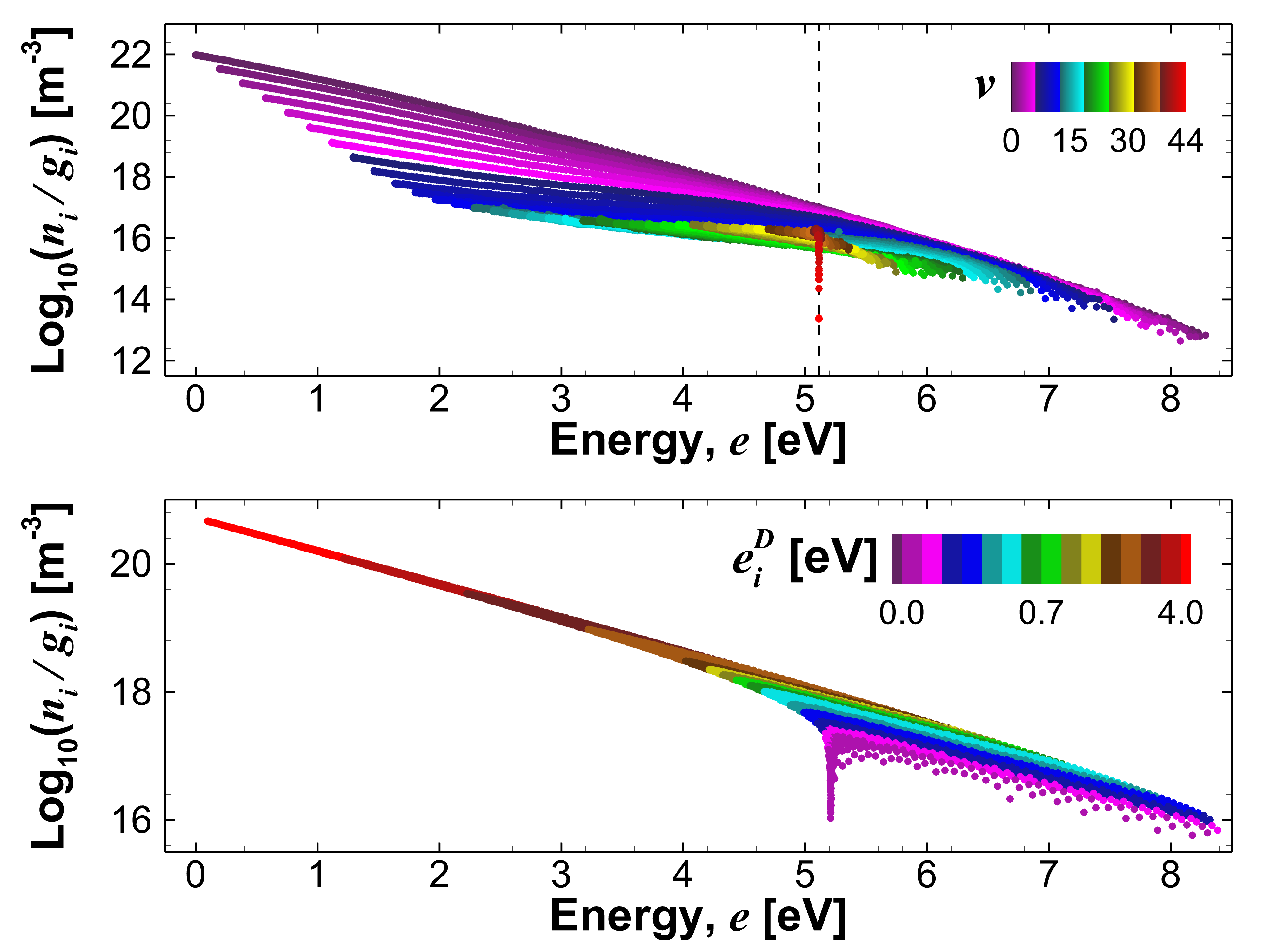}
		\label{fig:DMS_Case_Pop}
	}
	\subfigure[]
	{
		\includegraphics[width=0.48\textwidth]{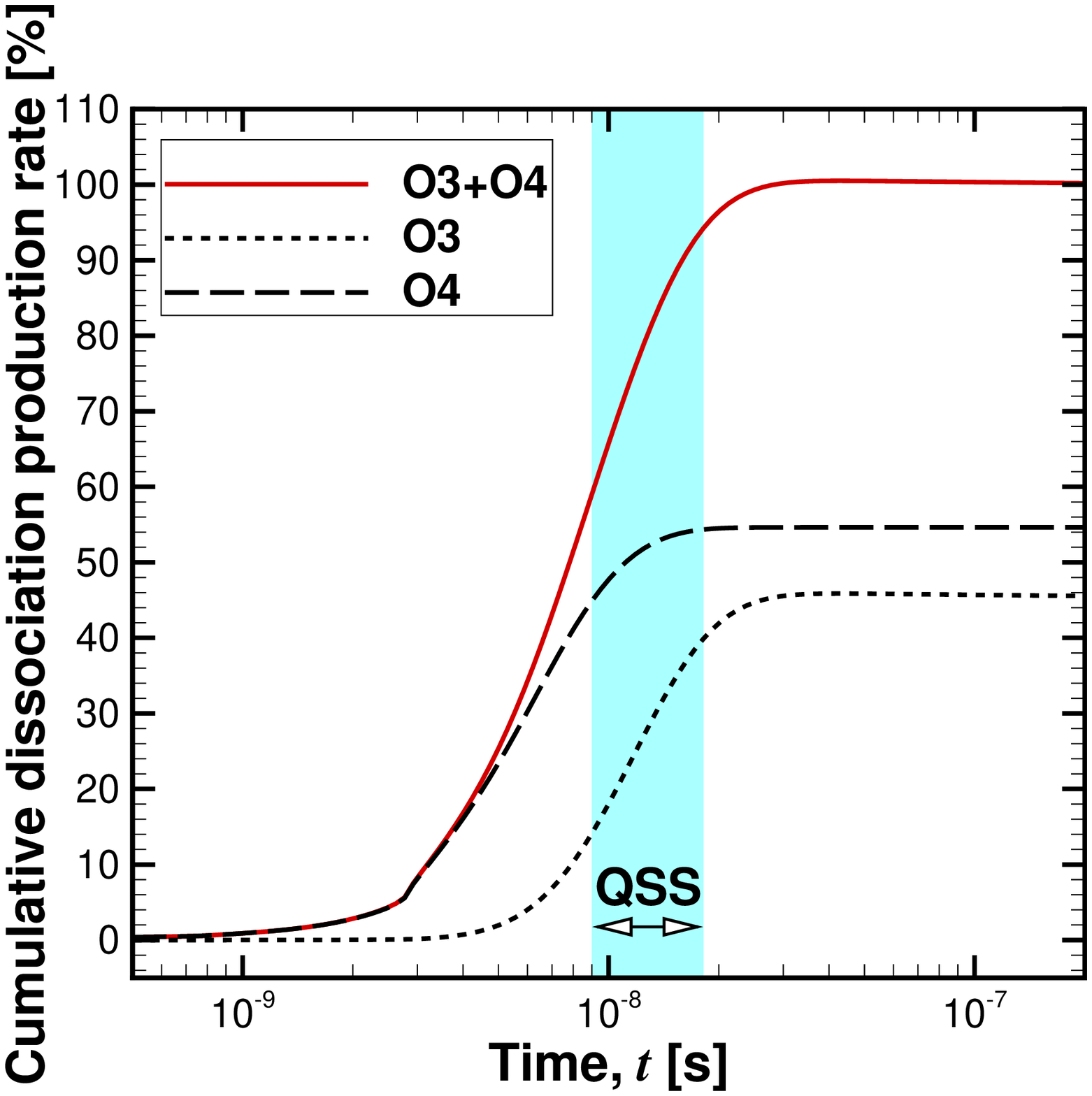}
		\label{fig:Cumulative_Diss_ProdRate_T10000K_wRecomb}
	}
	\caption{(a) Rovibrational distributions in $\text{O}_3$+$\text{O}_4$ system at 10\% of vibrational relaxation (top, colored by $v$) and QSS (bottom, colored by $e_i^D$). (b) Time-cumulative dissociation production rate. The cyan-shaded region indicates the QSS period of $\text{O}_2$.}
	\label{fig:DMS_Case_DecomposeComp}
\end{figure}

To the authors' best knowledge, no detailed rovibrational distribution has been presented for the complete oxygen system. In Fig. \ref{fig:DMS_Case_DecomposeComp}, the evolution of rovibrational distributions and the time-cumulative dissociation production rate are presented. It should be noted that the recombination is now turned on to simulate the results shown in Fig. \ref{fig:DMS_Case_DecomposeComp}. Figure \ref{fig:DMS_Case_Pop} confirms that the $\text{O}_2$+$\text{O}_2$ collision governs the early stage of energy transfer, since the rovibrational distribution at 10\% of vibrational relaxation represents very similar structure with that of isolated $\text{O}_2$+$\text{O}_2$ shown in Fig. \ref{fig:CompPop_O3O4_20PercVib_T10000K}. At the QSS however, the $\text{O}_2$+O collision takes control of the oxygen mixture. This can be inferred from the fact that the QSS distribution shown in Fig. \ref{fig:DMS_Case_Pop} is almost identical with that of isolated $\text{O}_2$+O shown in Fig. \ref{fig:QSS_Pop_O3_FullyStS_T10000K}. One more important takeaway from Fig. \ref{fig:DMS_Case_Pop} is the QSS distribution of the complete oxygen system (\emph{i.e.}, bottom) is clearly correlated with the each rovibrational level's energy deficit $e_i^D$. This implies that the coarse-grain strategy \cite{Venturi2020}, which was developed for the diatomic dissociation by correlating the energy deficit with the multi-group maximum entropy approach \cite{Yen_2015}, can also be applied to the complete oxygen mixture case.

The quantity presented in Fig. \ref{fig:Cumulative_Diss_ProdRate_T10000K_wRecomb} is defined by summing up and normalizing the mass production rate for the dissociation and recombination in Eq. \eqnref{eq:ME-TC_O2} at given time-step with the total amount of chemical production rate relevant to dissociation and recombination. This definition was taken from the previous study by Jo \emph{et al.} \cite{Jo_N2O}. From the figure, it is found that before the onset of QSS most of the dissociation is dominated by $\text{O}_2$+$\text{O}_2$ collision. During the QSS period, about 35\% of dissociation occurs in which $\text{O}_2$+O system has larger contribution than $\text{O}_2$+$\text{O}_2$. This observation is in line with the rovibrational distribution in Fig. \ref{fig:DMS_Case_Pop}. It is important to note that in total $\text{O}_2$+$\text{O}_2$ collision has larger contribution to the cumulative dissociation production rate. This is because the 4-atom system has more various dissociation pathways compared to the 3-atom system.

\subsection{Comparison of characteristic quantity against existing data}\label{sec:Comp_Literature}
In this section, the macroscopic quantities such as rovibrational relaxation time, dissociation rate coefficients, and average energy loss ratio obtained from the present analyses are compared with existing data. 

\begin{figure}[h!]
	\centering
	\includegraphics[width=0.55\textwidth]{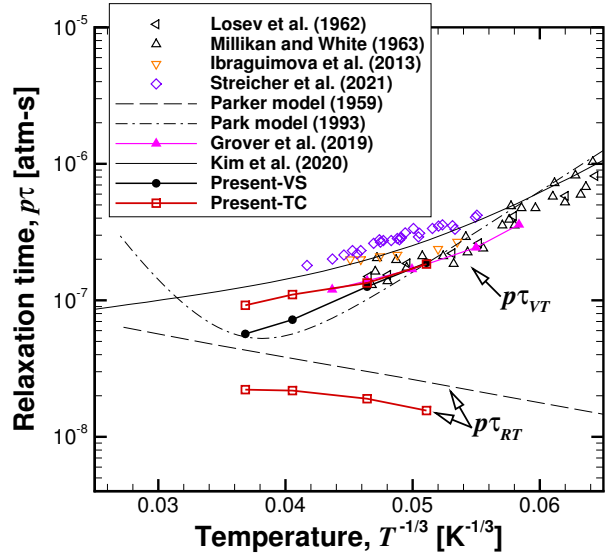}
	\caption{Comparison of predicted rotational and vibrational relaxation times in $\text{O}_2$+$\text{O}_2$ system with existing experimental \cite{losev1962study,Ibraguimova_2013,Streicher_2021} and theoretical \cite{Parker_1959,MW_1963,Park_1993_Earth,GROVER_O3O4,Kim_2020} data.}
	\label{fig:Tau}
\end{figure}

Figure \ref{fig:Tau} compares the rotational and vibrational relaxation times of $\text{O}_2$+$\text{O}_2$ system with data in the literature \cite{losev1962study,Ibraguimova_2013,Streicher_2021,Parker_1959,MW_1963,Park_1993_Earth,GROVER_O3O4,Kim_2020}. The symbol $p$ implies the pressure of collision partner. The present vibrational relaxation time by the TC model is in good agreement with the DMS data \cite{GROVER_O3O4}. The result from the VS approach however shows the distinctly shorter relaxation time at $T$=\SI{15000}{\kelvin} and \SI{20000}{\kelvin}, and represents closer agreement with the Park model including high-temperature correction \cite{Park_1993_Earth}. In comparison with the measured data \cite{losev1962study,Ibraguimova_2013,Streicher_2021}, the present TC result has slightly improved agreement than the VS model. It is worth mentioning that the $p\tau_{VT}$ data measured from the recent experiments \cite{Ibraguimova_2013,Streicher_2021} represents longer relaxation time than the present results. The rotational relaxation time $p\tau_{RT}$ by the present TC model shows significant discrepancy compared to the Parker model \cite{Parker_1959}, which was derived by classical mechanics along using the rigid-rotor approximation. From the present result by the TC model, it is found that in $\text{O}_2$+$\text{O}_2$ collision the convergence of the rotational and vibrational relaxation time to a common asymptote in high-temperature range \cite{Kim_2021} is relatively weak in comparison to other 3-atom systems, such as $\text{N}_2$+N \cite{PANESI_2013_BOXRVC,Kim_Boyd_Cphys_2013}, $\text{O}_2$+O \cite{VENTURI_JUN2019}, $\text{N}_2$+O and NO+N \cite{Jo_N2O}.

\begin{figure}[h!]
	\centering
	\includegraphics[width=0.60\textwidth]{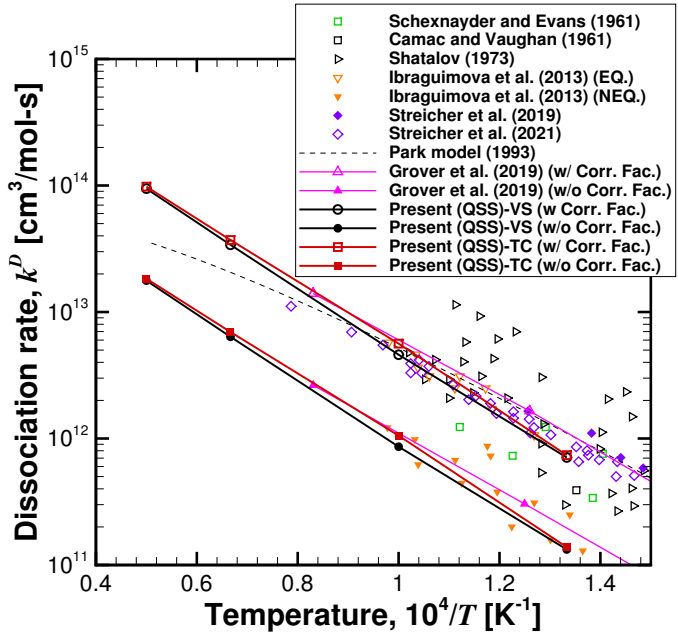}
	\caption{Comparison of predicted QSS dissociation rate coefficients in $\text{O}_2$+$\text{O}_2$ system with existing experimental \cite{Schexnayder_1961,Camac_1961,Shatalov_O2,Ibraguimova_2013,Streicher_2019,Streicher_2021} and theoretical \cite{Park_1993_Earth,GROVER_O3O4} data.}
	\label{fig:Diss_overall}
\end{figure}

Figure \ref{fig:Diss_overall} compares the present QSS dissociation rate coefficients in $\text{O}_2$+$\text{O}_2$ system with data from literature \cite{Schexnayder_1961,Camac_1961,Shatalov_O2,Ibraguimova_2013,Streicher_2019,Streicher_2021,Park_1993_Earth,GROVER_O3O4}. Since some of excited electronic states of $\text{O}_2$ are closely spaced near the ground level, the multi-surface correction factor \cite{DISS_CORR_FACTOR} 16/3 is multiplied to the present QSS rate coefficients to consider the dissociating contribution from the excited electronic states. Those results are labeled as \emph{w Corr. Fac.} in the figure. The present results from the TC and VS models are in very close agreement to each other, and they are well matched with the DMS data \cite{GROVER_O3O4}. In case that the multi-surface correction factor is considered, the predicted data by this work shows better agreement with the measurements \cite{Schexnayder_1961,Camac_1961,Shatalov_O2,Streicher_2019,Streicher_2021}, except for the non-equilibrium experiment by Ibraguimova \emph{et al.} \cite{Ibraguimova_2013}. This measured data is closer to the present result that does not include the correction factor. The trend of inconsistency implies the uncertainty of both measurement and modeling to consider dissociation from the excited electronic states. It is worth mentioning that the Park model \cite{Park_1993_Earth} represents somewhat different temperature dependency compared to the present data in the high-temperature region. In Fig. \ref{Suppfig:O4_QSS_DissRate_Decomp}, the overall QSS dissociation rates predicted by the TC model shown in Fig. \ref{fig:Diss_overall} are decomposed into the individual component by the two reaction channels in Eqs. \eqnref{eq:O4_TargetDiss_TC} and \eqnref{eq:O4_ColliderDiss_TC}. 

\begin{figure}[h!]
	\centering
	\subfigure[]
	{
		\includegraphics[width=0.48\textwidth]{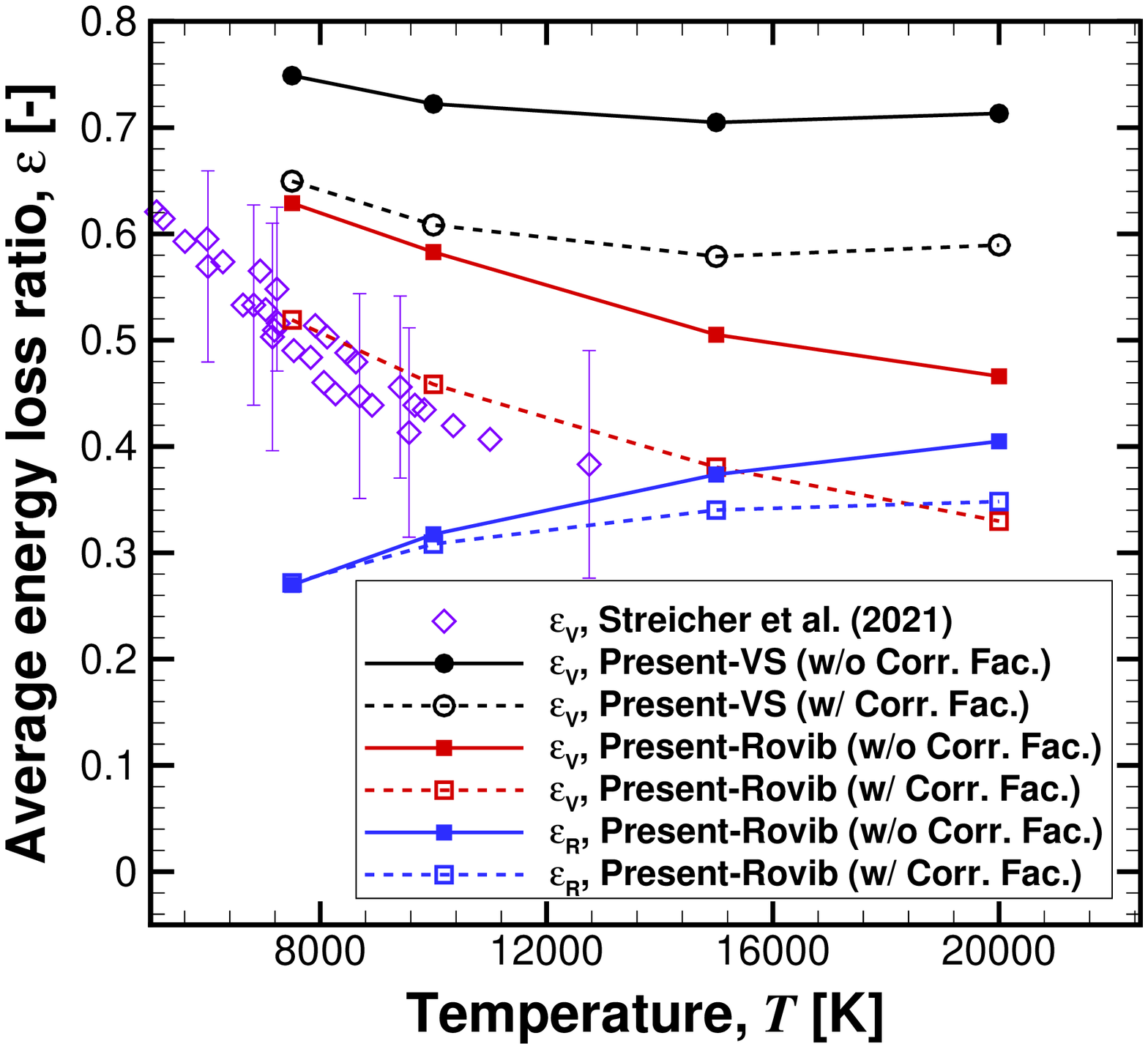}
		\label{fig:Avg_LossRatio}
	}
	\subfigure[]
	{
		\includegraphics[width=0.48\textwidth]{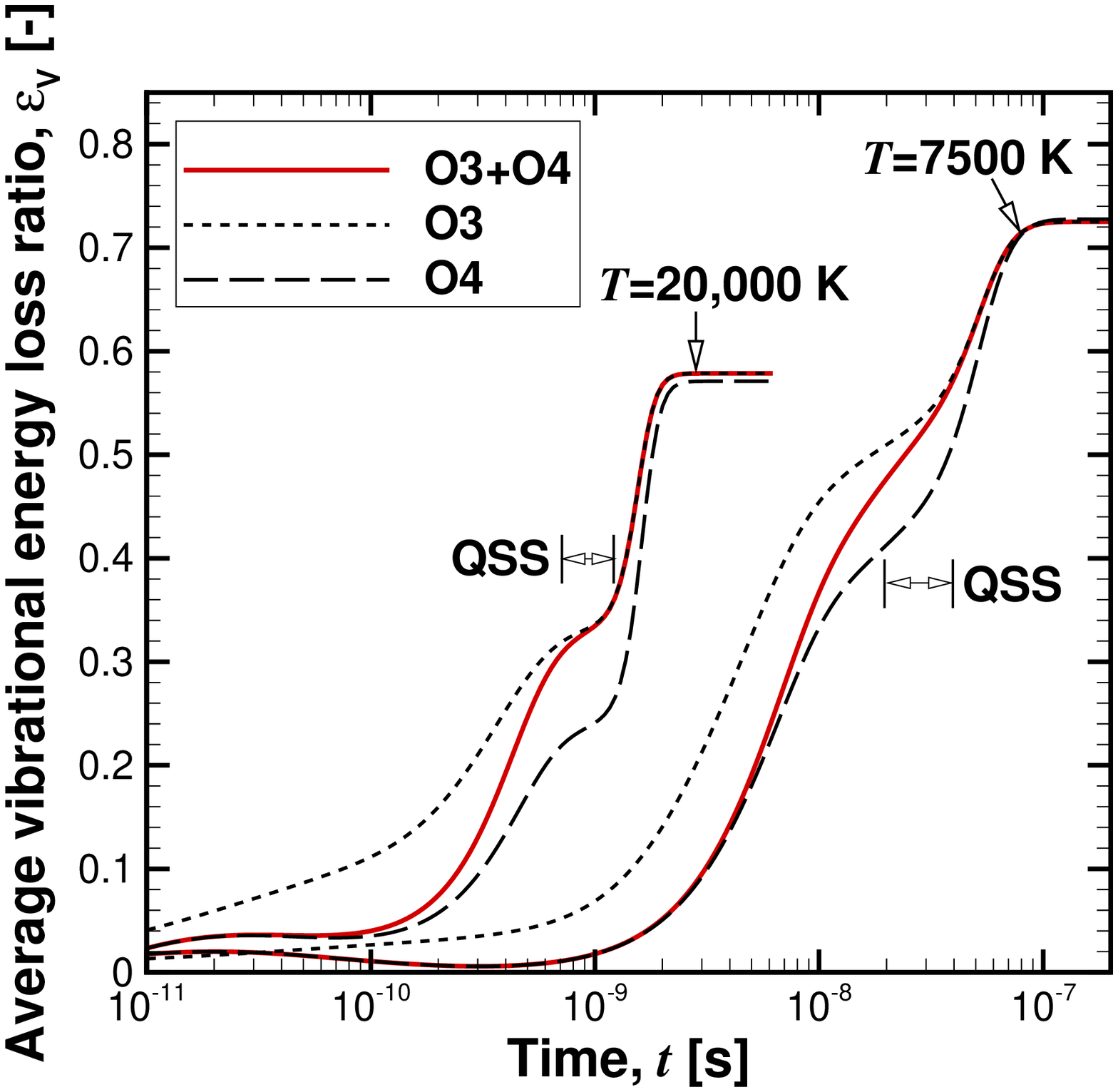}
		\label{fig:Vib_LossRatio}
	}
	\caption{Comparison of average energy loss ratio in $\text{O}_3$+$\text{O}_4$ system. (a) Comparison of the predicted QSS values with experimental data \cite{Streicher_2021}. (b) Temporal evolution at $T$=\SI{7500}{\kelvin} and \SI{20000}{\kelvin} that obtained from the \emph{Present-Rovib} approach with the multi-surface correction factor.}
	\label{fig:EnergyLoss}
\end{figure}

In Fig. \ref{fig:EnergyLoss}, comparison of the average energy loss ratio for the combined oxygen mixture (\emph{i.e.}, $\text{O}_3$+$\text{O}_4$) is presented. It is important to note that unlike in Fig. \ref{fig:Diss_overall} the multi-surface correction factor here is multiplied to the rovibrational-specific dissociation rate coefficients to consider the nonlinear nature in the master equations (See Eq. \eqnref{eq:ME-TC_O2}). In Fig. \ref{fig:Avg_LossRatio}, the predicted results are compared with the measured vibrational energy loss ratio by Streicher \emph{et al.} \cite{Streicher_2021}. The present rovibrational-specific result shows closer agreement with the measured data than the VS approach which overpredicts the measured quantity. Including the multi-surface correction factor improves the agreement between the experimental data and the present rovibrational-specific calculations by lowering the predicted value. This fact implies that the contribution of the excited electronic states to the dissociation is commonly captured by both the experiment and the numerical modeling. It is found that the average rotational energy loss ratio is relatively less sensitive to the multi-surface correction, especially at $T$=\SI{7500}{\kelvin} and \SI{10000}{\kelvin}. This might be attributed to the fact that the dissociation is mostly controlled by the high-$v$ and low-$J$ levels at those temperatures. Then as $T$ increases, the high-$J$ levels at low-$v$ region have larger contribution that ends up the greater sensitivity to the multi-surface correction. 
It should be noted that the QSS value of the average energy loss ratio of the combined $\text{O}_3$+$\text{O}_4$ is almost identical with that of $\text{O}_3$ (\emph{i.e.}, $\text{O}_2$+O) only. It is because the dissociation of $\text{O}_3$+$\text{O}_4$ during the QSS period is mostly governed by the O3 as observed from Fig. \ref{fig:DMS_Case_DecomposeComp}. This fact is supported by Fig. \ref{fig:Vib_LossRatio} showing the decomposition of the average vibrational energy loss ratio. During the QSS period, the overall curve in red is very close to that of $\text{O}_3$. From the present investigation, we concluded that the measured quantity by Streicher \emph{et al.} \cite{Streicher_2021} is most likely regarding the $\text{O}_2$+O collision rather than $\text{O}_2$+$\text{O}_2$.

\section{Conclusions}\label{sec:conc}
This work proposes a reduced-order rovibrational kinetic model of the $\text{O}_2$+$\text{O}_2$ system for applications relevant to high-enthalpy flows. To overcome the exponential number of possible reactive channels that characterizes the diatom-diatom systems, the population of internal levels of one of the reactants is assumed to follow a Boltzmann distribution at a prescribed internal temperature. Detailed QCT calculations are performed to determine the relevant inelastic and reactive reaction rate coefficients, carefully selecting between the endothermic and exothermic directions, the one characterized by the lower statistical error. The rovibrational-specific kinetic database constructed covers a wide temperature range (7500-\SI{20000}{\kelvin}) and is used in the master equation solver to study the kinetics of thermochemical relaxation in isothermal isochoric conditions. To the author's best knowledge, this work constitutes the first application of the rovibrational master equation approach to a diatom-diatom system (except for the $\text{H}_2$+$\text{H}_2$ system, which is far more straightforward). The results obtained have been compared with the DMS calculations showing excellent agreement in the population distribution and averaged energy and composition. 

The main findings are summarized below:
\begin{itemize}
	\item The rovibrational population distribution of the $\text{O}_2$ molecules is found to depart from equilibrium throughout the dissociation process. The extent of non-equilibrium is more pronounced if compared to the $\text{O}_2$+O case. Furthermore, the vibrational-rotational mixing is less efficient than in the $\text{O}_2$+O case leading to a more distinct separation of the rotational strands characterized by a constant vibrational quantum number. The strands exhibit significant curvature due to the less efficient inelastic and exchange energy processes. 
	
	\item The internal temperature used to prescribe the population of the collision partner has an impact on the results. It leads to a parametrization of the reaction rate parameters in the function of $T$ and $T_{\text{int}}$.
	
	\item The analysis of the combined $\text{O}_3$+$\text{O}_4$ system shows how $\text{O}_2$+$\text{O}_2$ collision governs the energy transfer in the early stage, while $\text{O}_2$+O takes control of the dissociation physics in correspondence to the QSS period. This is an important finding as it facilitates the construction of reduced-order models for oxygen chemistry. 
	
	\item Finally, the predicted macroscopic quantities, such as the rovibrational relaxation times, QSS dissociation rate coefficients, and dissociation-vibration coupling constants, agree with the available literature data. This fact provides validity to the physical model devised in the present work. 
\end{itemize}

\section*{Acknowledgments} This work is supported by AFOSR Grant No. FA9550-22-1-0039 with Dr. Sarah Popkin as Program Officer. The authors would like to thank Dr. R. L. Jaffe (NASA AMES Research Center) for the useful discussions about QCT calculations.

\newpage

\appendix
\renewcommand{\thefigure}{A\arabic{figure}}   
\setcounter{figure}{0}    
\section{Characteristics of sampled trajectories at $T_{\text{int}}$}\label{sec:Sampled_Trajectories}
\begin{figure}[h]
	\centering
	\subfigure[]
	{
		\includegraphics[width=0.44\textwidth]{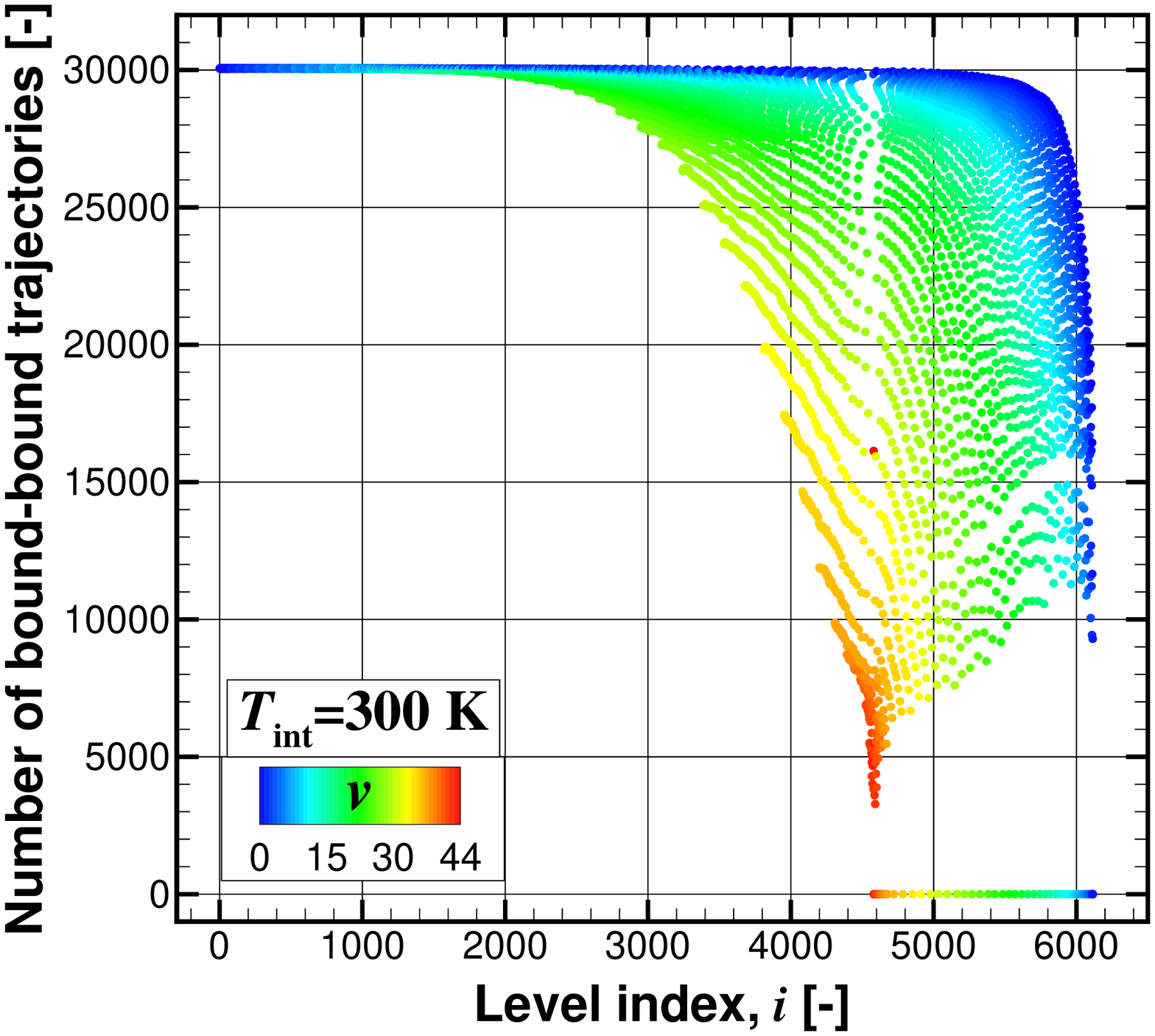}
		\label{fig:Bound-bound_Traj_Num}
	}
	\subfigure[]
	{
		\includegraphics[width=0.44\textwidth]{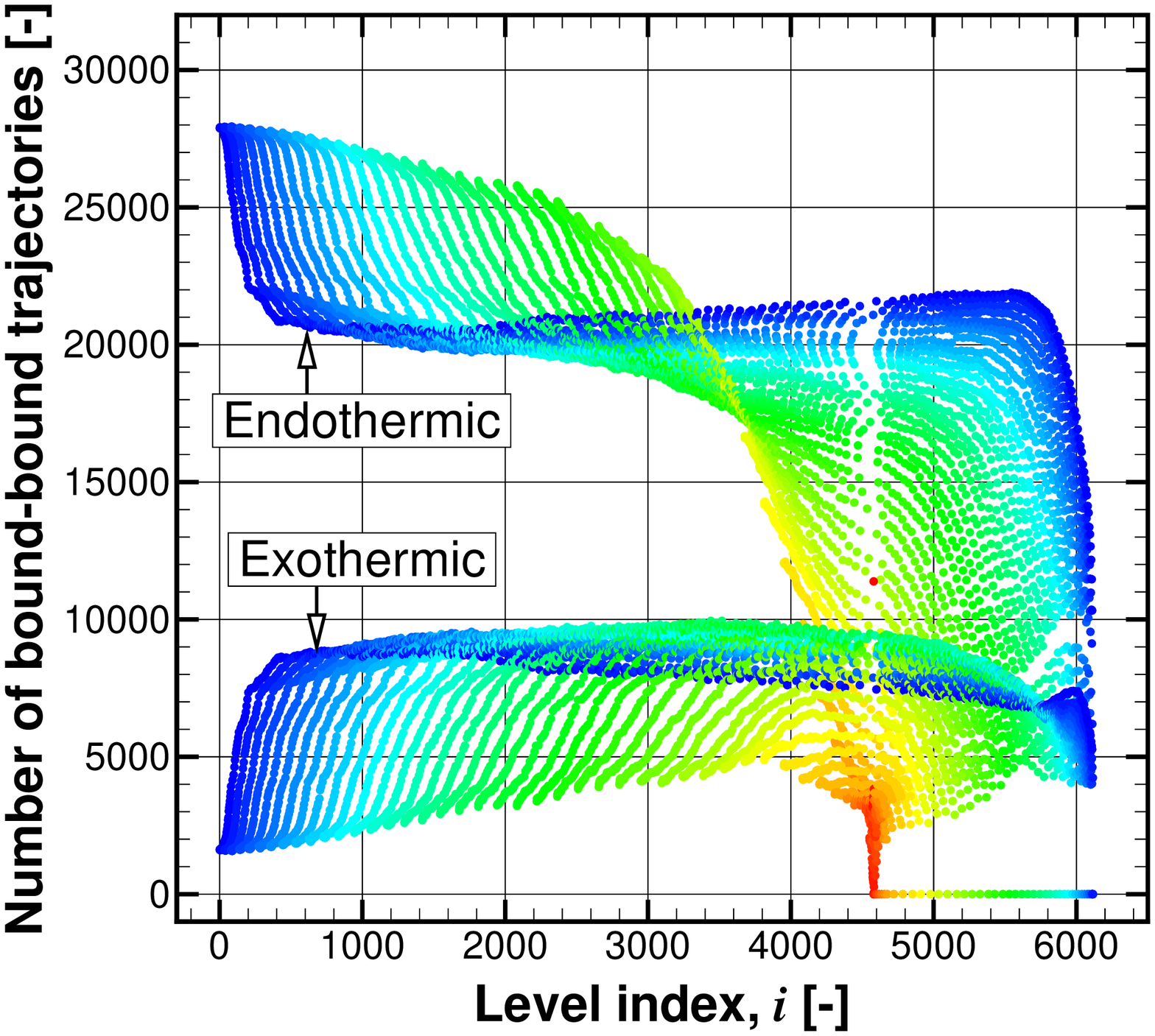}
		\label{fig:Bound-bound_Traj_Decomp}
	}
	\caption{Number of sampled bound-bound trajectories at $T$=\SI{10000}{\kelvin} and $T_{\text{int}}$=\SI{300}{\kelvin}. (a) Total and (b) decomposition into endothermic and exothermic trajectories. The distributions are colored by the vibrational quantum number.}
	\label{fig:Bound-bound_Traj}
\end{figure}

In Fig. \ref{fig:Bound-bound_Traj}(a), the number of sampled bound-bound trajectories are presented at $T$=\SI{10000}{\kelvin} and $T_{\text{int}}$=\SI{300}{\kelvin}. A bound-bound trajectory denotes that a given initial collision pair ends up to another one in which both target and collider are in the bound or quasi-bound states. For each energy level, 30,000 trajectories are simulated. Majority of the trajectories lead to bound final states for the low-lying initial target level (\emph{i.e.}, low level index $i$). As the initial target state is aligned to higher $v$ and $J$, the final energy level is more probable to end up in the dissociated state. 

Figure \ref{fig:Bound-bound_Traj}(b) shows the sampled bound-bound trajectories decomposed into the endothermic and exothermic forward directions. Determination of the energy direction is done using the individual trajectories that are not summed up for the collision partner's final state. As a result, all of the four energy levels in Eq. \eqnref{eq:O4_EnergyTransfer_StS} are considered. In the sampled trajectories, the endothermic direction overwhelms the exothermic one, especially in the low-lying energy levels that govern the rovibrational energy transfer. This also happens for the trajectory calculations at $T_{\text{int}}$=\SI{10000}{\kelvin} as shown in Fig. \ref{Suppfig:Bound-bound_Traj} in the Supplementary Material. The fact observed from Fig. \ref{fig:Bound-bound_Traj} implies that the endothermic transitions of the low-lying states are more probable to occur in description of the rovibrational-translational (RV-T) and RV-RV energy transfers when the initial states of collision partner are sampled from the Boltzmann distribution at a given $T_{\text{int}}$.

In a similar vein, Fig. \ref{Suppfig:Bound-bound_Traj_Ratio} shows ratio of the number of sampled trajectories by Eq. \eqnref{eq:O4_EnergyTransfer_TC} to the exact one from Eq. \eqnref{eq:O4_EnergyTransfer_StS} for both exothermic and endothermic directions. It is clear that the endothermic trajectories has better sampling quality than those of the exothermic direction, especially for the heads of each $v$ (\emph{i.e.}, $i{=}i(v, \, J{=}0)$) that governs the vibrational energy transfer.


\newpage

\renewcommand{\thefigure}{B\arabic{figure}}   
\setcounter{figure}{0}    
\section{Master equations for combined $\text{O}_3$+$\text{O}_4$ system}\label{sec:master_O3O4}
The set of master equations corresponding to the combined $\text{O}_3$+$\text{O}_4$ system leads to:
\begin{align}
	\frac{dn_i}{dt} &= \sum_j\left(-k_{i \rightarrow j}\left(T_{\text{int}}\right)n_in_{\text{O}_2}+k_{j \rightarrow i}\left(T_{\text{int}}\right)n_jn_{\text{O}_2}\right)\nonumber\\
	&- k_{i \rightarrow c}\left(T_{\text{int}}\right)n_in_{\text{O}_2}+k_{c \rightarrow i}\left(T_{\text{int}}\right)n_{\text{O}}^2n_{\text{O}_2}\nonumber\\
	&+ \sum_j\left(-k_{i \rightarrow j}^{\star}\left(T_{\text{int}}\right)n_in_{\text{O}_2}+k_{j \rightarrow i}^{\star}\left(T_{\text{int}}\right)n_jn_{\text{O}}^2\right)\nonumber\\
	&+ \sum_j\left(-\tilde{k}_{i \rightarrow j}n_in_{\text{O}}+\tilde{k}_{j \rightarrow i}n_jn_{\text{O}}\right)-\tilde{k}_{i \rightarrow c}n_in_{\text{O}}+\tilde{k}_{c \rightarrow i}n_{\text{O}}^3,
	\label{eq:ME-O4_TC-O3_StS_O2}
\end{align}
\begin{align}
	\frac{dn_{\text{O}}}{dt} &= 2\sum_i\left(k_{i \rightarrow c}\left(T_{\text{int}}\right)n_in_{\text{O}_2}-k_{c \rightarrow i}\left(T_{\text{int}}\right)n_{\text{O}}^2n_{\text{O}_2}\right)\nonumber\\
	&+2\sum_i\sum_j\left(k_{i \rightarrow j}^{\star}\left(T_{\text{int}}\right)n_in_{\text{O}_2}-k_{j \rightarrow i}^{\star}\left(T_{\text{int}}\right)n_jn_{\text{O}}^2\right)\nonumber\\
	&+ 2\sum_i\left(\tilde{k}_{i \rightarrow c}n_in_{\text{O}}-\tilde{k}_{c \rightarrow i}n_{\text{O}}^3\right),
	\label{eq:ME-O4_TC-O3_StS_O}
\end{align}
\noindent
where $\tilde{k}_{i \rightarrow j}$, $\tilde{k}_{j \rightarrow i}$, $\tilde{k}_{i \rightarrow c}$, and $\tilde{k}_{c \rightarrow i}$ are the excitation/de-excitation and dissociation/recombination rate coefficients of $\text{O}_2$+O system.

\newpage

\section{Supplementary Material}\label{sec:SuppMat}

\renewcommand{\thefigure}{C\arabic{figure}}

\begin{figure}[h]
	\centering
	\includegraphics[width=0.6\textwidth]{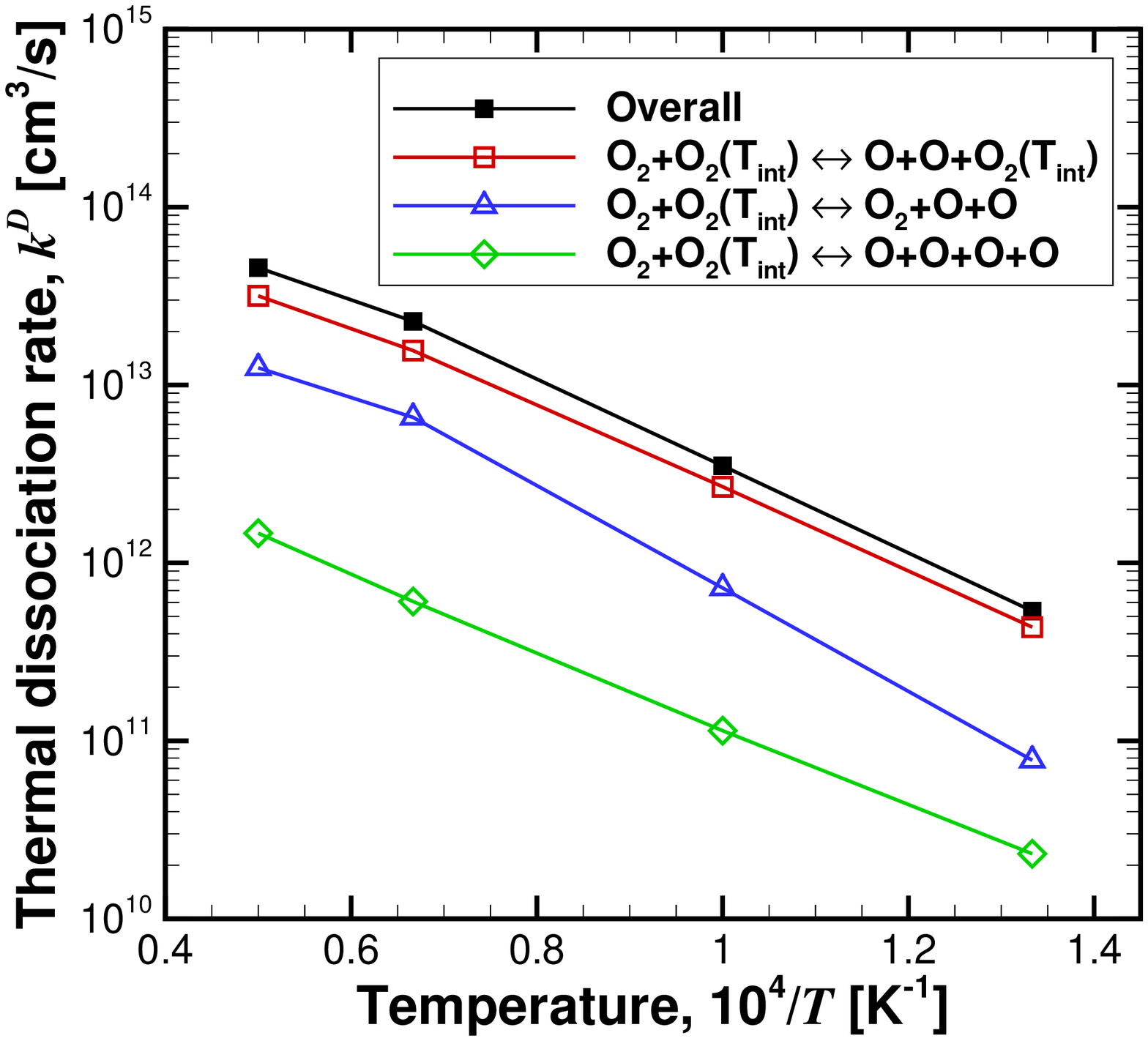}
	\caption{Distributions of thermal dissociation rate coefficients for $\text{O}_2$+$\text{O}_2$ system.}
	\label{Suppfig:O4_Thermal_DissRate_Decomp}
\end{figure}

\begin{figure}[h]
	\centering
	\subfigure[]
	{
		\includegraphics[width=0.44\textwidth]{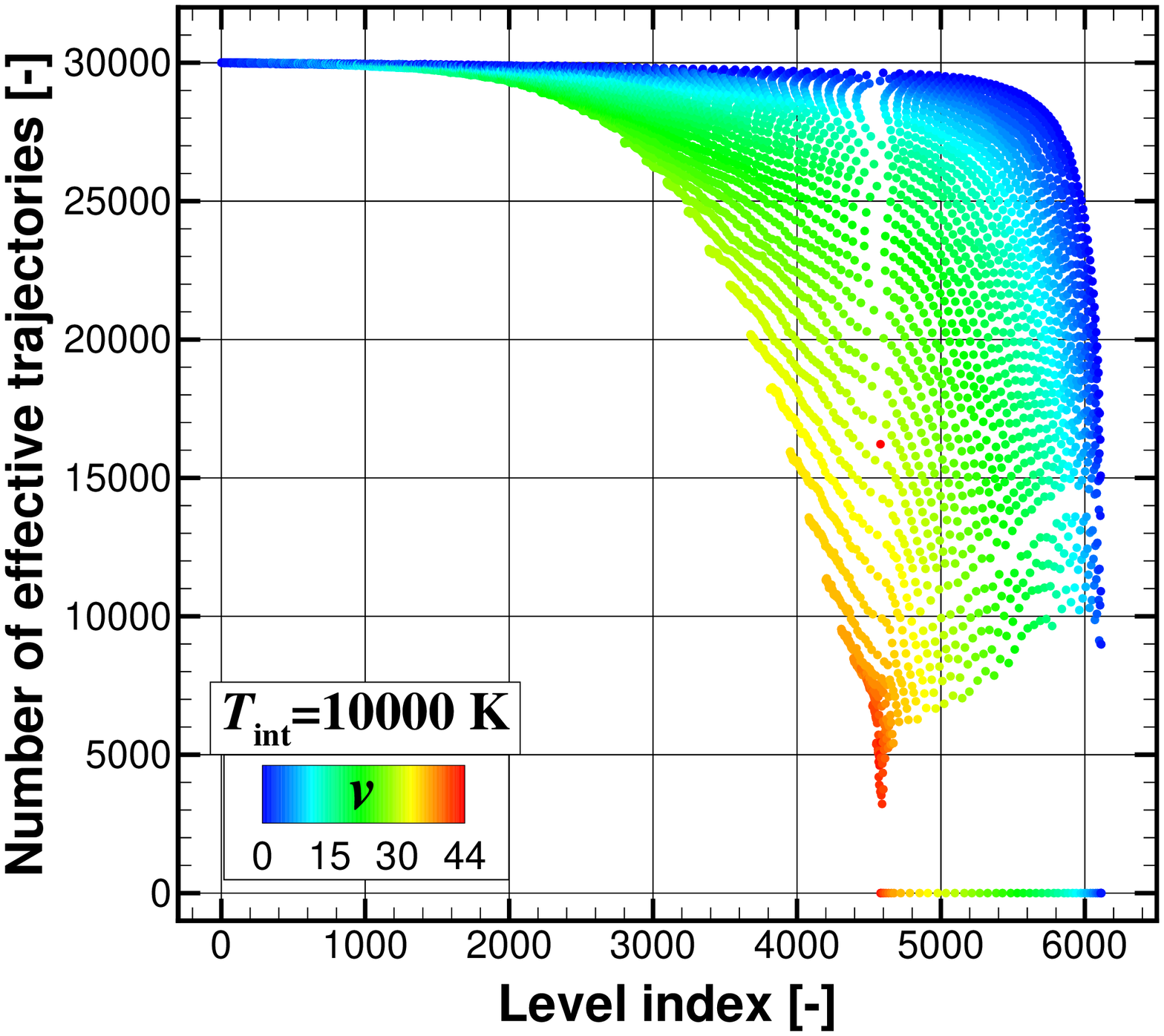}
		\label{Suppfig:Bound-bound_Traj_Num}
	}
	\subfigure[]
	{
		\includegraphics[width=0.44\textwidth]{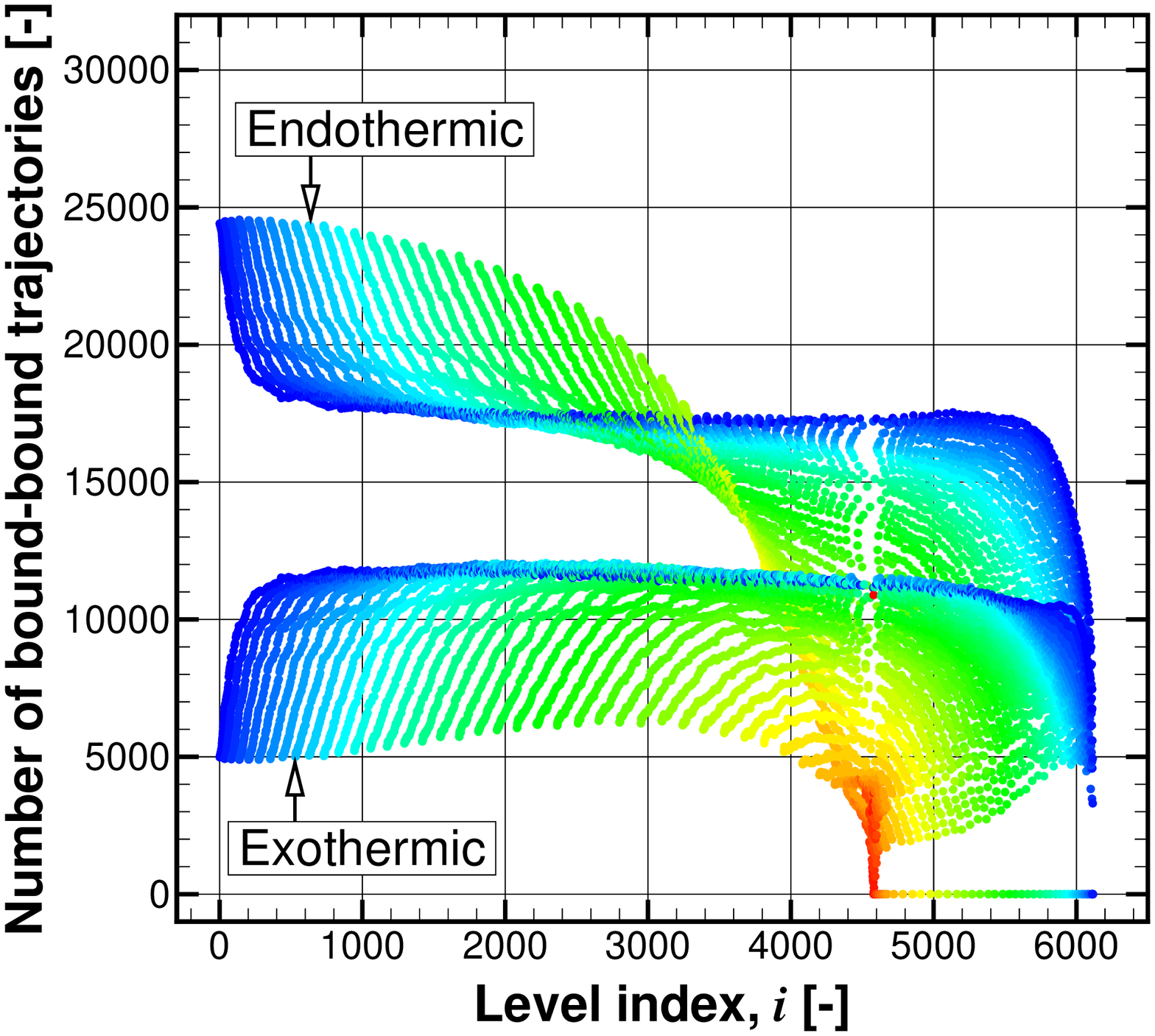}
		\label{Suppfig:Bound-bound_Traj_Decomp}
	}
	\caption{Number of sampled bound-bound trajectories at $T$=\SI{10000}{\kelvin} and $T_{\text{int}}$=\SI{10000}{\kelvin}. (a) Total and (b) decomposition into endothermic and exothermic trajectories. The distributions are colored by the vibrational quantum number.}
	\label{Suppfig:Bound-bound_Traj}
\end{figure}
\begin{figure}[h]
	\centering
	\subfigure[]
	{
		\includegraphics[width=0.44\textwidth]{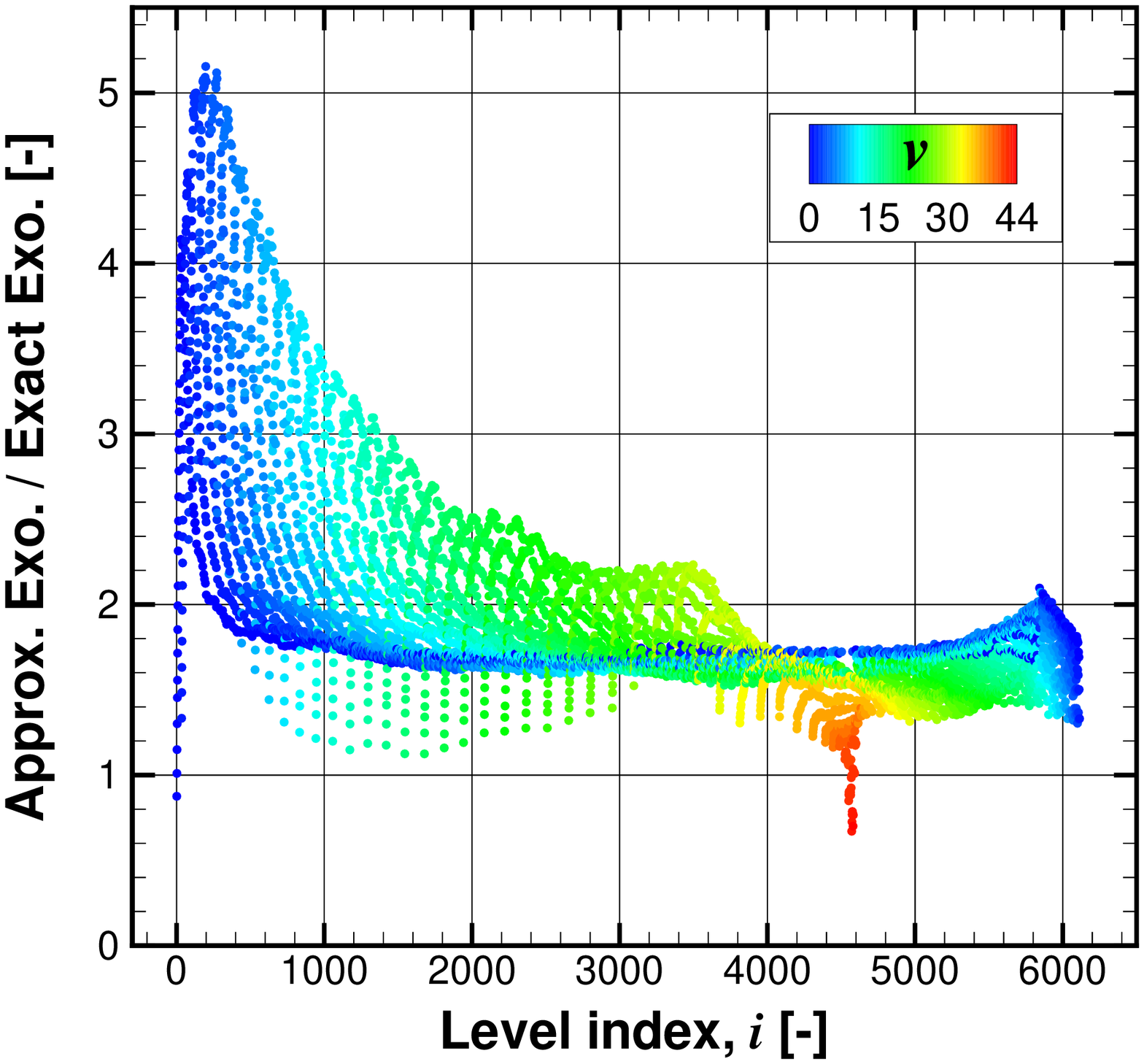}
		\label{Suppfig:Bound-bound_Traj_Ratio_Exo}
	}
	\subfigure[]
	{
		\includegraphics[width=0.44\textwidth]{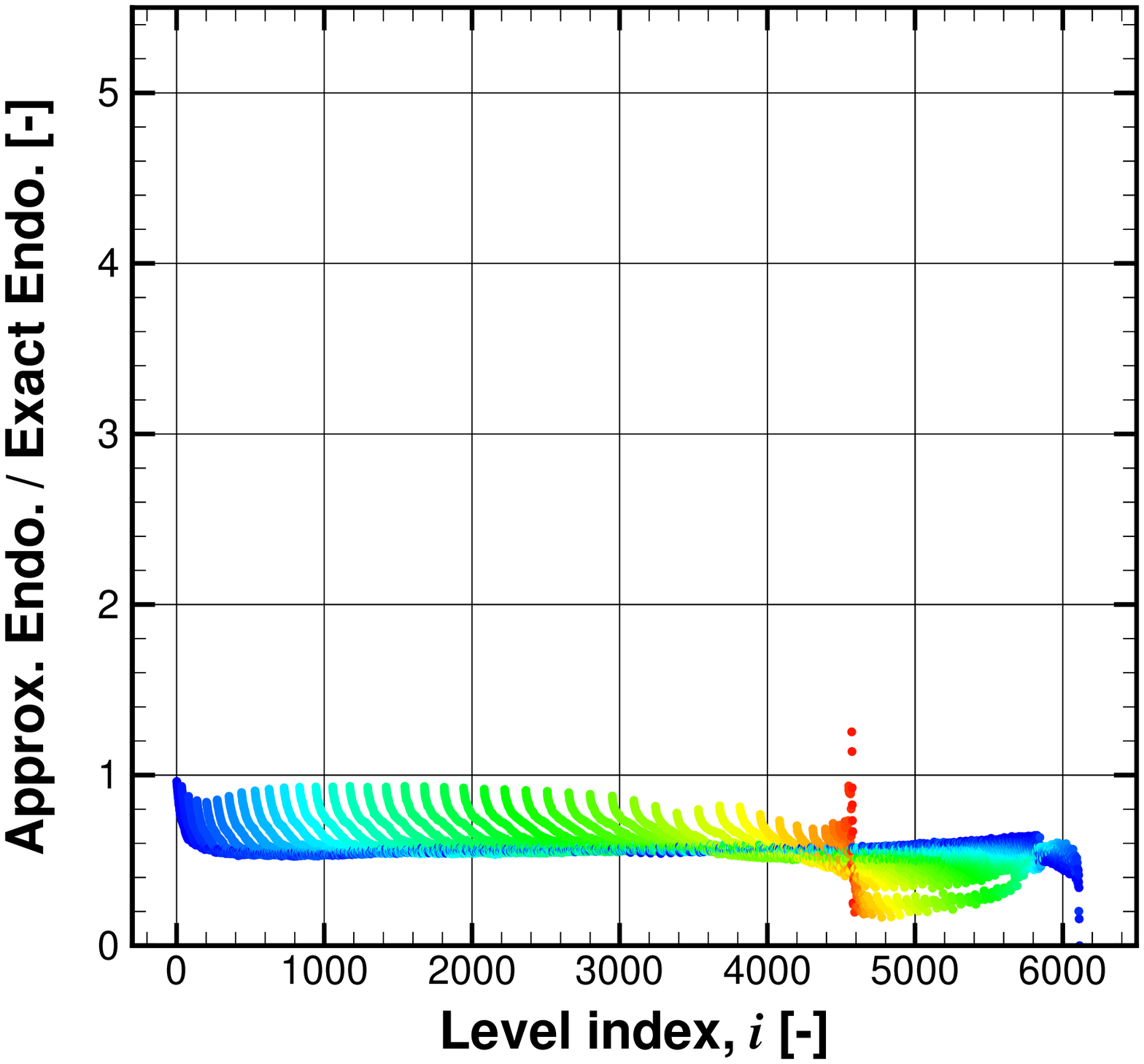}
		\label{Suppfig:Bound-bound_Traj_Ratio_Endo}
	}
	\caption{Ratio of sampled bound-bound trajectories by Eq. \eqnref{eq:O4_EnergyTransfer_TC} to Eq. \eqnref{eq:O4_EnergyTransfer_StS} at $T$=\SI{10000}{\kelvin} and $T_{\text{int}}$=\SI{300}{\kelvin}. (a) Exothermic and (b) endothermic trajectories. The distributions are colored by the vibrational quantum number. The value in $y$-ordinate indicates the number of trajectories that are misinterpreted in deciding the forward direction by relying on the sampling approach in Eq. \eqnref{eq:O4_EnergyTransfer_TC}. As this quantity is closer to one, the quality of the sampled trajectories is more reliable. It should be noted that, by introducing the sampling strategy, large number of trajectories are misinterpreted as the exothermic one that is actually in the reverse direction as shown in Fig. \ref{Suppfig:Bound-bound_Traj_Ratio_Exo}.}
	\label{Suppfig:Bound-bound_Traj_Ratio}
\end{figure}

\begin{figure}[h]
	\centering
	\subfigure[]
	{
		\includegraphics[width=0.48\textwidth]{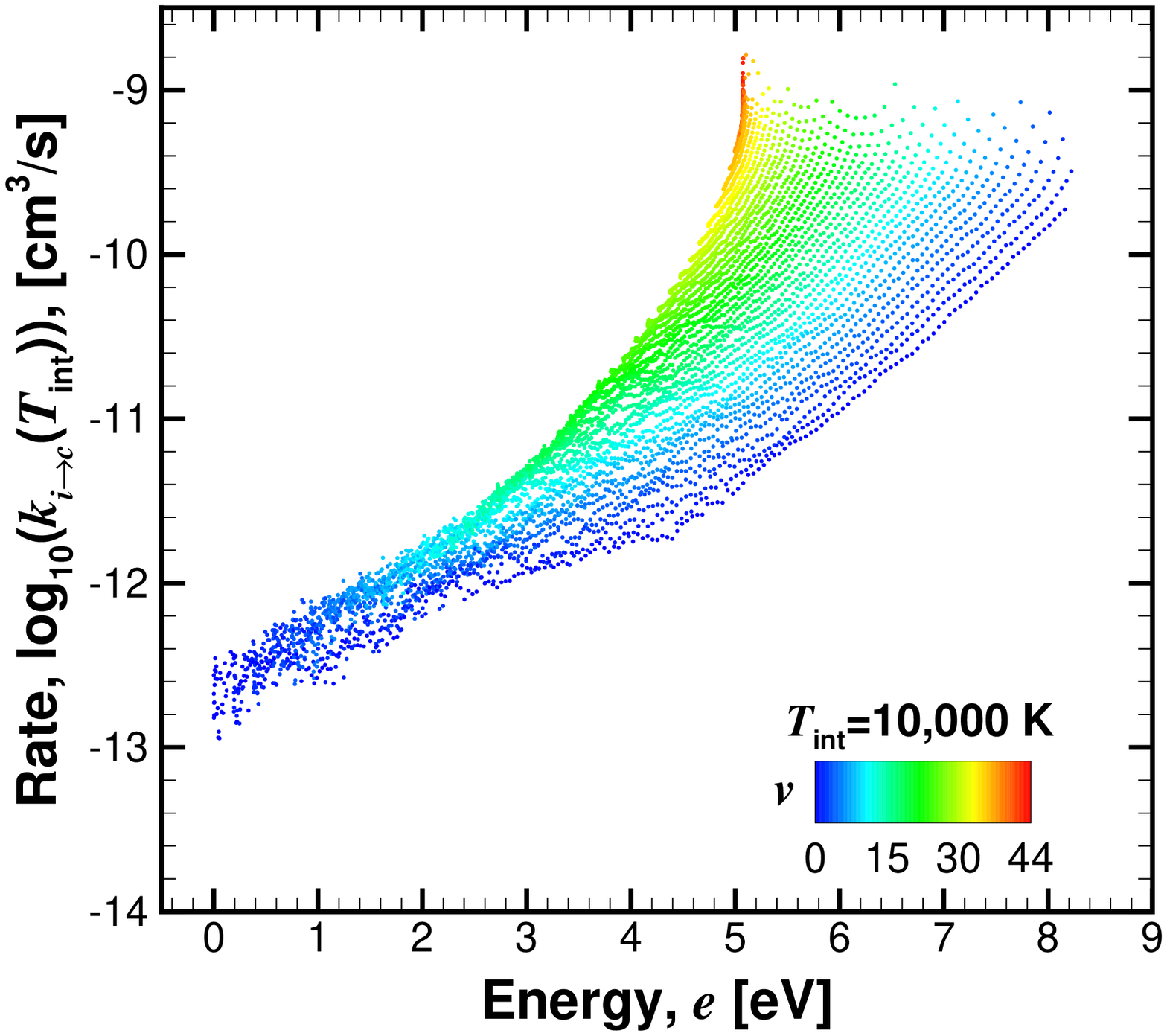}
		\label{fig:DissTar_Dist_T10000K_Tint10000K}
	}
	\subfigure[]
	{
		\includegraphics[width=0.48\textwidth]{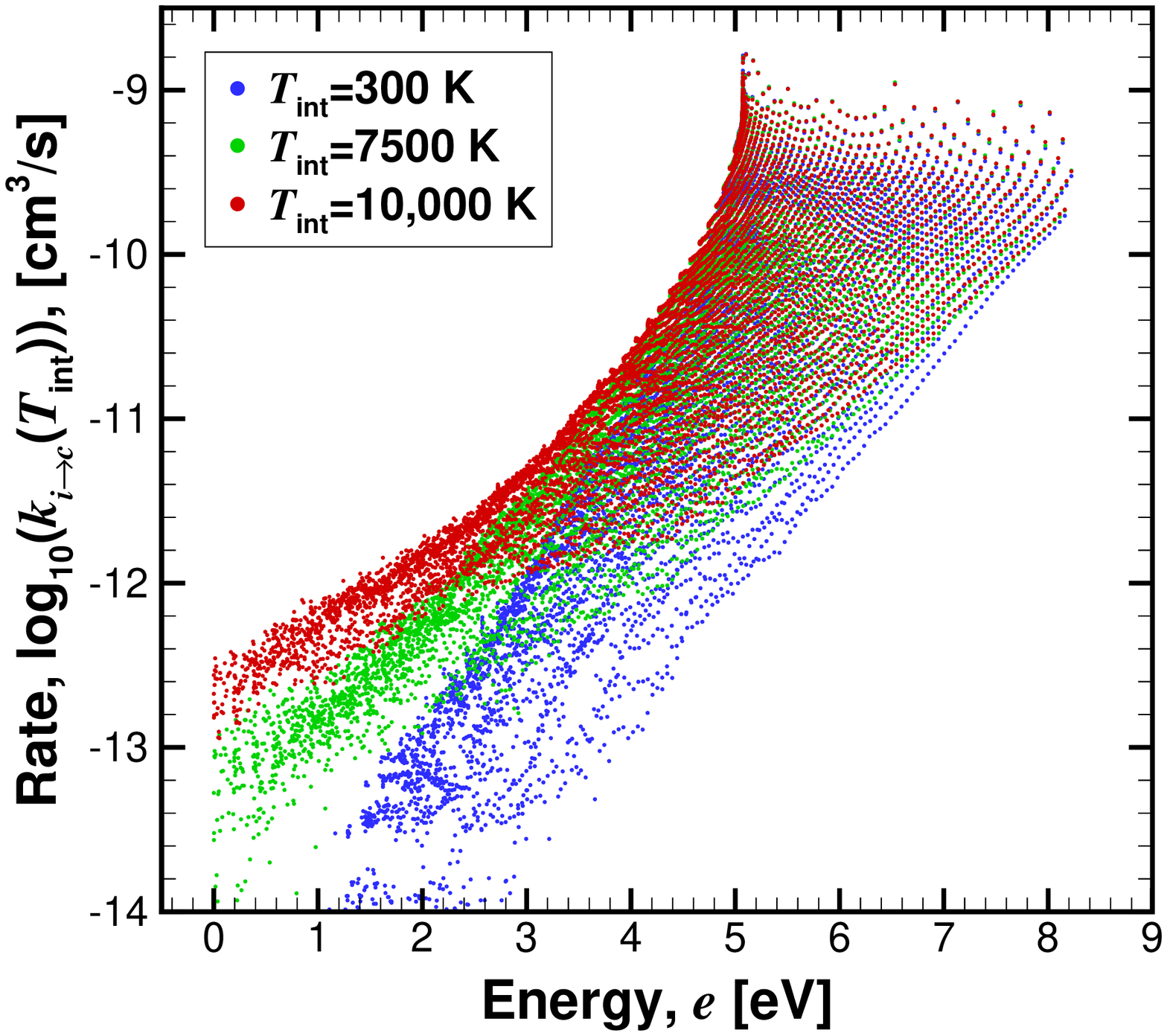}
		\label{fig:DissTar_Dist_T10000K}
	}
	\caption{Distributions of the dissociation rate coefficients in Eq. \eqnref{eq:Bound-free_Rate_TC} at $T$=\SI{10000}{\kelvin}. (a) $T_{\text{int}}$=\SI{10000}{\kelvin} colored by the vibrational quantum number, (b) $T_{\text{int}}$=\SI{300}{\kelvin}, \SI{7500}{\kelvin}, and \SI{10000}{\kelvin}.} 
	\label{fig:DissTar_Dist}
\end{figure}

\begin{figure}[h]
	\centering
	\includegraphics[width=0.6\textwidth]{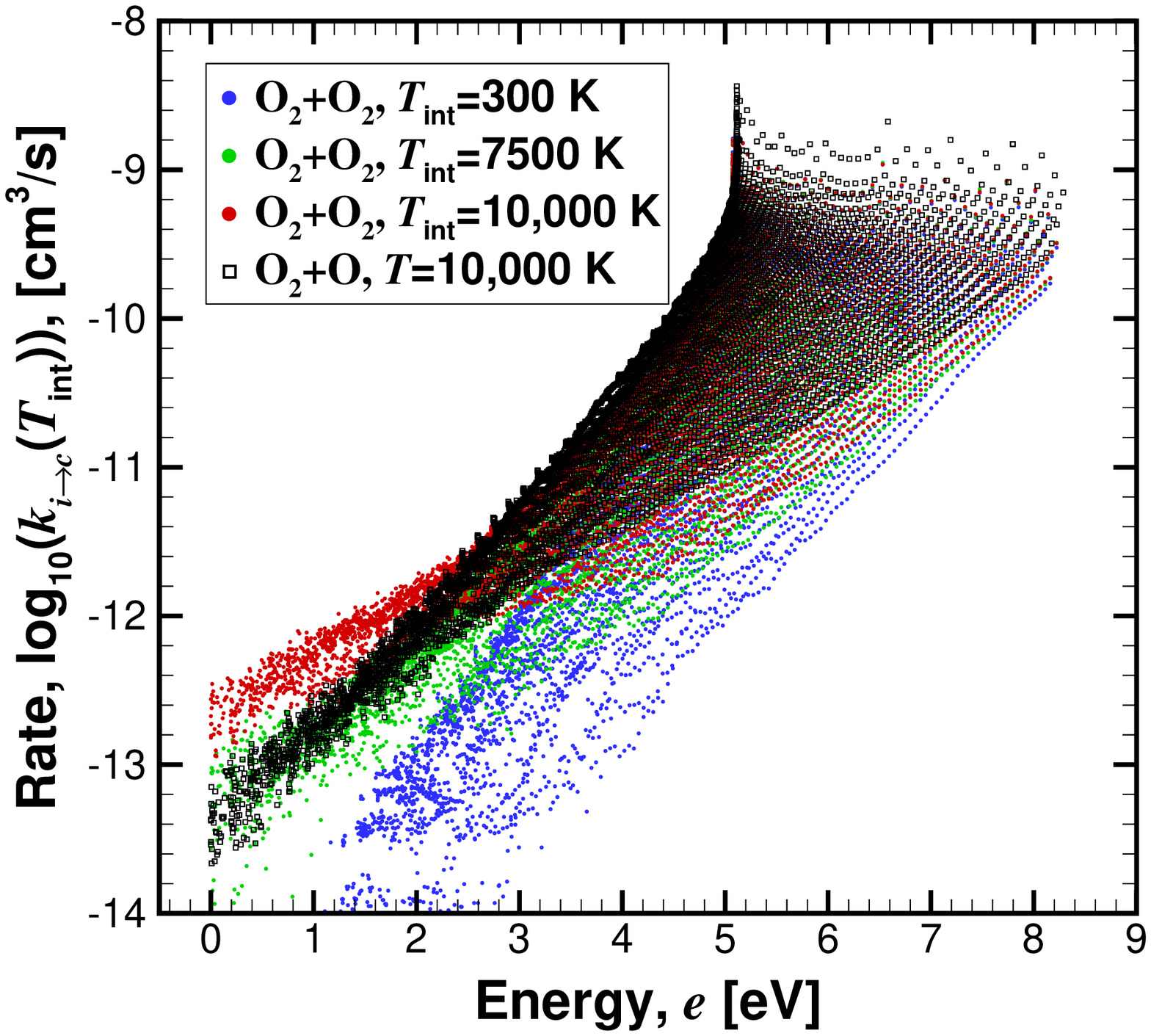}
	\caption{Comparison between the dissociation rate coefficients of $\text{O}_2$+$\text{O}_2$ in Eq. \eqnref{eq:Bound-free_Rate_TC} and that of $\text{O}_2$+O $\rightarrow$ 3O at $T$=\SI{10000}{\kelvin}.}
	\label{Suppfig:DissRate_O3_O4_T10000K}
\end{figure}

\begin{figure}[h]
	\centering
	\subfigure[]
	{
		\includegraphics[width=0.48\textwidth]{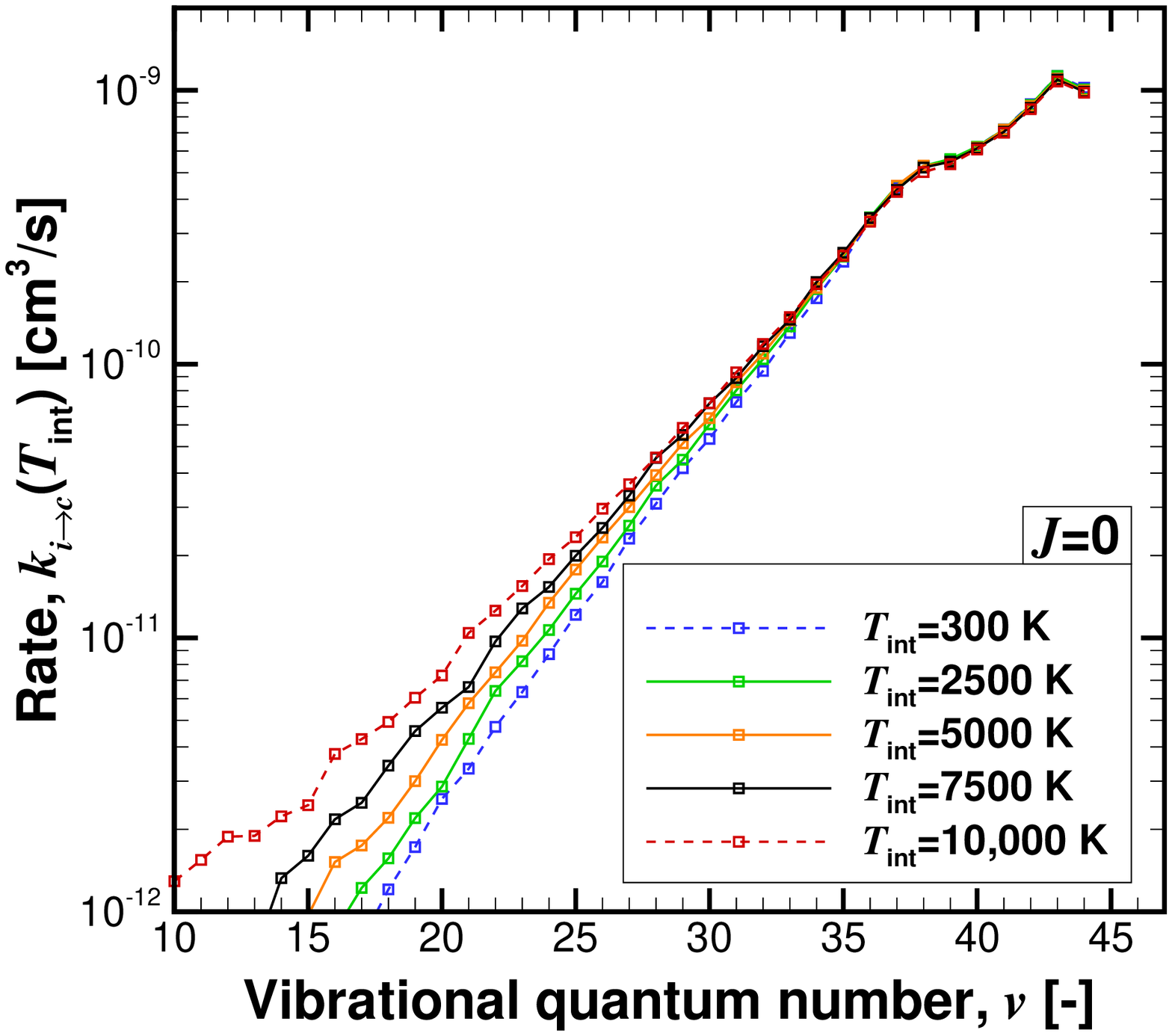}
		\label{fig:DissTar_along-v}
	}
	\subfigure[]
	{
		\includegraphics[width=0.48\textwidth]{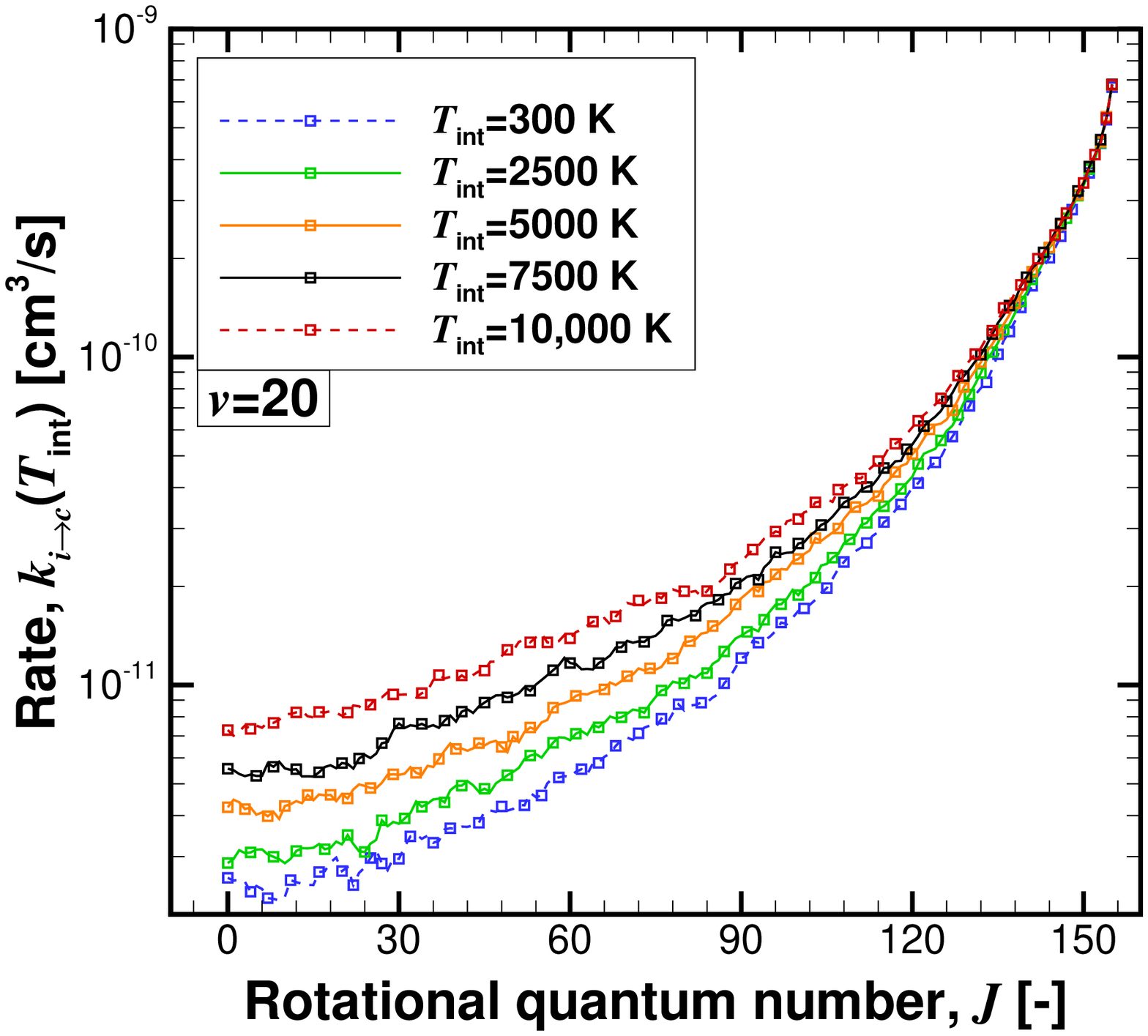}
		\label{fig:DissTar_along-J_at_v20}
	}
	\subfigure[]
	{
		\includegraphics[width=0.48\textwidth]{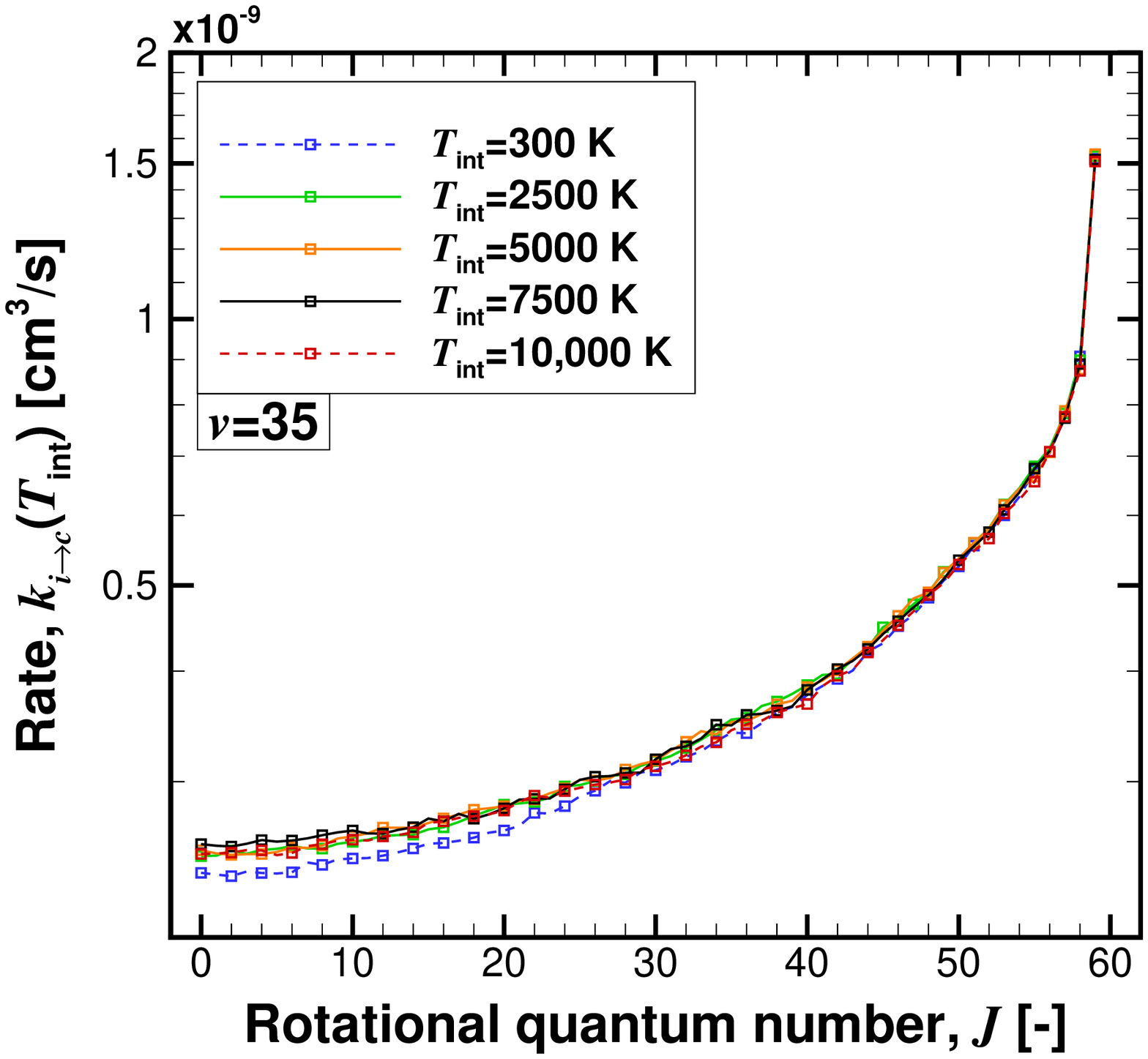}
		\label{fig:DissTar_along-J_at_v35}
	}
	\caption{Distributions of the dissociation rate coefficients in Eq. \eqnref{eq:O4_TargetDiss_TC} at $T$=\SI{10000}{\kelvin}. The initial target states are: (a) $\text{O}_2(v, J{=}0)$, (b) $\text{O}_2(v{=}20, J)$, (c) $\text{O}_2(v{=}35, J)$.}
	\label{Suppfig:DissTar_line}
\end{figure}

\begin{figure}[h]
	\centering
	\includegraphics[width=0.75\textwidth]{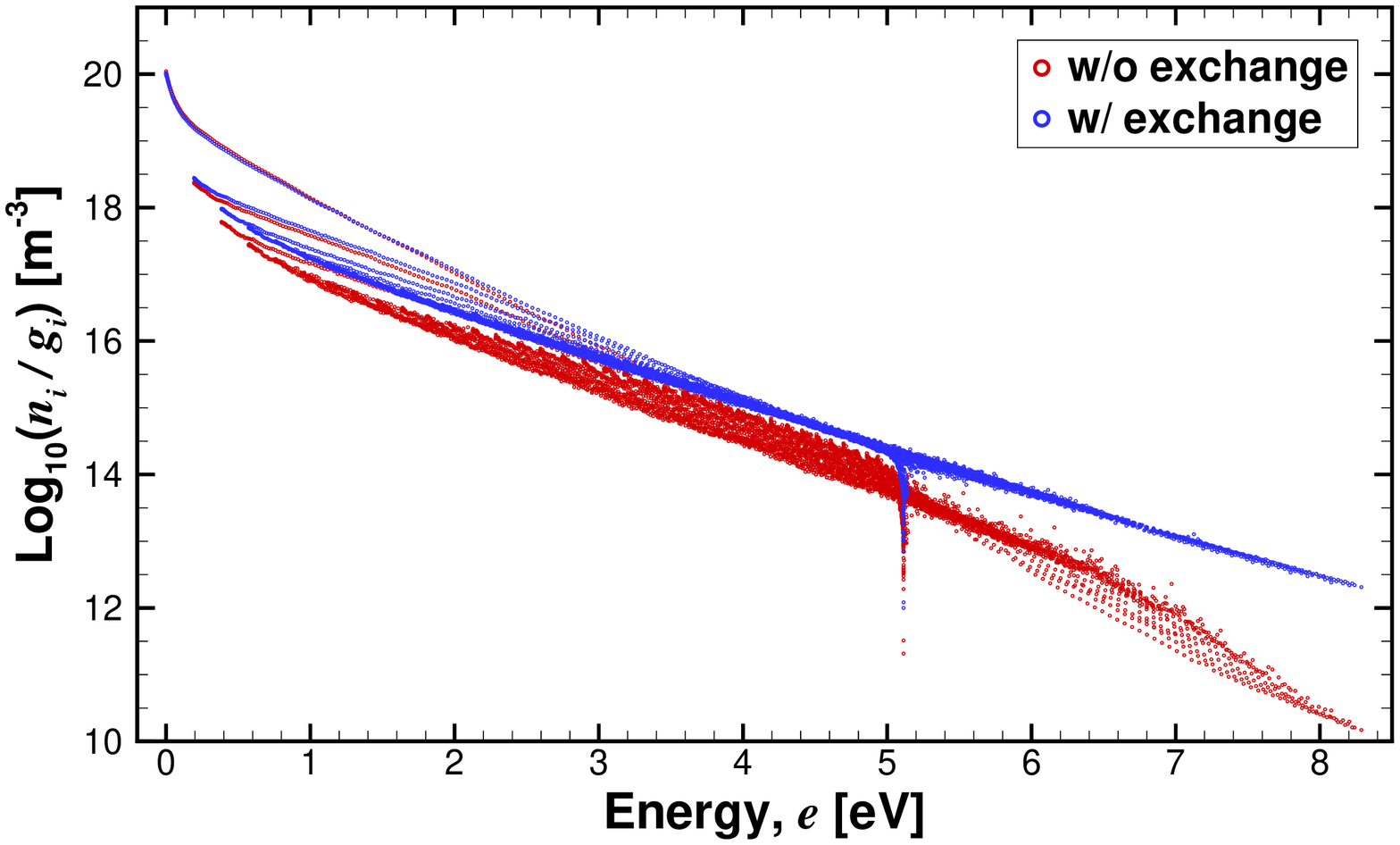}
	\caption{Influence of the exchange kinetics on the rovibrational distributions of $\text{O}_2$+O system at $t$=2.95$\times$10$^{-8}$ s of $T$=\SI{10000}{\kelvin} in which 8.8\% of vibrational relxation occurs in case that the exchange process is considered. The initial mole fraction was set to $\chi_{\text{O}_2}$=$\chi_{\text{O}}$=0.5 and the dissociation process was excluded.}
	\label{Suppfig:O3_Effect_of_Exch_Pop}
\end{figure}
\begin{figure}[h]
	\centering
	\subfigure[]
	{
		\includegraphics[width=0.44\textwidth]{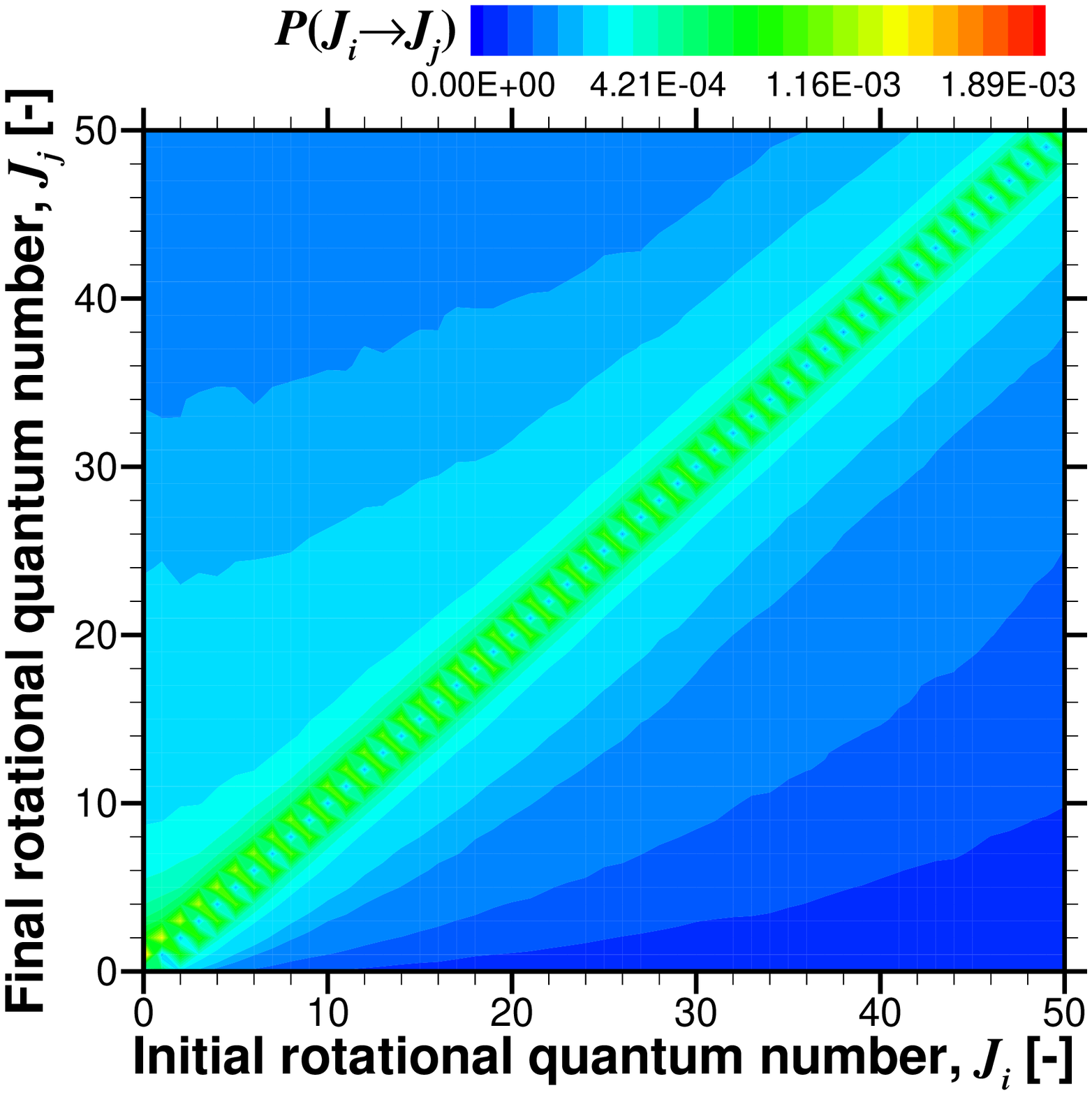}
		\label{Suppfig:Rotational_Transition_Probability_O4}
	}
	\subfigure[]
	{
		\includegraphics[width=0.44\textwidth]{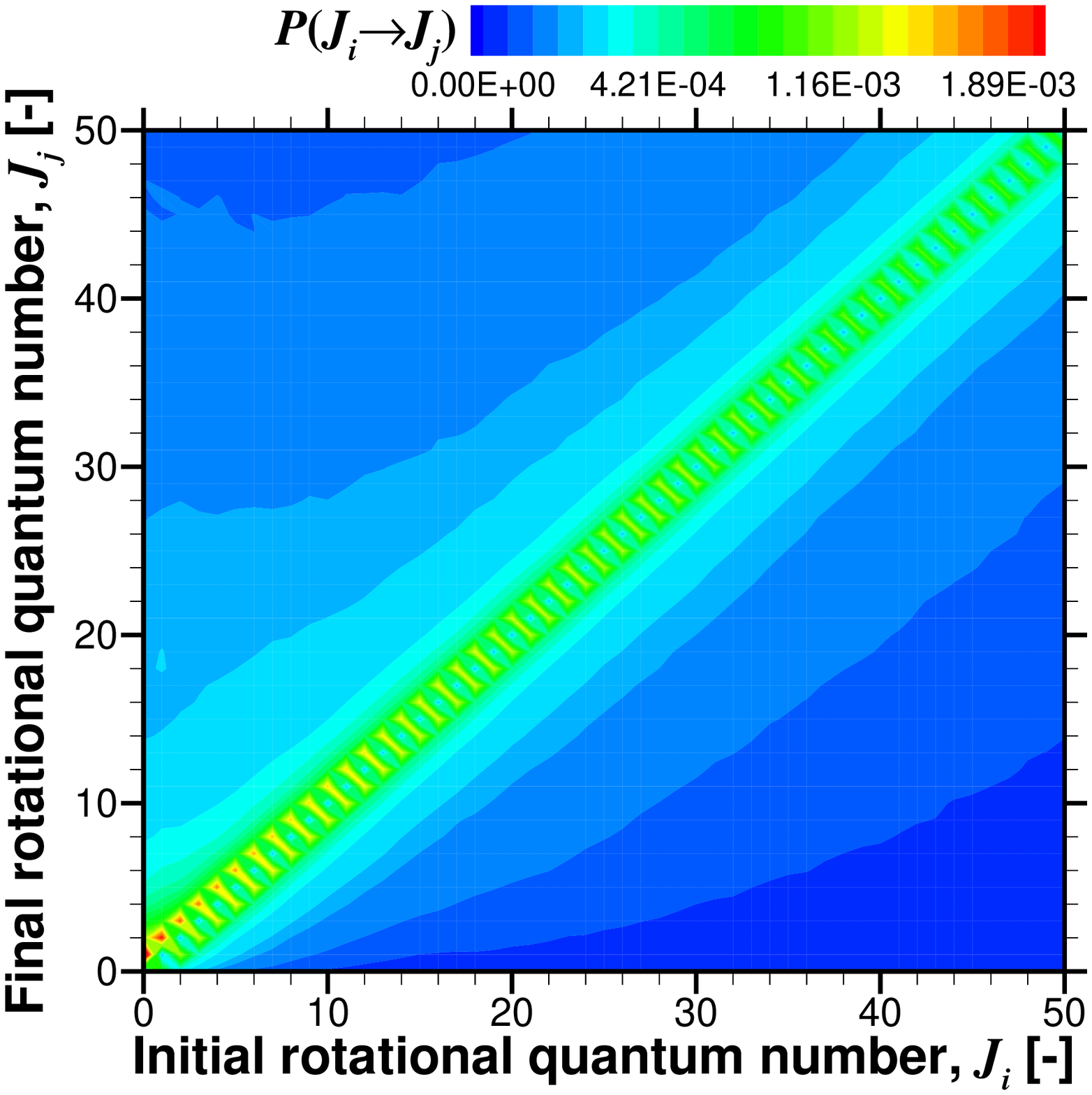}
		\label{Suppfig:Rotational_Transition_Probability_O3}
	}
	\caption{Rotational transition probability distributions due to the inelastic process for (a) $\text{O}_2$+$\text{O}_2$ at $T_{\text{int}}$=\SI{300}{\kelvin} and (b) $\text{O}_2$+O at $T$=\SI{10000}{\kelvin}.}
	\label{Suppfig:Rotational_Transition_Probability_10000K}
\end{figure}

\begin{figure}[h]
	\centering
	\includegraphics[width=0.6\textwidth]{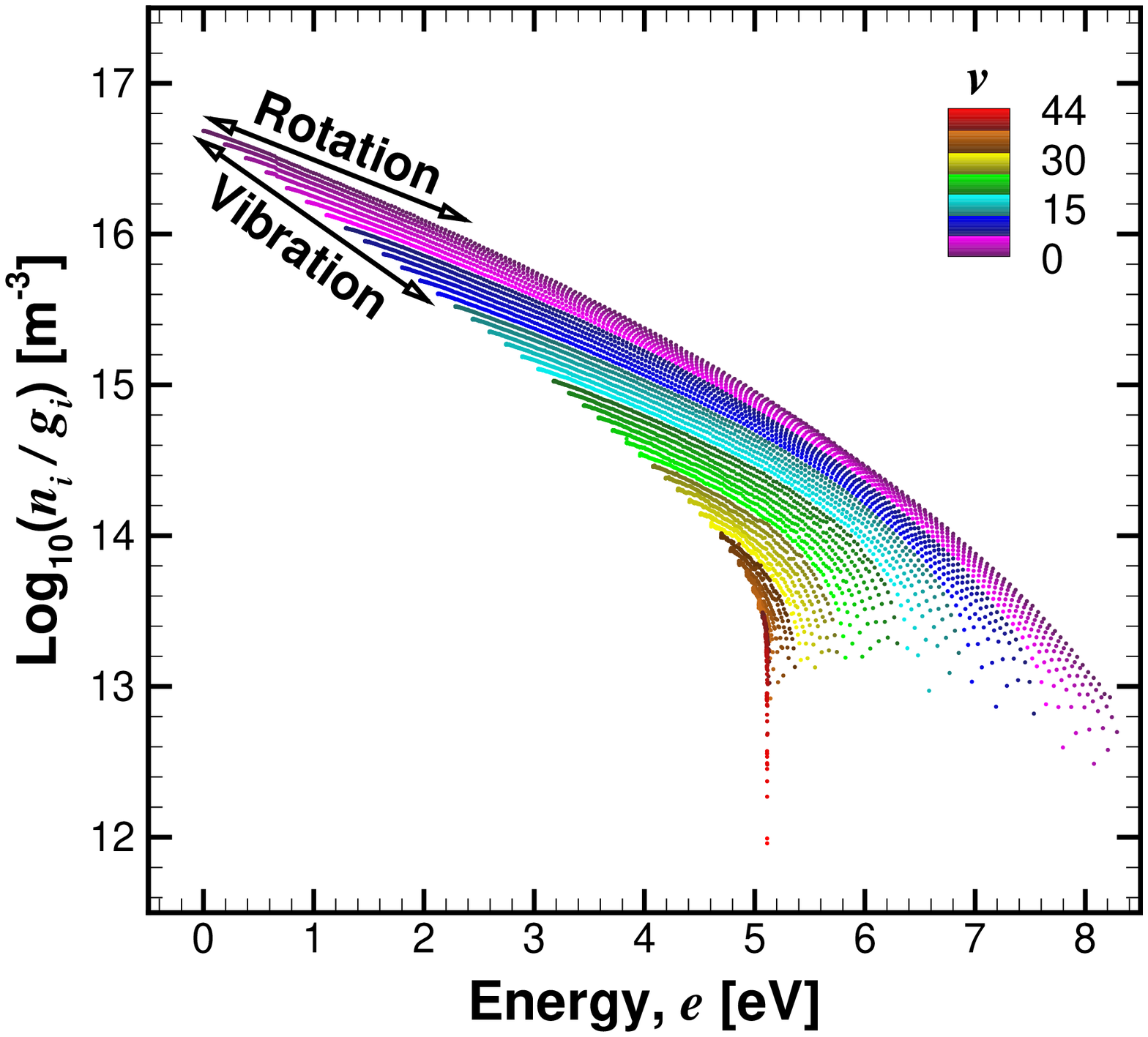}
	\caption{Rovibrational distributions at QSS of $T$=\SI{20000}{\kelvin}. The black arrows depict the gradients of rotational and vibrational distributions that imply their characteristic temperatures.}
	\label{Suppfig:O4_QSS_Pop_T20000K}
\end{figure}
\begin{figure}[h]
	\centering
	\includegraphics[width=0.6\textwidth]{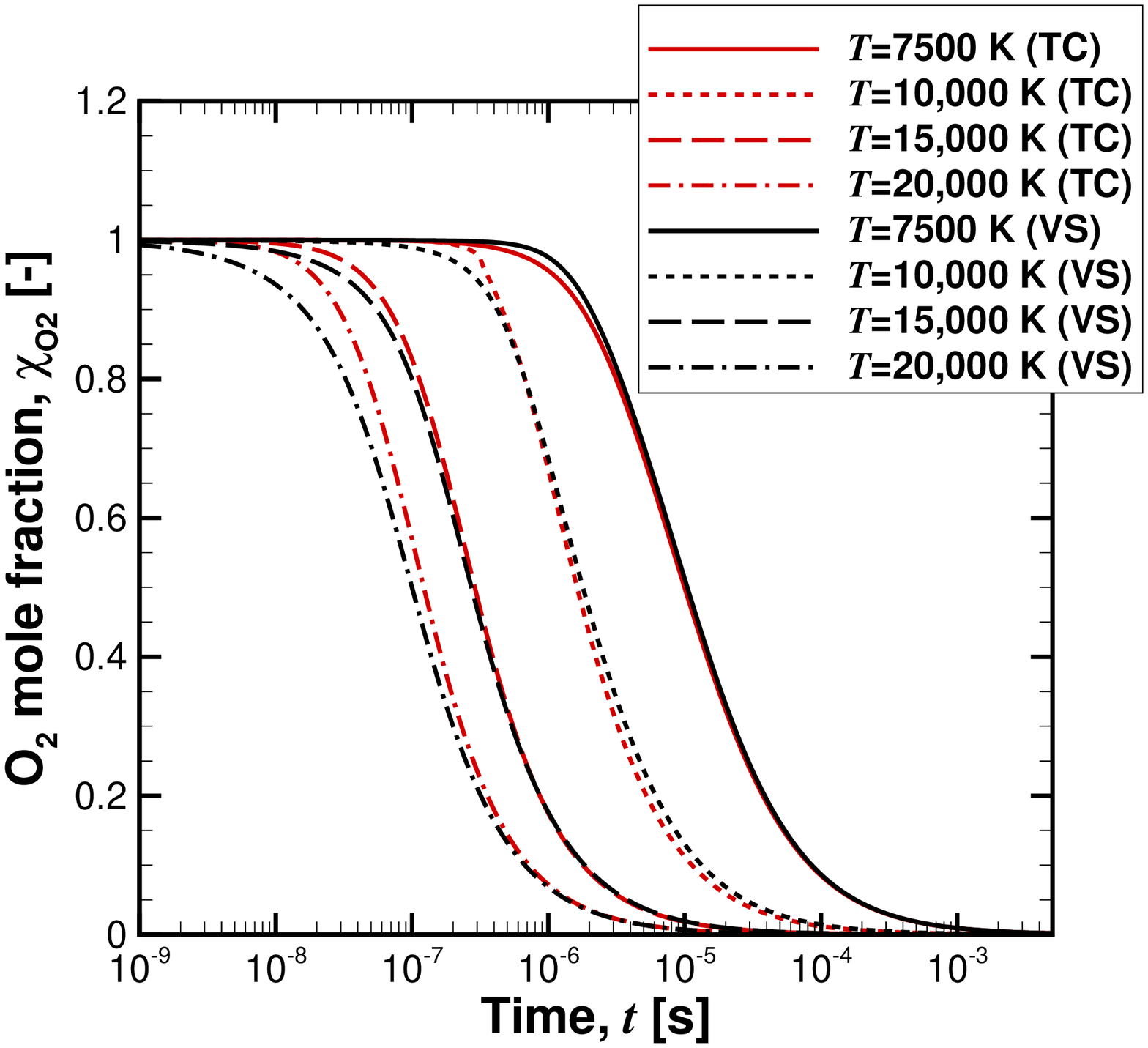}
	\caption{Comparison of mole fraction profiles of $\text{O}_2$ between the TC and VS models.}
	\label{Suppfig:Temp_All_TC_VS}
\end{figure}

\begin{figure}[h]
	\centering
	\includegraphics[width=0.6\textwidth]{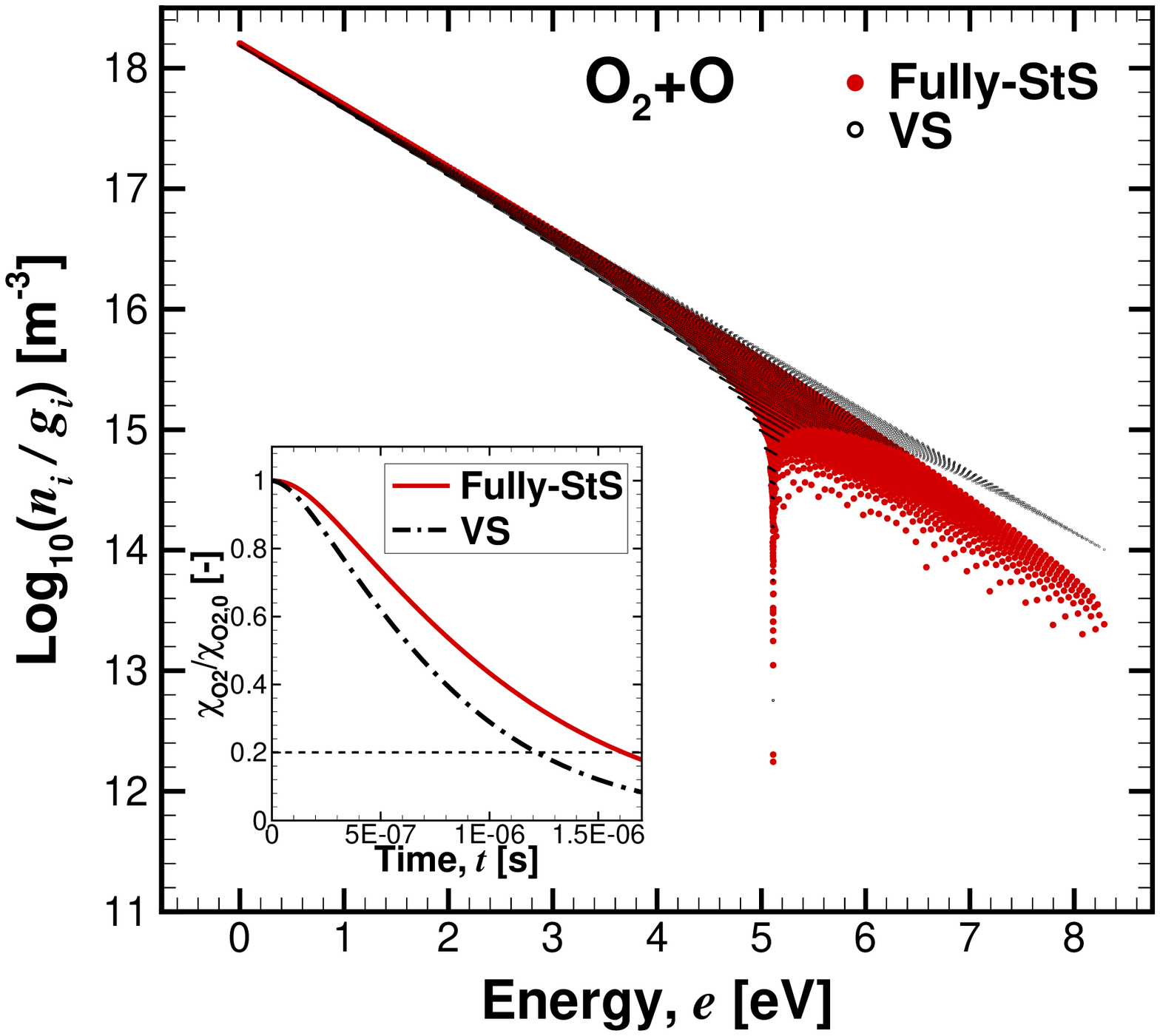}
	\caption{Comparison of the rovibrational distributions of $\text{O}_2$+O system between the fully-StS and VS models at $T$=\SI{10000}{\kelvin}. The distributions are extracted at $\chi_{\text{O}_2}$/$\chi_{\text{O}_2,0}$=0.2.}
	\label{Suppfig:QSS_Pop_O3_FullyStS_T10000K_CompVS_20PercO2}
\end{figure}
\begin{figure}[h]
	\centering
	\includegraphics[width=0.6\textwidth]{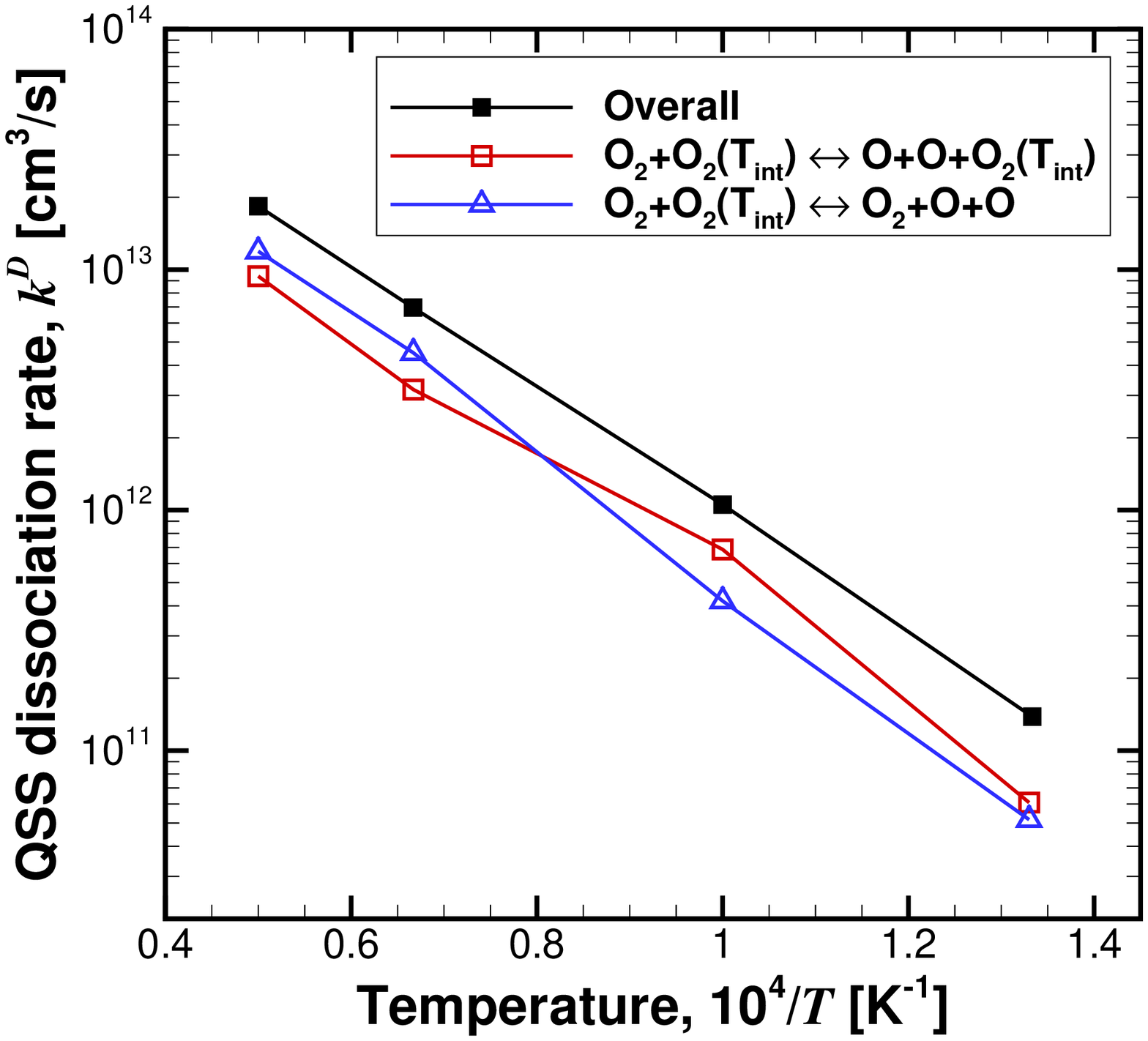}
	\caption{Distributions of QSS dissociation rate coefficients for $\text{O}_2$+$\text{O}_2$ system.}
	\label{Suppfig:O4_QSS_DissRate_Decomp}
\end{figure}

\clearpage
\newpage
\bibliography{ref}

\begin{thebibliography}{46}%
\makeatletter
\providecommand \@ifxundefined [1]{%
 \@ifx{#1\undefined}
}%
\providecommand \@ifnum [1]{%
 \ifnum #1\expandafter \@firstoftwo
 \else \expandafter \@secondoftwo
 \fi
}%
\providecommand \@ifx [1]{%
 \ifx #1\expandafter \@firstoftwo
 \else \expandafter \@secondoftwo
 \fi
}%
\providecommand \natexlab [1]{#1}%
\providecommand \enquote  [1]{``#1''}%
\providecommand \bibnamefont  [1]{#1}%
\providecommand \bibfnamefont [1]{#1}%
\providecommand \citenamefont [1]{#1}%
\providecommand \href@noop [0]{\@secondoftwo}%
\providecommand \href [0]{\begingroup \@sanitize@url \@href}%
\providecommand \@href[1]{\@@startlink{#1}\@@href}%
\providecommand \@@href[1]{\endgroup#1\@@endlink}%
\providecommand \@sanitize@url [0]{\catcode `\\12\catcode `\$12\catcode
  `\&12\catcode `\#12\catcode `\^12\catcode `\_12\catcode `\%12\relax}%
\providecommand \@@startlink[1]{}%
\providecommand \@@endlink[0]{}%
\providecommand \url  [0]{\begingroup\@sanitize@url \@url }%
\providecommand \@url [1]{\endgroup\@href {#1}{\urlprefix }}%
\providecommand \urlprefix  [0]{URL }%
\providecommand \Eprint [0]{\href }%
\providecommand \doibase [0]{http://dx.doi.org/}%
\providecommand \selectlanguage [0]{\@gobble}%
\providecommand \bibinfo  [0]{\@secondoftwo}%
\providecommand \bibfield  [0]{\@secondoftwo}%
\providecommand \translation [1]{[#1]}%
\providecommand \BibitemOpen [0]{}%
\providecommand \bibitemStop [0]{}%
\providecommand \bibitemNoStop [0]{.\EOS\space}%
\providecommand \EOS [0]{\spacefactor3000\relax}%
\providecommand \BibitemShut  [1]{\csname bibitem#1\endcsname}%
\let\auto@bib@innerbib\@empty
\bibitem [{\citenamefont {Panesi}\ \emph {et~al.}(2013)\citenamefont {Panesi},
  \citenamefont {Jaffe}, \citenamefont {Schwenke},\ and\ \citenamefont
  {Magin}}]{PANESI_2013_BOXRVC}%
  \BibitemOpen
  \bibfield  {author} {\bibinfo {author} {\bibfnamefont {M.}~\bibnamefont
  {Panesi}}, \bibinfo {author} {\bibfnamefont {R.~L.}\ \bibnamefont {Jaffe}},
  \bibinfo {author} {\bibfnamefont {D.~W.}\ \bibnamefont {Schwenke}}, \ and\
  \bibinfo {author} {\bibfnamefont {T.~E.}\ \bibnamefont {Magin}},\ }\bibfield
  {title} {\enquote {\bibinfo {title} {Rovibrational internal energy transfer
  and dissociation of $\rm{N_2}(\rm{{}^1\Sigma^+_g})- \rm{N}(\rm{{}^4S_u})$
  system in hypersonic flows},}\ }\href {\doibase 10.1063/1.4774412} {\bibfield
   {journal} {\bibinfo  {journal} {Journal of Chemical Physics}\ }\textbf
  {\bibinfo {volume} {138}},\ \bibinfo {pages} {044312} (\bibinfo {year}
  {2013})}\BibitemShut {NoStop}%
\bibitem [{\citenamefont {Venturi}\ \emph {et~al.}(2020)\citenamefont
  {Venturi}, \citenamefont {Sharma}, \citenamefont {Lopez},\ and\ \citenamefont
  {Panesi}}]{Venturi2020}%
  \BibitemOpen
  \bibfield  {author} {\bibinfo {author} {\bibfnamefont {S.}~\bibnamefont
  {Venturi}}, \bibinfo {author} {\bibfnamefont {M.~P.}\ \bibnamefont {Sharma}},
  \bibinfo {author} {\bibfnamefont {B.}~\bibnamefont {Lopez}}, \ and\ \bibinfo
  {author} {\bibfnamefont {M.}~\bibnamefont {Panesi}},\ }\bibfield  {title}
  {\enquote {\bibinfo {title} {Data-inspired and physics-driven model reduction
  for dissociation: Application to the $\text{O}_2$ + $\text{O}$ system},}\
  }\href {\doibase 10.1021/acs.jpca.0c04516} {\bibfield  {journal} {\bibinfo
  {journal} {Journal of Physical Chemistry A}\ }\textbf {\bibinfo {volume}
  {124}},\ \bibinfo {pages} {8359--8372} (\bibinfo {year} {2020})}\BibitemShut
  {NoStop}%
\bibitem [{\citenamefont {Jo}\ \emph {et~al.}(2022)\citenamefont {Jo},
  \citenamefont {Venturi}, \citenamefont {Sharma}, \citenamefont {Munafò},\
  and\ \citenamefont {Panesi}}]{Jo_N2O}%
  \BibitemOpen
  \bibfield  {author} {\bibinfo {author} {\bibfnamefont {S.~M.}\ \bibnamefont
  {Jo}}, \bibinfo {author} {\bibfnamefont {S.}~\bibnamefont {Venturi}},
  \bibinfo {author} {\bibfnamefont {M.~P.}\ \bibnamefont {Sharma}}, \bibinfo
  {author} {\bibfnamefont {A.}~\bibnamefont {Munafò}}, \ and\ \bibinfo
  {author} {\bibfnamefont {M.}~\bibnamefont {Panesi}},\ }\bibfield  {title}
  {\enquote {\bibinfo {title} {Rovibrational-specific qct and master equation
  study on $\text{N}_2({X}^1\sigma_g^+)$+{O}(${}^3p$) and
  $\text{NO}({X}^2\pi)$+{N}(${}^4s$) systems in high-energy collisions},}\
  }\href {\doibase 10.1021/acs.jpca.1c10346} {\bibfield  {journal} {\bibinfo
  {journal} {The Journal of Physical Chemistry A}\ } (\bibinfo {year} {2022}),\
  10.1021/acs.jpca.1c10346}\BibitemShut {NoStop}%
\bibitem [{\citenamefont {Priyadarshini}\ \emph {et~al.}(0)\citenamefont
  {Priyadarshini}, \citenamefont {Jo}, \citenamefont {Venturi}, \citenamefont
  {Schwenke}, \citenamefont {Jaffe},\ and\ \citenamefont
  {Panesi}}]{Sharma_HCN}%
  \BibitemOpen
  \bibfield  {author} {\bibinfo {author} {\bibfnamefont {M.~S.}\ \bibnamefont
  {Priyadarshini}}, \bibinfo {author} {\bibfnamefont {S.~M.}\ \bibnamefont
  {Jo}}, \bibinfo {author} {\bibfnamefont {S.}~\bibnamefont {Venturi}},
  \bibinfo {author} {\bibfnamefont {D.~W.}\ \bibnamefont {Schwenke}}, \bibinfo
  {author} {\bibfnamefont {R.~L.}\ \bibnamefont {Jaffe}}, \ and\ \bibinfo
  {author} {\bibfnamefont {M.}~\bibnamefont {Panesi}},\ }\bibfield  {title}
  {\enquote {\bibinfo {title} {Comprehensive study of {HCN}: Potential energy
  surfaces, state-to-state kinetics, and master equation analysis},}\ }\href
  {\doibase 10.1021/acs.jpca.2c03959} {\bibfield  {journal} {\bibinfo
  {journal} {The Journal of Physical Chemistry A}\ }\textbf {\bibinfo {volume}
  {0}},\ \bibinfo {pages} {null} (\bibinfo {year} {0})},\ \bibinfo {note}
  {pMID: 36288431}\BibitemShut {NoStop}%
\bibitem [{\citenamefont {Jo}\ \emph {et~al.}()\citenamefont {Jo},
  \citenamefont {Munafò}, \citenamefont {Priyadarshini}, \citenamefont
  {Venturi},\ and\ \citenamefont {Panesi}}]{Jo_2022_Scitech}%
  \BibitemOpen
  \bibfield  {author} {\bibinfo {author} {\bibfnamefont {S.~M.}\ \bibnamefont
  {Jo}}, \bibinfo {author} {\bibfnamefont {A.}~\bibnamefont {Munafò}},
  \bibinfo {author} {\bibfnamefont {M.~S.}\ \bibnamefont {Priyadarshini}},
  \bibinfo {author} {\bibfnamefont {S.}~\bibnamefont {Venturi}}, \ and\
  \bibinfo {author} {\bibfnamefont {M.}~\bibnamefont {Panesi}},\ }\enquote
  {\bibinfo {title} {Rovibrational-specific master equation analysis of
  high-temperature air mixture},}\ in\ \href {\doibase 10.2514/6.2022-0342}
  {\emph {\bibinfo {booktitle} {AIAA SCITECH 2022 Forum}}},\ \Eprint
  {http://arxiv.org/abs/https://arc.aiaa.org/doi/pdf/10.2514/6.2022-0342}
  {https://arc.aiaa.org/doi/pdf/10.2514/6.2022-0342} \BibitemShut {NoStop}%
\bibitem [{\citenamefont {Grover}, \citenamefont {Torres},\ and\ \citenamefont
  {Schwartzentruber}(2019)}]{GROVER_O3O4}%
  \BibitemOpen
  \bibfield  {author} {\bibinfo {author} {\bibfnamefont {M.~S.}\ \bibnamefont
  {Grover}}, \bibinfo {author} {\bibfnamefont {E.}~\bibnamefont {Torres}}, \
  and\ \bibinfo {author} {\bibfnamefont {T.~E.}\ \bibnamefont
  {Schwartzentruber}},\ }\bibfield  {title} {\enquote {\bibinfo {title} {Direct
  molecular simulation of internal energy relaxation and dissociation in
  oxygen},}\ }\href {\doibase 10.1063/1.5108666} {\bibfield  {journal}
  {\bibinfo  {journal} {Physics of Fluids}\ }\textbf {\bibinfo {volume} {31}},\
  \bibinfo {pages} {076107} (\bibinfo {year} {2019})}\BibitemShut {NoStop}%
\bibitem [{\citenamefont {Streicher}, \citenamefont {Krish},\ and\
  \citenamefont {Hanson}(2021)}]{Streicher_2021}%
  \BibitemOpen
  \bibfield  {author} {\bibinfo {author} {\bibfnamefont {J.~W.}\ \bibnamefont
  {Streicher}}, \bibinfo {author} {\bibfnamefont {A.}~\bibnamefont {Krish}}, \
  and\ \bibinfo {author} {\bibfnamefont {R.~K.}\ \bibnamefont {Hanson}},\
  }\bibfield  {title} {\enquote {\bibinfo {title} {Coupled
  vibration-dissociation time-histories and rate measurements in shock-heated,
  nondilute $\text{O}_2$ and $\text{O}_2$-{Ar} mixtures from 6000 to 14000
  {K}},}\ }\href {\doibase 10.1063/5.0048059} {\bibfield  {journal} {\bibinfo
  {journal} {Physics of Fluids}\ }\textbf {\bibinfo {volume} {33}},\ \bibinfo
  {pages} {056107} (\bibinfo {year} {2021})}\BibitemShut {NoStop}%
\bibitem [{\citenamefont {Torres}, \citenamefont {Geistfeld},\ and\
  \citenamefont {Schwartzentruber}()}]{Torres_2022_Scitech}%
  \BibitemOpen
  \bibfield  {author} {\bibinfo {author} {\bibfnamefont {E.}~\bibnamefont
  {Torres}}, \bibinfo {author} {\bibfnamefont {E.~C.}\ \bibnamefont
  {Geistfeld}}, \ and\ \bibinfo {author} {\bibfnamefont {T.~E.}\ \bibnamefont
  {Schwartzentruber}},\ }\enquote {\bibinfo {title} {Direct molecular
  simulation of rovibrational relaxation and chemical reactions in air
  mixtures},}\ in\ \href {\doibase 10.2514/6.2022-1010} {\emph {\bibinfo
  {booktitle} {AIAA SCITECH 2022 Forum}}},\ \Eprint
  {http://arxiv.org/abs/https://arc.aiaa.org/doi/pdf/10.2514/6.2022-1010}
  {https://arc.aiaa.org/doi/pdf/10.2514/6.2022-1010} \BibitemShut {NoStop}%
\bibitem [{\citenamefont {Andrienko}\ and\ \citenamefont
  {Boyd}(2018)}]{Andrienko_O2N2}%
  \BibitemOpen
  \bibfield  {author} {\bibinfo {author} {\bibfnamefont {D.~A.}\ \bibnamefont
  {Andrienko}}\ and\ \bibinfo {author} {\bibfnamefont {I.~D.}\ \bibnamefont
  {Boyd}},\ }\bibfield  {title} {\enquote {\bibinfo {title} {Vibrational energy
  transfer and dissociation in o2–n2 collisions at hyperthermal
  temperatures},}\ }\href {\doibase 10.1063/1.5007069} {\bibfield  {journal}
  {\bibinfo  {journal} {The Journal of Chemical Physics}\ }\textbf {\bibinfo
  {volume} {148}},\ \bibinfo {pages} {084309} (\bibinfo {year} {2018})},\
  \Eprint {http://arxiv.org/abs/https://doi.org/10.1063/1.5007069}
  {https://doi.org/10.1063/1.5007069} \BibitemShut {NoStop}%
\bibitem [{\citenamefont {Macdonald}\ \emph
  {et~al.}(2018{\natexlab{a}})\citenamefont {Macdonald}, \citenamefont {Jaffe},
  \citenamefont {Schwenke},\ and\ \citenamefont {Panesi}}]{MACDONALD_MEQCT}%
  \BibitemOpen
  \bibfield  {author} {\bibinfo {author} {\bibfnamefont {R.~L.}\ \bibnamefont
  {Macdonald}}, \bibinfo {author} {\bibfnamefont {R.~L.}\ \bibnamefont
  {Jaffe}}, \bibinfo {author} {\bibfnamefont {D.~W.}\ \bibnamefont {Schwenke}},
  \ and\ \bibinfo {author} {\bibfnamefont {M.}~\bibnamefont {Panesi}},\
  }\bibfield  {title} {\enquote {\bibinfo {title} {Construction of a
  coarse-grain quasi-classical trajectory method. $\text{I}$. {T}heory and
  application to $\text{N}_2$–$\text{N}_2$ system},}\ }\href {\doibase
  10.1063/1.5011331} {\bibfield  {journal} {\bibinfo  {journal} {Journal of
  Chemical Physics}\ }\textbf {\bibinfo {volume} {148}},\ \bibinfo {pages}
  {054309} (\bibinfo {year} {2018}{\natexlab{a}})}\BibitemShut {NoStop}%
\bibitem [{\citenamefont {Ibraguimova}\ \emph {et~al.}(2013)\citenamefont
  {Ibraguimova}, \citenamefont {Sergievskaya}, \citenamefont {Levashov},
  \citenamefont {Shatalov}, \citenamefont {Tunik},\ and\ \citenamefont
  {Zabelinskii}}]{Ibraguimova_2013}%
  \BibitemOpen
  \bibfield  {author} {\bibinfo {author} {\bibfnamefont {L.~B.}\ \bibnamefont
  {Ibraguimova}}, \bibinfo {author} {\bibfnamefont {A.~L.}\ \bibnamefont
  {Sergievskaya}}, \bibinfo {author} {\bibfnamefont {V.~Y.}\ \bibnamefont
  {Levashov}}, \bibinfo {author} {\bibfnamefont {O.~P.}\ \bibnamefont
  {Shatalov}}, \bibinfo {author} {\bibfnamefont {Y.~V.}\ \bibnamefont {Tunik}},
  \ and\ \bibinfo {author} {\bibfnamefont {I.~E.}\ \bibnamefont
  {Zabelinskii}},\ }\bibfield  {title} {\enquote {\bibinfo {title}
  {Investigation of oxygen dissociation and vibrational relaxation at
  temperatures 4000–10\,800 k},}\ }\href {\doibase 10.1063/1.4813070}
  {\bibfield  {journal} {\bibinfo  {journal} {Journal of Chemical Physics}\
  }\textbf {\bibinfo {volume} {139}},\ \bibinfo {pages} {034317} (\bibinfo
  {year} {2013})}\BibitemShut {NoStop}%
\bibitem [{\citenamefont {{Lino da Silva}}, \citenamefont {Loureiro},\ and\
  \citenamefont {Guerra}(2012)}]{LINODASILVA201228}%
  \BibitemOpen
  \bibfield  {author} {\bibinfo {author} {\bibfnamefont {M.}~\bibnamefont
  {{Lino da Silva}}}, \bibinfo {author} {\bibfnamefont {J.}~\bibnamefont
  {Loureiro}}, \ and\ \bibinfo {author} {\bibfnamefont {V.}~\bibnamefont
  {Guerra}},\ }\bibfield  {title} {\enquote {\bibinfo {title} {A multiquantum
  dataset for vibrational excitation and dissociation in high-temperature
  $\text{O}_2$–$\text{O}_2$ collisions},}\ }\href {\doibase
  https://doi.org/10.1016/j.cplett.2012.01.074} {\bibfield  {journal} {\bibinfo
   {journal} {Chemical Physics Letters}\ }\textbf {\bibinfo {volume} {531}},\
  \bibinfo {pages} {28--33} (\bibinfo {year} {2012})}\BibitemShut {NoStop}%
\bibitem [{\citenamefont {Andrienko}\ and\ \citenamefont
  {Boyd}(2017)}]{ANDRIENKO201774}%
  \BibitemOpen
  \bibfield  {author} {\bibinfo {author} {\bibfnamefont {D.~A.}\ \bibnamefont
  {Andrienko}}\ and\ \bibinfo {author} {\bibfnamefont {I.~D.}\ \bibnamefont
  {Boyd}},\ }\bibfield  {title} {\enquote {\bibinfo {title} {State-specific
  dissociation in $\text{O}_2$–$\text{O}_2$ collisions by quasiclassical
  trajectory method},}\ }\href {\doibase
  https://doi.org/10.1016/j.chemphys.2017.05.005} {\bibfield  {journal}
  {\bibinfo  {journal} {Chemical Physics}\ }\textbf {\bibinfo {volume} {491}},\
  \bibinfo {pages} {74--81} (\bibinfo {year} {2017})}\BibitemShut {NoStop}%
\bibitem [{\citenamefont {Varandas}\ and\ \citenamefont
  {Pais}(1991)}]{Varandas_O4_Chapter}%
  \BibitemOpen
  \bibfield  {author} {\bibinfo {author} {\bibfnamefont {A.~J.~C.}\
  \bibnamefont {Varandas}}\ and\ \bibinfo {author} {\bibfnamefont {A.~A.
  C.~C.}\ \bibnamefont {Pais}},\ }\enquote {\bibinfo {title} {Double many-body
  expansion potential energy surface for $\text{O}_4$(${}^3\text{A}$), dynamics
  of the $\text{O}$($^3\text{P}$) + $\text{O}_3$($^1\text{A}_1$) reaction, and
  second virial coefficients of molecular oxygen},}\ in\ \href {\doibase
  10.1007/978-94-011-3584-9_4} {\emph {\bibinfo {booktitle} {Theoretical and
  Computational Models for Organic Chemistry}}},\ \bibinfo {editor} {edited by\
  \bibinfo {editor} {\bibfnamefont {S.~J.}\ \bibnamefont {Formosinho}},
  \bibinfo {editor} {\bibfnamefont {I.~G.}\ \bibnamefont {Csizmadia}}, \ and\
  \bibinfo {editor} {\bibfnamefont {L.~G.}\ \bibnamefont {Arnaut}}}\ (\bibinfo
  {publisher} {Springer},\ \bibinfo {address} {Dordrecht},\ \bibinfo {year}
  {1991})\ pp.\ \bibinfo {pages} {55--78}\BibitemShut {NoStop}%
\bibitem [{\citenamefont {Paukku}\ \emph {et~al.}(2017)\citenamefont {Paukku},
  \citenamefont {Yang}, \citenamefont {Varga}, \citenamefont {Song},
  \citenamefont {Bender},\ and\ \citenamefont {Truhlar}}]{Paukku_O4_1}%
  \BibitemOpen
  \bibfield  {author} {\bibinfo {author} {\bibfnamefont {Y.}~\bibnamefont
  {Paukku}}, \bibinfo {author} {\bibfnamefont {K.~R.}\ \bibnamefont {Yang}},
  \bibinfo {author} {\bibfnamefont {Z.}~\bibnamefont {Varga}}, \bibinfo
  {author} {\bibfnamefont {G.}~\bibnamefont {Song}}, \bibinfo {author}
  {\bibfnamefont {J.~D.}\ \bibnamefont {Bender}}, \ and\ \bibinfo {author}
  {\bibfnamefont {D.~G.}\ \bibnamefont {Truhlar}},\ }\bibfield  {title}
  {\enquote {\bibinfo {title} {Potential energy surfaces of quintet and singlet
  $\text{O}_4$},}\ }\href {\doibase 10.1063/1.4993624} {\bibfield  {journal}
  {\bibinfo  {journal} {Journal of Chemical Physics}\ }\textbf {\bibinfo
  {volume} {147}},\ \bibinfo {pages} {034301} (\bibinfo {year}
  {2017})}\BibitemShut {NoStop}%
\bibitem [{\citenamefont {Paukku}, \citenamefont {Varga},\ and\ \citenamefont
  {Truhlar}(2018)}]{Paukku_O4_2}%
  \BibitemOpen
  \bibfield  {author} {\bibinfo {author} {\bibfnamefont {Y.}~\bibnamefont
  {Paukku}}, \bibinfo {author} {\bibfnamefont {Z.}~\bibnamefont {Varga}}, \
  and\ \bibinfo {author} {\bibfnamefont {D.~G.}\ \bibnamefont {Truhlar}},\
  }\bibfield  {title} {\enquote {\bibinfo {title} {Potential energy surface of
  triplet $\text{O}_4$},}\ }\href {\doibase 10.1063/1.5017489} {\bibfield
  {journal} {\bibinfo  {journal} {Journal of Chemical Physics}\ }\textbf
  {\bibinfo {volume} {148}},\ \bibinfo {pages} {124314} (\bibinfo {year}
  {2018})}\BibitemShut {NoStop}%
\bibitem [{\citenamefont {Jo}, \citenamefont {Panesi},\ and\ \citenamefont
  {Kim}(2021)}]{Jo_2021}%
  \BibitemOpen
  \bibfield  {author} {\bibinfo {author} {\bibfnamefont {S.~M.}\ \bibnamefont
  {Jo}}, \bibinfo {author} {\bibfnamefont {M.}~\bibnamefont {Panesi}}, \ and\
  \bibinfo {author} {\bibfnamefont {J.~G.}\ \bibnamefont {Kim}},\ }\bibfield
  {title} {\enquote {\bibinfo {title} {Prediction of shock standoff distance
  with modified rotational relaxation time of air mixture},}\ }\href {\doibase
  10.1063/5.0045842} {\bibfield  {journal} {\bibinfo  {journal} {Physics of
  Fluids}\ }\textbf {\bibinfo {volume} {33}},\ \bibinfo {pages} {047102}
  (\bibinfo {year} {2021})}\BibitemShut {NoStop}%
\bibitem [{\citenamefont {Sahai}\ \emph {et~al.}(2017)\citenamefont {Sahai},
  \citenamefont {Lopez}, \citenamefont {Johnston},\ and\ \citenamefont
  {Panesi}}]{SAHAI_ADAPTIVE}%
  \BibitemOpen
  \bibfield  {author} {\bibinfo {author} {\bibfnamefont {A.}~\bibnamefont
  {Sahai}}, \bibinfo {author} {\bibfnamefont {B.}~\bibnamefont {Lopez}},
  \bibinfo {author} {\bibfnamefont {C.~O.}\ \bibnamefont {Johnston}}, \ and\
  \bibinfo {author} {\bibfnamefont {M.}~\bibnamefont {Panesi}},\ }\bibfield
  {title} {\enquote {\bibinfo {title} {Adaptive coarse graining method for
  energy transfer and dissociation kinetics of polyatomic species},}\ }\href
  {\doibase 10.1063/1.4996654} {\bibfield  {journal} {\bibinfo  {journal}
  {Journal of Chemical Physics}\ }\textbf {\bibinfo {volume} {147}},\ \bibinfo
  {pages} {054107} (\bibinfo {year} {2017})}\BibitemShut {NoStop}%
\bibitem [{\citenamefont {Kim}\ and\ \citenamefont
  {Boyd}(2012)}]{Kim_POF_2012}%
  \BibitemOpen
  \bibfield  {author} {\bibinfo {author} {\bibfnamefont {J.~G.}\ \bibnamefont
  {Kim}}\ and\ \bibinfo {author} {\bibfnamefont {I.~D.}\ \bibnamefont {Boyd}},\
  }\bibfield  {title} {\enquote {\bibinfo {title} {State-resolved
  thermochemical nonequilibrium analysis of hydrogen mixture flows},}\ }\href
  {\doibase 10.1063/1.4747340} {\bibfield  {journal} {\bibinfo  {journal}
  {Physics of Fluids}\ }\textbf {\bibinfo {volume} {24}},\ \bibinfo {pages}
  {086102} (\bibinfo {year} {2012})}\BibitemShut {NoStop}%
\bibitem [{\citenamefont {Kim}, \citenamefont {Kwon},\ and\ \citenamefont
  {Park}(2010)}]{Kim_JTHT_2010}%
  \BibitemOpen
  \bibfield  {author} {\bibinfo {author} {\bibfnamefont {J.~G.}\ \bibnamefont
  {Kim}}, \bibinfo {author} {\bibfnamefont {O.~J.}\ \bibnamefont {Kwon}}, \
  and\ \bibinfo {author} {\bibfnamefont {C.}~\bibnamefont {Park}},\ }\bibfield
  {title} {\enquote {\bibinfo {title} {Master equation study and nonequilibrium
  chemical reactions for hydrogen molecule},}\ }\href {\doibase
  10.2514/1.45283} {\bibfield  {journal} {\bibinfo  {journal} {Journal of
  Thermophysics and Heat Transfer}\ }\textbf {\bibinfo {volume} {24}},\
  \bibinfo {pages} {281--290} (\bibinfo {year} {2010})}\BibitemShut {NoStop}%
\bibitem [{\citenamefont {Baluckram}\ and\ \citenamefont
  {Andrienko}()}]{Andrienko_2021}%
  \BibitemOpen
  \bibfield  {author} {\bibinfo {author} {\bibfnamefont {V.~T.}\ \bibnamefont
  {Baluckram}}\ and\ \bibinfo {author} {\bibfnamefont {D.}~\bibnamefont
  {Andrienko}},\ }\enquote {\bibinfo {title} {First-principle simulation of
  vibrational activation and dissociation in oxygen shock flows},}\ in\ \href
  {\doibase 10.2514/6.2021-0447} {\emph {\bibinfo {booktitle} {AIAA Scitech
  2021 Forum}}}\BibitemShut {NoStop}%
\bibitem [{\citenamefont {Munaf\`{o}}\ \emph {et~al.}(2012)\citenamefont
  {Munaf\`{o}}, \citenamefont {Panesi}, \citenamefont {Jaffe}, \citenamefont
  {Colonna}, \citenamefont {Bourdon},\ and\ \citenamefont
  {Magin}}]{Munafo_EPJD_2012}%
  \BibitemOpen
  \bibfield  {author} {\bibinfo {author} {\bibfnamefont {A.}~\bibnamefont
  {Munaf\`{o}}}, \bibinfo {author} {\bibfnamefont {M.}~\bibnamefont {Panesi}},
  \bibinfo {author} {\bibfnamefont {R.~L.}\ \bibnamefont {Jaffe}}, \bibinfo
  {author} {\bibfnamefont {G.}~\bibnamefont {Colonna}}, \bibinfo {author}
  {\bibfnamefont {A.}~\bibnamefont {Bourdon}}, \ and\ \bibinfo {author}
  {\bibfnamefont {T.~E.}\ \bibnamefont {Magin}},\ }\bibfield  {title} {\enquote
  {\bibinfo {title} {{QCT}-based vibrational collisional models applied to
  nonequilibrium nozzle flows},}\ }\href {\doibase 10.1140/epjd/e2012-30079-3}
  {\bibfield  {journal} {\bibinfo  {journal} {European Physical Journal D}\
  }\textbf {\bibinfo {volume} {66}},\ \bibinfo {pages} {188} (\bibinfo {year}
  {2012})}\BibitemShut {NoStop}%
\bibitem [{\citenamefont {Sharma}, \citenamefont {Liu},\ and\ \citenamefont
  {Panesi}(2020)}]{Sharma_2020}%
  \BibitemOpen
  \bibfield  {author} {\bibinfo {author} {\bibfnamefont {M.~P.}\ \bibnamefont
  {Sharma}}, \bibinfo {author} {\bibfnamefont {Y.}~\bibnamefont {Liu}}, \ and\
  \bibinfo {author} {\bibfnamefont {M.}~\bibnamefont {Panesi}},\ }\bibfield
  {title} {\enquote {\bibinfo {title} {Coarse-grained modeling of
  thermochemical nonequilibrium using the multigroup maximum entropy quadratic
  formulation},}\ }\href {\doibase 10.1103/PhysRevE.101.013307} {\bibfield
  {journal} {\bibinfo  {journal} {Physical Review E}\ }\textbf {\bibinfo
  {volume} {101}},\ \bibinfo {pages} {013307} (\bibinfo {year}
  {2020})}\BibitemShut {NoStop}%
\bibitem [{\citenamefont {Sahai}\ \emph {et~al.}(2019)\citenamefont {Sahai},
  \citenamefont {Johnston}, \citenamefont {Lopez},\ and\ \citenamefont
  {Panesi}}]{Sahai_2019_PRF}%
  \BibitemOpen
  \bibfield  {author} {\bibinfo {author} {\bibfnamefont {A.}~\bibnamefont
  {Sahai}}, \bibinfo {author} {\bibfnamefont {C.~O.}\ \bibnamefont {Johnston}},
  \bibinfo {author} {\bibfnamefont {B.}~\bibnamefont {Lopez}}, \ and\ \bibinfo
  {author} {\bibfnamefont {M.}~\bibnamefont {Panesi}},\ }\bibfield  {title}
  {\enquote {\bibinfo {title} {Flow-radiation coupling in {CO}${}_{2}$
  hypersonic wakes using reduced-order non-boltzmann models},}\ }\href
  {\doibase 10.1103/PhysRevFluids.4.093401} {\bibfield  {journal} {\bibinfo
  {journal} {Physical Review Fluids}\ }\textbf {\bibinfo {volume} {4}},\
  \bibinfo {pages} {093401} (\bibinfo {year} {2019})}\BibitemShut {NoStop}%
\bibitem [{\citenamefont {Munafò}\ \emph {et~al.}(2020)\citenamefont
  {Munafò}, \citenamefont {Alberti}, \citenamefont {Pantano}, \citenamefont
  {Freund},\ and\ \citenamefont {Panesi}}]{MUNAFO_HEGEL}%
  \BibitemOpen
  \bibfield  {author} {\bibinfo {author} {\bibfnamefont {A.}~\bibnamefont
  {Munafò}}, \bibinfo {author} {\bibfnamefont {A.}~\bibnamefont {Alberti}},
  \bibinfo {author} {\bibfnamefont {C.}~\bibnamefont {Pantano}}, \bibinfo
  {author} {\bibfnamefont {J.~B.}\ \bibnamefont {Freund}}, \ and\ \bibinfo
  {author} {\bibfnamefont {M.}~\bibnamefont {Panesi}},\ }\bibfield  {title}
  {\enquote {\bibinfo {title} {A computational model for nanosecond pulse
  laser-plasma interactions},}\ }\href {\doibase
  https://doi.org/10.1016/j.jcp.2019.109190} {\bibfield  {journal} {\bibinfo
  {journal} {Journal of Computational Physics}\ }\textbf {\bibinfo {volume}
  {406}},\ \bibinfo {pages} {109190} (\bibinfo {year} {2020})}\BibitemShut
  {NoStop}%
\bibitem [{\citenamefont {Venturi}, \citenamefont {Jaffe},\ and\ \citenamefont
  {Panesi}(2020)}]{Venturi2020_ML}%
  \BibitemOpen
  \bibfield  {author} {\bibinfo {author} {\bibfnamefont {S.}~\bibnamefont
  {Venturi}}, \bibinfo {author} {\bibfnamefont {R.~L.}\ \bibnamefont {Jaffe}},
  \ and\ \bibinfo {author} {\bibfnamefont {M.}~\bibnamefont {Panesi}},\
  }\bibfield  {title} {\enquote {\bibinfo {title} {Bayesian machine learning
  approach to the quantification of uncertainties on ab initio potential energy
  surfaces},}\ }\href {\doibase 10.1021/acs.jpca.0c02395} {\bibfield  {journal}
  {\bibinfo  {journal} {Journal of Physical Chemistry A}\ }\textbf {\bibinfo
  {volume} {124}},\ \bibinfo {pages} {5129--5146} (\bibinfo {year}
  {2020})}\BibitemShut {NoStop}%
\bibitem [{\citenamefont {Schwenke}(1988)}]{SCHWENKE_VVTC_1988}%
  \BibitemOpen
  \bibfield  {author} {\bibinfo {author} {\bibfnamefont {D.~W.}\ \bibnamefont
  {Schwenke}},\ }\bibfield  {title} {\enquote {\bibinfo {title} {Calculations
  of rate constants for three-body recombination of $\rm {H}_2$ in the presence
  of $\rm {H}_2$},}\ }\href {\doibase 10.1063/1.455104} {\bibfield  {journal}
  {\bibinfo  {journal} {Journal of Chemical Physics}\ }\textbf {\bibinfo
  {volume} {89}},\ \bibinfo {pages} {2076--2091} (\bibinfo {year}
  {1988})}\BibitemShut {NoStop}%
\bibitem [{\citenamefont {Truhlar}\ and\ \citenamefont
  {Muckerman}(1979)}]{Truhlar1979}%
  \BibitemOpen
  \bibfield  {author} {\bibinfo {author} {\bibfnamefont {D.~G.}\ \bibnamefont
  {Truhlar}}\ and\ \bibinfo {author} {\bibfnamefont {J.~T.}\ \bibnamefont
  {Muckerman}},\ }\bibfield  {title} {\enquote {\bibinfo {title} {Reactive
  {S}cattering {C}ross {S}ections {III}: {Q}uasiclassical and {S}emiclassical
  {M}ethods},}\ }in\ \href {\doibase 10.1007/978-1-4613-2913-8_16} {\emph
  {\bibinfo {booktitle} {Atom-molecule Collision Theory. {A} Guide for the
  Experimentalist}}},\ \bibinfo {editor} {edited by\ \bibinfo {editor}
  {\bibfnamefont {R.~B.}\ \bibnamefont {Bernstein}}}\ (\bibinfo  {publisher}
  {Springer US},\ \bibinfo {year} {1979})\ pp.\ \bibinfo {pages}
  {505--566}\BibitemShut {NoStop}%
\bibitem [{\citenamefont {Park}(1990)}]{Park_book}%
  \BibitemOpen
  \bibfield  {author} {\bibinfo {author} {\bibfnamefont {C.}~\bibnamefont
  {Park}},\ }\href@noop {} {\emph {\bibinfo {title} {Nonequilibrium Hypersonic
  Aerothermodynamics}}}\ (\bibinfo  {publisher} {Wiley},\ \bibinfo {address}
  {New York, NY},\ \bibinfo {year} {1990})\BibitemShut {NoStop}%
\bibitem [{\citenamefont {Landau}\ and\ \citenamefont
  {Teller}(1936)}]{LT_1936}%
  \BibitemOpen
  \bibfield  {author} {\bibinfo {author} {\bibfnamefont {L.}~\bibnamefont
  {Landau}}\ and\ \bibinfo {author} {\bibfnamefont {E.}~\bibnamefont
  {Teller}},\ }\bibfield  {title} {\enquote {\bibinfo {title} {Theory of sound
  dispersion},}\ }\href@noop {} {\bibfield  {journal} {\bibinfo  {journal}
  {Phys. Z Sowjetunion}\ }\textbf {\bibinfo {volume} {10}},\ \bibinfo {pages}
  {34--43} (\bibinfo {year} {1936})},\ \bibinfo {note} {in German}\BibitemShut
  {NoStop}%
\bibitem [{\citenamefont {Macdonald}\ \emph
  {et~al.}(2018{\natexlab{b}})\citenamefont {Macdonald}, \citenamefont
  {Grover}, \citenamefont {Schwartzentruber},\ and\ \citenamefont
  {Panesi}}]{MACDONALD_MEQCT_Comp}%
  \BibitemOpen
  \bibfield  {author} {\bibinfo {author} {\bibfnamefont {R.~L.}\ \bibnamefont
  {Macdonald}}, \bibinfo {author} {\bibfnamefont {M.~S.}\ \bibnamefont
  {Grover}}, \bibinfo {author} {\bibfnamefont {T.~E.}\ \bibnamefont
  {Schwartzentruber}}, \ and\ \bibinfo {author} {\bibfnamefont
  {M.}~\bibnamefont {Panesi}},\ }\bibfield  {title} {\enquote {\bibinfo {title}
  {Construction of a coarse-grain quasi-classical trajectory method.
  $\text{II}$. {C}omparison against the direct molecular simulation method},}\
  }\href {\doibase 10.1063/1.5011332} {\bibfield  {journal} {\bibinfo
  {journal} {Journal of Chemical Physics}\ }\textbf {\bibinfo {volume} {148}},\
  \bibinfo {pages} {054310} (\bibinfo {year} {2018}{\natexlab{b}})}\BibitemShut
  {NoStop}%
\bibitem [{\citenamefont {Macdonald}\ \emph {et~al.}(2020)\citenamefont
  {Macdonald}, \citenamefont {Torres}, \citenamefont {Schwartzentruber},\ and\
  \citenamefont {Panesi}}]{Macdonald_2020}%
  \BibitemOpen
  \bibfield  {author} {\bibinfo {author} {\bibfnamefont {R.~L.}\ \bibnamefont
  {Macdonald}}, \bibinfo {author} {\bibfnamefont {E.}~\bibnamefont {Torres}},
  \bibinfo {author} {\bibfnamefont {T.~E.}\ \bibnamefont {Schwartzentruber}}, \
  and\ \bibinfo {author} {\bibfnamefont {M.}~\bibnamefont {Panesi}},\
  }\bibfield  {title} {\enquote {\bibinfo {title} {State-to-state master
  equation and direct molecular simulation study of energy transfer and
  dissociation for the $\text{N}_2$-{N} system},}\ }\href {\doibase
  10.1021/acs.jpca.0c04029} {\bibfield  {journal} {\bibinfo  {journal} {Journal
  of Physical Chemistry A}\ }\textbf {\bibinfo {volume} {124}},\ \bibinfo
  {pages} {6986--7000} (\bibinfo {year} {2020})}\BibitemShut {NoStop}%
\bibitem [{\citenamefont {Liu}\ \emph {et~al.}(2015)\citenamefont {Liu},
  \citenamefont {Panesi}, \citenamefont {Sahai},\ and\ \citenamefont
  {Vinokur}}]{Yen_2015}%
  \BibitemOpen
  \bibfield  {author} {\bibinfo {author} {\bibfnamefont {Y.}~\bibnamefont
  {Liu}}, \bibinfo {author} {\bibfnamefont {M.}~\bibnamefont {Panesi}},
  \bibinfo {author} {\bibfnamefont {A.}~\bibnamefont {Sahai}}, \ and\ \bibinfo
  {author} {\bibfnamefont {M.}~\bibnamefont {Vinokur}},\ }\bibfield  {title}
  {\enquote {\bibinfo {title} {General multi-group macroscopic modeling for
  thermo-chemical non-equilibrium gas mixtures},}\ }\href@noop {} {\bibfield
  {journal} {\bibinfo  {journal} {Journal of Chemical Physics}\ }\textbf
  {\bibinfo {volume} {142}},\ \bibinfo {pages} {134109} (\bibinfo {year}
  {2015})}\BibitemShut {NoStop}%
\bibitem [{\citenamefont {Losev}\ and\ \citenamefont
  {Generalov}(1962)}]{losev1962study}%
  \BibitemOpen
  \bibfield  {author} {\bibinfo {author} {\bibfnamefont {S.~A.}\ \bibnamefont
  {Losev}}\ and\ \bibinfo {author} {\bibfnamefont {N.~A.}\ \bibnamefont
  {Generalov}},\ }\bibfield  {title} {\enquote {\bibinfo {title} {A study of
  the excitation of vibrations and dissociation of oxygen molecules at high
  temperatures},}\ }in\ \href@noop {} {\emph {\bibinfo {booktitle} {Soviet
  Physics Doklady}}},\ Vol.~\bibinfo {volume} {6}\ (\bibinfo {year} {1962})\
  p.\ \bibinfo {pages} {1081}\BibitemShut {NoStop}%
\bibitem [{\citenamefont {Parker}(1959)}]{Parker_1959}%
  \BibitemOpen
  \bibfield  {author} {\bibinfo {author} {\bibfnamefont {J.~G.}\ \bibnamefont
  {Parker}},\ }\bibfield  {title} {\enquote {\bibinfo {title} {{R}otational and
  vibrational relaxation in diatomic gases},}\ }\href {\doibase
  10.1063/1.1724417} {\bibfield  {journal} {\bibinfo  {journal} {Physics of
  Fluids}\ }\textbf {\bibinfo {volume} {2}},\ \bibinfo {pages} {449--462}
  (\bibinfo {year} {1959})}\BibitemShut {NoStop}%
\bibitem [{\citenamefont {Millikan}\ and\ \citenamefont
  {White}(1963)}]{MW_1963}%
  \BibitemOpen
  \bibfield  {author} {\bibinfo {author} {\bibfnamefont {R.~C.}\ \bibnamefont
  {Millikan}}\ and\ \bibinfo {author} {\bibfnamefont {D.~R.}\ \bibnamefont
  {White}},\ }\bibfield  {title} {\enquote {\bibinfo {title} {{S}ystematics of
  vibrational relaxation},}\ }\href {\doibase 10.1063/1.1734182} {\bibfield
  {journal} {\bibinfo  {journal} {Journal of Chemical Physics}\ }\textbf
  {\bibinfo {volume} {39}},\ \bibinfo {pages} {3209--3214} (\bibinfo {year}
  {1963})}\BibitemShut {NoStop}%
\bibitem [{\citenamefont {Park}(1993)}]{Park_1993_Earth}%
  \BibitemOpen
  \bibfield  {author} {\bibinfo {author} {\bibfnamefont {C.}~\bibnamefont
  {Park}},\ }\bibfield  {title} {\enquote {\bibinfo {title} {Review of
  chemical-kinetic problems of future {NASA} missions, {I}: {E}arth entries},}\
  }\href {\doibase 10.2514/3.431} {\bibfield  {journal} {\bibinfo  {journal}
  {Journal of Thermophysics and Heat Transfer}\ }\textbf {\bibinfo {volume}
  {7}},\ \bibinfo {pages} {385--398} (\bibinfo {year} {1993})}\BibitemShut
  {NoStop}%
\bibitem [{\citenamefont {Kim}, \citenamefont {Kang},\ and\ \citenamefont
  {Park}(2020)}]{Kim_2020}%
  \BibitemOpen
  \bibfield  {author} {\bibinfo {author} {\bibfnamefont {J.~G.}\ \bibnamefont
  {Kim}}, \bibinfo {author} {\bibfnamefont {S.~H.}\ \bibnamefont {Kang}}, \
  and\ \bibinfo {author} {\bibfnamefont {S.~H.}\ \bibnamefont {Park}},\
  }\bibfield  {title} {\enquote {\bibinfo {title} {Thermochemical
  nonequilibrium modeling of oxygen in hypersonic air flows},}\ }\href
  {\doibase 10.1016/j.ijheatmasstransfer.2019.119059} {\bibfield  {journal}
  {\bibinfo  {journal} {International Journal of Heat and Mass Transfer}\
  }\textbf {\bibinfo {volume} {148}},\ \bibinfo {pages} {119059} (\bibinfo
  {year} {2020})}\BibitemShut {NoStop}%
\bibitem [{\citenamefont {Kim}\ and\ \citenamefont {Jo}(2021)}]{Kim_2021}%
  \BibitemOpen
  \bibfield  {author} {\bibinfo {author} {\bibfnamefont {J.~G.}\ \bibnamefont
  {Kim}}\ and\ \bibinfo {author} {\bibfnamefont {S.~M.}\ \bibnamefont {Jo}},\
  }\bibfield  {title} {\enquote {\bibinfo {title} {Modification of
  chemical-kinetic parameters for 11-air species in re-entry flows},}\ }\href
  {\doibase 10.1016/j.ijheatmasstransfer.2021.120950} {\bibfield  {journal}
  {\bibinfo  {journal} {International Journal of Heat and Mass Transfer}\
  }\textbf {\bibinfo {volume} {169}},\ \bibinfo {pages} {120950} (\bibinfo
  {year} {2021})}\BibitemShut {NoStop}%
\bibitem [{\citenamefont {Kim}\ and\ \citenamefont
  {Boyd}(2013)}]{Kim_Boyd_Cphys_2013}%
  \BibitemOpen
  \bibfield  {author} {\bibinfo {author} {\bibfnamefont {J.~G.}\ \bibnamefont
  {Kim}}\ and\ \bibinfo {author} {\bibfnamefont {I.~D.}\ \bibnamefont {Boyd}},\
  }\bibfield  {title} {\enquote {\bibinfo {title} {State-resolved master
  equation analysis of thermochemical nonequilibrium of nitrogen},}\ }\href
  {\doibase 10.1016/j.chemphys.2013.01.027} {\bibfield  {journal} {\bibinfo
  {journal} {Chemical Physics}\ }\textbf {\bibinfo {volume} {415}},\ \bibinfo
  {pages} {237--246} (\bibinfo {year} {2013})}\BibitemShut {NoStop}%
\bibitem [{\citenamefont {Venturi}\ \emph {et~al.}()\citenamefont {Venturi},
  \citenamefont {Sharma~Priyadarshini}, \citenamefont {Racca},\ and\
  \citenamefont {Panesi}}]{VENTURI_JUN2019}%
  \BibitemOpen
  \bibfield  {author} {\bibinfo {author} {\bibfnamefont {S.}~\bibnamefont
  {Venturi}}, \bibinfo {author} {\bibfnamefont {M.}~\bibnamefont
  {Sharma~Priyadarshini}}, \bibinfo {author} {\bibfnamefont {A.}~\bibnamefont
  {Racca}}, \ and\ \bibinfo {author} {\bibfnamefont {M.}~\bibnamefont
  {Panesi}},\ }\enquote {\bibinfo {title} {Effects of ab-initio potential
  energy surfaces on $\text{O}_2$-{O} non-equilibrium kinetics},}\ in\ \href
  {\doibase 10.2514/6.2019-3358} {\emph {\bibinfo {booktitle} {AIAA Aviation
  2019 Forum}}}\BibitemShut {NoStop}%
\bibitem [{\citenamefont {Schexnayder}\ and\ \citenamefont
  {Evans}(1961)}]{Schexnayder_1961}%
  \BibitemOpen
  \bibfield  {author} {\bibinfo {author} {\bibfnamefont {C.~J.}\ \bibnamefont
  {Schexnayder}}\ and\ \bibinfo {author} {\bibfnamefont {J.~S.}\ \bibnamefont
  {Evans}},\ }\href@noop {} {\enquote {\bibinfo {title} {Measurements of the
  dissociation rate of molecular oxygen},}\ }\bibinfo {type} {NASA TR-R-108}\
  (\bibinfo {year} {1961})\BibitemShut {NoStop}%
\bibitem [{\citenamefont {Camac}\ and\ \citenamefont
  {Vaughan}(1961)}]{Camac_1961}%
  \BibitemOpen
  \bibfield  {author} {\bibinfo {author} {\bibfnamefont {M.}~\bibnamefont
  {Camac}}\ and\ \bibinfo {author} {\bibfnamefont {A.}~\bibnamefont
  {Vaughan}},\ }\bibfield  {title} {\enquote {\bibinfo {title} {$\text{O}_2$
  dissociation rates $\text{O}_2$‐{Ar} mixtures},}\ }\href {\doibase
  10.1063/1.4757209} {\bibfield  {journal} {\bibinfo  {journal} {Journal of
  Chemical Physics}\ }\textbf {\bibinfo {volume} {34}},\ \bibinfo {pages}
  {460--470} (\bibinfo {year} {1961})}\BibitemShut {NoStop}%
\bibitem [{\citenamefont {Shatalov}(1973)}]{Shatalov_O2}%
  \BibitemOpen
  \bibfield  {author} {\bibinfo {author} {\bibfnamefont {O.}~\bibnamefont
  {Shatalov}},\ }\bibfield  {title} {\enquote {\bibinfo {title} {Molecular
  dissociation of oxygen in the absence of vibrational equilibrium},}\ }\href
  {\doibase 10.1007/BF00742888} {\bibfield  {journal} {\bibinfo  {journal}
  {Combustion, Explosion and Shock Waves}\ }\textbf {\bibinfo {volume} {9}},\
  \bibinfo {pages} {610–613} (\bibinfo {year} {1973})}\BibitemShut {NoStop}%
\bibitem [{\citenamefont {Streicher}\ \emph {et~al.}()\citenamefont
  {Streicher}, \citenamefont {Krish}, \citenamefont {Wang}, \citenamefont
  {Davidson},\ and\ \citenamefont {Hanson}}]{Streicher_2019}%
  \BibitemOpen
  \bibfield  {author} {\bibinfo {author} {\bibfnamefont {J.~W.}\ \bibnamefont
  {Streicher}}, \bibinfo {author} {\bibfnamefont {A.}~\bibnamefont {Krish}},
  \bibinfo {author} {\bibfnamefont {S.}~\bibnamefont {Wang}}, \bibinfo {author}
  {\bibfnamefont {D.~F.}\ \bibnamefont {Davidson}}, \ and\ \bibinfo {author}
  {\bibfnamefont {R.~K.}\ \bibnamefont {Hanson}},\ }\enquote {\bibinfo {title}
  {Measurements of oxygen vibrational relaxation and dissociation using
  ultraviolet laser absorption in shock tube experiments},}\ in\ \href
  {\doibase 10.2514/6.2019-0795} {\emph {\bibinfo {booktitle} {AIAA Scitech
  2019 Forum}}}\BibitemShut {NoStop}%
\bibitem [{\citenamefont {Nikitin}, \citenamefont {Kearsley},\ and\
  \citenamefont {Press}(1974)}]{DISS_CORR_FACTOR}%
  \BibitemOpen
  \bibfield  {author} {\bibinfo {author} {\bibfnamefont {E.}~\bibnamefont
  {Nikitin}}, \bibinfo {author} {\bibfnamefont {M.}~\bibnamefont {Kearsley}}, \
  and\ \bibinfo {author} {\bibfnamefont {O.~U.}\ \bibnamefont {Press}},\ }\href
  {https://books.google.com/books?id=3rjQAAAAMAAJ} {\emph {\bibinfo {title}
  {Theory of Elementary Atomic and Molecular Processes in Gases}}}\ (\bibinfo
  {publisher} {Clarendon Press},\ \bibinfo {year} {1974})\BibitemShut {NoStop}%
\end{thebibliography}%

\end{document}